# Features of Two-Quasiparticle Rotational Bands in Deformed Odd-Odd Nuclei, 156≤A≤168

Pinky[1], Sushil Kumar[1*], Sukhjeet Singh[1] and A.K. Jain[1,2]
[1]*Department of Physics, Akal University Talwandi Sabo, Bathinda, Punjab-151302, India*
[2]*AINST, Amity University, Noida- 201313, India*
[*]*email:* sushil.rathi179@gmail.com

**Abstract**

In present work, we evaluated the experimental data pertains to two-quasiparticle (2-qp) rotational structures in deformed odd–odd nuclei having Z = 67–71 and N = 89–97 and A = 156–168 mass region. The compilation includes total 234 rotational bands/states among which 173 are rotational bands and 61 bandhead states. Gallagher-Moszkowski (GM) doublets are identified for 63 two quasiparticle configurations which provide a good testing ground for neutron–proton residual-interaction systematics. The highest excitation energies reach approximately 18 MeV for $\pi i 13/2 \otimes \nu h 9/2$ and $\pi 9/2[514] \otimes \nu 5/2[642]$ configuration observed in $^{164}_{71}Lu$ and in $^{168}_{71}Lu$. The triplet configuration become ground state in case of 22 nuclides ( $^{156-170}_{67}Ho$, $^{156-176}_{69}Tm$ and $^{162-168}_{71}Lu$). The maximum angular momentum populated in $50\hbar$ in case of $\pi 9/2[514] \otimes \nu 5/2[642]$ configuration observed in $^{168}_{71}Lu$. The bandhead values ranges from 0 to 9 which is consistent with strong-coupling expectations for high-j orbitals and indicate the possibility of high-K isomers. Signature splitting is reported for 76 bands and signature inversion is identified in 29 bands which indicate the presence of Coriolis couplings and evolving configuration mixing with spin. The average signature-splitting amplitudes $((|\Delta E|)$ lies between $\approx 12 – 300$ keV. In a given 2-qp configuration in isotopes of Ho or Tb or Dy, the average splitting tends to increase slightly with neutron number. However, in an isotonic chain the trend is even weaker, the splittings remain nearly constant (±10 keV). Band crossings have been observed in case of 10 bands, often accompanying changes in the kinematic and dynamic moments of inertia as paired high-j quasiparticles align with the rotational axis. Half-life information is available for 58 bandhead states (≈28%) of total rotational bands. The reduced-transition-probability ratios B(M1)/B(E2) and effective gyromagnetic-factor differences $g_k − g_r$ remain scares. In total, 137 bands display regular level sequences of energy levels whereas 8 exhibit irregular patterns. The present evaluation of two-quasiparticle rotational bands reveals substantial gaps in the experimental data and highlight the need of fresh measurements to support rigorous calculations and reliable systematics.

# 1. Introduction

Over the past two decades, sophisticated accelerator facilities and high-efficiency of γ-ray detection arrays have greatly improved the quality and statistics of nuclear spectroscopic data. As a result, an extensive level schemes have been established for one-quasiparticle (1-qp), two-quasiparticle (2-qp), three-quasiparticle (3-qp), and multi-quasiparticle excitations in even–even, odd–odd, and odd-A nuclides [1–8] which reveals number of characteristic rotational phenomena, including back-bending [9], bandcrossing [10-12], chiral doublets [13–14], signature effects [15-20], and high-K isomerism [21–25]. The analysis of the experimental data at high angular-momentum states highlights how intrinsic single-particle motion and collective rotation compete each other and evolves with spin. At higher angular momentum, centrifugal and Coriolis forces become more important and thus modify key nuclear properties, notably, the equilibrium shape (e.g., changes in $β_2$ and triaxiality) and the pairing correlations (pair breaking and alignment), the patterns of angular-momentum generation (single-particle alignments vs collective rotor contributions). The Coriolis and centrifugal forces affect key observables such as moments of inertia, level spacings, and electromagnetic transition rates. In present work, we critically review the experimental data of two quasiparticle rotational bands in odd-odd nuclei (Z = 67–71 and N = 89–97) lying in A = 156–168 mass region.

The two-quasiparticle rotational bands in deformed odd–odd nuclei provide a theoretical laboratory for investigation of the interplay between the neutron–proton (n–p) residual interaction and the rotational degrees of freedom of the deformed core. This interplay governs not only the energy ordering but also causes various nuclear structure phenomena such as signature splitting, signature inversion, *K*-dependent anomalies such as the Newby shift in K=0 bands [26] and isomerism.

# 2. Formation of Two-Quasiparticle Rotational Bands: Gallagher & Moszkowski Doublets

In odd-odd axially symmetric nuclei, the projection of proton angular momenta $Ω_p$ and of neutron angular momenta $Ω_n$ coupled in parallel or anti-parallel manner to form high-*K* ($K_+ = |Ω_p + Ω_n|$) or low-*K* ($K_- = |Ω_p - Ω_n|$) state respectively. The energy ordering of these states is governed by the competition of neutron-proton residual *n-p* interaction and rotational term. In the absence of *n-p* interactions, the rotational term pushes the energy state with smaller value of *K* to have lower in energy as compare to the energy state with high value of *K*. However, the magnitude of the *n-p* interactions is generally more than the rotational term and hence the energy ordering among low-*K* and high-*K* states is determined by the magnitude and nature of the *n-p* interactions.

Depending upon the orientations of proton and neutron intrinsic spins, form high-$K$ ($K_+ = |\Omega_p + \Omega_n|$) or low-$K$ ($K_- = |\Omega_p - \Omega_n|$) states are of two types namely singlet (↑↓ or ↓↑) or triplet (↑↑ or ↓↓). In order to fix the energy ordering among singlet and triplet states of deformed odd-odd nuclei, Gallagher & Moszkowski devised an empirical rule, which states that, the triplet energy state always lie lower in energy as compared to its singlet counterpart and energy difference between singlet and triplet states is known as Gallagher Moszkowski (GM) splitting energy and this couple is termed GM doublet [27]. In the analogy with two quasiparticle excitations observed in odd-odd nuclides, Gallagher and Moszkowski also analyzed the empirical data of even-even nuclides and devised one more rule for fixing the energy ordering of singlet and triplet states pertaining to deformed even-even nuclides. According to this rule, a singlet state (↑↓ or ↓↑) always lie lower in energy as compared to triplet (↑↑ or ↓↓) state [28].

**3. Features of Two Quasiparticle Rotational Bands**

3.1 Signature Splitting and Signature Inversion

Signature splitting refers to the energy staggering between the two signature branches of the same rotational band [29-30]. This splitting is a consequence of the signature quantum number arises from the invariance of the nuclear wavefunction under a rotation by angle $\pi$ about an axis perpendicular to the symmetry axis. For even–even or odd-odd nuclei, the rotational band is typically separate into even-I and odd-I sequences with signatures $\alpha=0,1$. In some 2-qp bands, the favored signature at low spin becomes unfavored beyond a critical spin I and the underlaying phenomena is called as signature inversion. Microscopically, signature inversion can be understood with Coriolis coupling (first and higher orders), evolving deformation or triaxiality, n-p interactions and configuration mixing among the nearby bands close in the energy.

In this Table 1, we present the 2qp rotational bands which show signature splitting and sometime signature inversion in odd-odd deformed nuclei in A = 156–168 mass region. The average signature-splitting amplitudes (($|\Delta E|$) lies between $\approx 12-300$ keV. The configurations based on the high-$j$ coupling $1/2[541]\pi \otimes 5/2[642]\nu$ and $\pi 7/2[523] \otimes \nu 5/2[642]$ generally show the largest average splitting (> 100 keV), whereas low-$j$ pairs such as $\pi 7/2[404] \otimes \nu 1/2[521]$ or $\pi 7/2[523] \otimes \nu 5/2[523]$ fall below 25 keV. The inversion point, when observed, ranges from I ≈ 10ℏ up to I ≈ 32ℏ, most frequently around 17–20ℏ. In a given 2-qp configuration for example $\pi 7/2[523] \otimes \nu 5/2[642]$ in isotopes of Ho or Tb or Dy, the average splitting tends to increase slightly with neutron number. However, in an isotonic chain the trend is even weaker, the splittings remain nearly constant (±10 keV). The experimental data is insufficient to have a significant analysis of point of inversion in isotonic and isotopic chain. Concludingly, high-j proton i.e. 7/2[523] coupled with neutron $i_{13/2}$ or $\nu 5/2[642]$ type orbital favor larger staggering and the amplitude of signature splitting amplified with increase in neutron number in given isotopic chain.

**Table 1:** Compilation of 2qp rotational bands with signature splitting and signature inversion

| Nuclei | Configuration | Signature splitting $|\Delta E|$ | Signature inversion |
|---|---|---|---|
| $^{156}_{67}Ho_{89}$ | $\pi h_{11/2} \otimes \nu i_{13/2}$ | 50.96 | 19 ℏ |
| | $\pi 7/2[523] \otimes \nu 1/2[660]$ | 67.81 | 12 ℏ and 19 ℏ |
| | $\pi 7/2[404] \otimes \nu 1/2[660]$ | 92.72 | 20 ℏ |
| | $\pi 7/2[404] \otimes \nu 1/2[660]$ | 39.14 | 23 ℏ |
| $^{158}_{67}Ho_{91}$ | $\pi 7/2[523] \otimes \nu 3/2[521]$ | 70.35 | - |
| | $\pi 7/2[523] \otimes \nu 3/2[402]$ $\pi 7/2[404] \otimes \nu 3/2[521]$ $\pi 7/2[523] \otimes \nu 5/2[642]$ | 26.00 | 14 ℏ |
| | $\pi 7/2[404] \otimes \nu 3/2[651]$ | 12.30 | - |
| | $\pi 7/2[523] \otimes \nu 3/2[651]$ | 20.48 | - |
| $^{160}_{67}Ho_{93}$ | $\pi 7/2[523] \otimes \nu 3/2[521]$ | 34.45 | 20 ℏ |
| | $\pi 7/2[523] \otimes \nu 5/2[642]$ | 44.95 | - |
| | $\pi 7/2[523] \otimes \nu 11/2[505]$ | 11.00 | - |
| | $\pi 7/2[404] \otimes \nu 5/2[523]$ | 21.30 | |
| $^{162}_{67}Ho_{95}$ | $\pi 7/2[523] \otimes \nu 5/2[523]$ | 21.07 | - |
| | $\pi 7/2[523] \otimes \nu 5/2[642]$ | 52.90 | - |
| | $\pi 7/2[523] \otimes \nu 5/2[642]$ | 23.67 | - |
| | $\pi 7/2[523] \otimes \nu 5/2[523]$ | 16.90 | - |
| | $\pi 7/2[523] \otimes \nu 3/2[521]$ | 21.47 | - |
| $^{164}_{67}Ho_{97}$ | $\pi 7/2[523] \otimes \nu 5/2[523]$ | 19.57 | - |
| | $\pi 7/2[523] \otimes \nu 5/2[642]$ | 35.80 | - |
| | $\pi 7/2[523] \otimes \nu 5/2[642]$ | 20.74 | - |
| $^{166}_{67}Ho_{99}$ | $\pi 3/2[411] \otimes \nu 1/2[521]$ | 21.92 | - |
| | $\pi 1/2[411] \otimes \nu 1/2[521]$ | 105.28 | - |
| | $\pi 7/2[523] \otimes \nu 5/2[523]$ | 18.22 | - |
| | $\pi 7/2[523] \otimes \nu 7/2[633]$ | 122.71 | - |
| $^{158}_{69}Tm_{89}$ | $\pi h_{11/2} \otimes \nu i_{13/2}$ | 69.75 | - |
| $^{160}_{69}Tm_{91}$ | $\pi h_{11/2} \otimes h_{9/2}$ | 27.60 | 19 ℏ and 24 ℏ |
| | $\pi h_{11/2} \otimes \nu i_{13/2}$ | 76.11 | 19 ℏ |
| $^{162}_{69}Tm_{93}$ | $\pi 7/2[523] \otimes \nu 3/2[521]$ | 37.50 | - |
| | $\pi 7/2[523] \otimes \nu 5/2[642]$ | 118.08 | 17 ℏ |
| | $\pi 7/2[404] \otimes \nu 5/2[642]$ | 35.16 | 21 ℏ and 32 ℏ |
| | $\pi 1/2[541] \otimes \nu 5/2[642]$ | 78.54 | 18 ℏ |
| | $\pi 1/2[411] \otimes \nu 5/2[642]$ | 125.29 | - |
| | $\pi 7/2[523] \otimes \nu 5/2[523]$ | 16.76 | - |
| | $\pi 7/2[404] \otimes \nu 3/2[521]$ | 39.43 | - |
| | $\pi 5/2[402] \otimes \nu 5/2[642]$ | 21.92 | 10 ℏ |
| | $\pi 1/2[411] \otimes \nu 5/2[642]$ | 64.63 | 11 ℏ |

| Nuclei | Configuration | Signature splitting $|\Delta E|$ | Signature inversion |
|---|---|---|---|
| $^{164}_{69}Tm_{95}$ | π7/2[404]⊗ν5/2[642] | 20.68 | 12ℏ and 18ℏ |
| | π7/2[523]⊗ν5/2[642] | 115.18 | - |
| | π7/2[404]⊗ν5/2[642] | 21.01 | - |
| | π1/2[541]⊗ν5/2[642] | 56.73 | 20 ℏ |
| | π7/2[523]⊗ν5/2[642] | 35.44 | 18 ℏ |
| | π7/2[523]⊗ν3/2[521] | 15.74 | |
| $^{166}_{69}Tm_{97}$ | π1/2[411]⊗ν5/2[642] | 76.34 | 11 ℏ and 16 ℏ |
| | π1/2[411]⊗ν5/2[523] | 104.61 | - |
| | π7/2[523]⊗ν5/2[642] | 95.63 | - |
| | π7/2[404]⊗ν1/2[521] | 18.92 | - |
| | π7/2[404]⊗ν5/2[642] | 99.94 | - |
| | π1/2[541]⊗ν5/2[642] | 117.43 | 24 ℏ |
| | π1/2[541]⊗ν5/2[523] | 93.38 | - |
| | π1/2[541]⊗ν3/2[521] | 86.72 | - |
| | π7/2[404]⊗ν5/2[642] | 72.39 | - |
| | π1/2[411]⊗ν5/2[642] | 78.45 | - |
| $^{168}_{69}Tm_{99}$ | π1/2[411]⊗ν7/2[633] | 48.16 | - |
| | π7/2[404]⊗ν7/2[633] | 112.76 | - |
| | π1/2[411]⊗ν1/2[521] | 87 | - |
| | π$h_{9/2}$⊗ν$i_{13/2}$ | 57.03 | - |
| | π7/2[523]⊗ν$i_{13/2}$ | 24.58 | - |
| $^{170}_{69}Tm_{101}$ | π1/2[411]⊗ν1/2[521] | 55.00 | - |
| | π1/2[411]⊗ν1/2[521] | 122.14 | - |
| $^{160}_{71}Lu_{89}$ | π$h_{11/2}$⊗ν$i_{13/2}$ | 92.12 | - |
| | π5/2[402]⊗ν1/2[660] | 90.69 | - |
| $^{162}_{71}Lu_{91}$ | π$h_{11/2}$⊗ν$i_{13/2}$ | 43.19 | 21 ℏ |
| $^{164}_{71}Lu_{93}$ | π7/2[523]⊗ν5/2[642] | 42.29 | 19 ℏ |
| | π$h_{11/2}$⊗ν(3/2[521] or 5/2[523]) | 22.57 | - |
| | π$g_{7/2}$⊗ν$i_{13/2}$ | 31.45 | 10 ℏ |
| | π$h_{11/2}$⊗ν$i_{13/2}$ | 28.34 | 12 ℏ and 19 ℏ |
| $^{166}_{71}Lu_{95}$ | π7/2[404]⊗ν5/2[642] | 21.36 | 19 ℏ |
| | π9/2[514]⊗ν5/2[642] | 29.87 | 17 ℏ |
| | π1/2[541]⊗ν5/2[642] | 93.05 | 18 ℏ |
| | π5/2[402]⊗ν5/2[642] | 24.73 | 16 ℏ |
| $^{168}_{71}Lu_{97}$ | π7/2[404]⊗ν5/2[642] | 47.66 | |
| | π1/2[541]⊗ν5/2[642] | 272.4 | 22 ℏ |
| | π9/2[514]⊗ν5/2[642] | 165.42 | |
| | π1/2[541]⊗ν5/2[642] | 295.43 | 23 ℏ |
| | π1/2[411]⊗ν5/2[642] | 133.27 | |
| | π5/2[402]⊗ν5/2[642] | 35.96 | 17 ℏ |

## 3.2 Backbending and Bandcrossing

Backbending is a well-known phenomenon in nuclear structure where the moment of inertia of a rotational band shows a sudden increase (or "bend") as a function of angular momentum. In two-quasiparticle (2qp) bands, this is typically attributed to the alignment of high-j nucleon pairs with the rotational axis, which results in a rapid gain in angular momentum without a corresponding increase in rotational frequency. This behaviour indicates a transition from collective to single-particle dominated rotation, often interpreted as a pair-breaking effect that reduces pairing correlations. A classic example is seen in the nucleus $^{164}$Er, where the first backbending occurs around spin I=$12\hbar$, attributed to the alignment of a pair of neutrons in the $i_{13/2}$ orbital.

Band crossing refers to the interaction between two rotational bands that come close in energy at certain angular momenta, often leading to a change in the dominant configuration of the yrast (lowest energy) band. In 2qp bands, such crossings typically involve a change in the quasiparticle structure for instance, a ground-state band crossing with a band involving aligned high-j orbitals. When the energies of the two bands are comparable, they may interact and exchange character, a phenomenon referred to as band mixing or avoided crossing. An example is observed in $^{156}$Dy, where the ground-state rotational band crosses with a two-quasiparticle band involving the $i_{13/2}$ configuration, resulting in a change in the slope of the energy versus spin plot and a change in the moment of inertia beyond the crossing point. Band crossings have been observed in case of 10 bands, often accompanying changes in the kinematic and dynamic moments of inertia as paired high-j quasiparticles align with the rotational axis.

## 4. Ambiguous Configuration assignments

In Table 2, we present the ambiguous configuration (competing/tentative) proposed for two quasiparticle rotational bands. Most of the competing configuration involve *high-j* couplings such as $\pi h_{11/2} \otimes \nu f_{7/2}/\nu h_{9/2}$ and $\pi 7/2[523] \otimes \nu 5/2[512]/\nu 5/2[642]$, where several proton–neutron combinations lie close in energy and are easily mixed by Coriolis and affected by residual *n–p* interactions. The tentative configurations are dominated by *low-j* Nilsson orbitals containing $\pi 1/2[411]$ paired with $\nu 3/2[521]$, $\nu 1/2[510]$ $\nu 1/2[521]$, or $\nu 5/2[523]$; here the absence of decisive B(M1)/B(E2) ratios or g-factor data leaves the preferred assignment unconfirmed. Overall, we need for lifetime measurements and polarization analyses to resolve competing assignments, especially for bands built on the frequently mixed $\pi 1/2[411]$ proton orbital.

Table 2: Compilation of ambiguous configuration assignments

| Nuclei | Configuration | Tentative/Competing |
|---|---|---|
| $^{156}_{67}Ho_{89}$ | $\pi h_{11/2} \otimes \nu f_{7/2}$ or $\pi h_{11/2} \otimes \nu h_{9/2}$ | C |
| $^{158}_{67}Ho_{91}$ | $\pi 7/2[523] \otimes \nu 5/2[642]$ or $\pi 1/2[411] \otimes \nu 3/2[521]$ or $\pi 5/2[402] \otimes \nu 3/2[521]$ | C |
| | $\pi 7/2[523] \otimes \nu 3/2[402]$ or $\pi 7/2[404] \otimes \nu 3/2[521]$ or $\pi 7/2[523] \otimes \nu 5/2[642]$ | C |
| $^{160}_{67}Ho_{93}$ | $\pi 7/2[404] \otimes \nu 3/2[521]$ or $\pi 1/2[411] \otimes \nu 3/2[521]$ | C |
| $^{166}_{67}Ho_{99}$ | $\pi 7/2[523] \otimes \nu 3/2[521]$ | T |
| $^{166}_{67}Ho_{99}$ | $\pi 7/2[523] \otimes \nu 3/2[521]$ | T |
| | $\pi 3/2[411] \otimes \nu 7/2[633]$ or $\pi 7/2[523] \otimes \nu 3/2[521]$ | C |
| | $\pi 3/2[411] \otimes \nu 7/2[633]$ or $\pi 7/2[523] \otimes \nu 3/2[521]$ | C |
| | $\pi 1/2[541] \otimes \nu 7/2[633]$ or $\pi 1/2[411] \otimes \nu 5/2[512]$ | C |
| $^{168}_{67}Ho_{101}$ | $\pi 1/2[411] \otimes \nu 1/2[521]$ or $\pi 3/2[411] \otimes \nu 1/2[521]$ | C |
| $^{156}_{69}Tm_{87}$ | $\pi d_{3/2}(h_{11/2})^4_{0+} \otimes (f_{7/2})^5_{7/2}$ or $\pi 7/2[404] \otimes \nu 3/2[532]$ or $\pi 5/2[402] \otimes \nu 1/2[530]$ | C |
| | $h_{11/2} \otimes f_{7/2}$ or $\pi 5/2[532] \otimes \nu 3/2[532]$ or $\pi 3/2[541] \otimes \nu 1/2[530]$ | C |
| | $\pi d_{3/2} \otimes \nu f_{7/2}$ or $\pi s_{1/2} \otimes \nu f_{7/2}$ | C |
| | $\pi d_{3/2} \otimes \nu f_{7/2}$ or $\pi s_{1/2} \otimes \nu f_{7/2}$ | C |
| $^{168}_{69}Tm_{99}$ | $\pi 1/2[411] \otimes \nu 3/2[521]$, $\pi 1/2[411] \otimes \nu 3/2[521]$, $\pi 1/2[411] \otimes \nu 5/2[642]$ $\pi 1/2[411] \otimes \nu 1/2[510]$, $\pi 1/2[411] \otimes \nu 5/2[523]$, $\pi 1/2[411] \otimes \nu 1/2[510]$, $\pi 1/2[411] \otimes \nu 5/2[523]$, | T |
| | $\pi 5/2[402] \otimes \nu 7/2[633]$ or $\pi 7/2[523] \otimes \nu 5/2[512]$ | C |
| $^{170}_{69}Tm_{101}$ | $\pi 1/2[411] \otimes \nu 3/2[521]$ | T |
| | $\pi 7/2[404] \otimes \nu 5/2[512]$ or $\pi 3/2[411] \otimes \nu 1/2[521]$ | C |
| | $\pi 1/2[411] \otimes \nu 3/2[521]$ or $\pi 7/2[404] \otimes \nu 5/2[512]$ or $\pi 3/2[411] \otimes \nu 1/2[521]$ | C |
| | $\pi 1/2[411] \otimes \nu 3/2[521]$ or $\pi 3/2[411] \otimes \nu 1/2[521]$ or $\pi 1/2[411] \otimes \nu 1/2[521]$-$Q_{22}$ | C |
| | $\pi 3/2[411] \otimes \nu 5/2[512]$ or $\pi 1/2[411] \otimes \nu 5/2[512]$-$Q_{22}$ or $\pi 1/2[411] \otimes \nu 1/2[510]$ | C |
| | $\pi 3/2[411] \otimes \nu 1/2[521]$+$Q_{22}$ or $\pi 1/2[411] \otimes \nu 1/2[521]$ (T) | C/T |
| $^{172}_{69}Tm_{103}$ | $\pi 1/2[411] \otimes \nu 1/2[521]$ or $\pi 1/2[411] \otimes \nu 5/2[512]$-$Q_{22}$ | C |
| $^{174}_{69}Tm_{105}$ | $\pi 7/2[523] \otimes \nu 5/2[512]$ or $\pi 7/2[523] \otimes \nu 9/2[624]$ | C |
| $^{162}_{71}Lu_{91}$ | $\pi 1/2[411] \otimes \nu 3/2[521]$ or $\pi 7/2[523] \otimes \nu 5/2[523]$ | C |

## 5. Conclusions

A comprehensive survey of two-quasiparticle (2-qp) rotational structures in deformed odd-odd nuclei with $67 \leq Z \leq 71$ and $89 \leq N \leq 97$ (A = 156–168) yields 234 distinct rotational bands/states. The listed 63 Gallagher–Moszkowski doublets provides data for the study of neutron–proton residual-interaction systematics. Excitation energies reach ≈ *18 MeV* for the $\pi i_{13/2} \otimes \nu h_{9/2}$ and $\pi 9/2[514] \otimes \nu 5/2[642]$ configurations observed in in $^{164}_{71}Lu$ and in $^{168}_{71}Lu$ nuclides while the maximum spin of *50 ℏ* is populated in the latter. Triplet members constitute the ground state in 22 nuclides across the Ho, Tm, and Lu isotopic chains. Band-head K values between 0 and 9 accord with strong-coupling expectations for high-*j* orbitals and imply widespread high-K isomerism. Signature effects are pronounced: 76 bands exhibit splitting (|ΔE| ≈ 12–300 keV) and 29 bands display inversion indicating the presence of Coriolis mixing that increases. A total of 10 bands show clear band crossings associated with quasiparticle alignments. Half-life information is available for 58 bandheads, whereas electromagnetic observables remain scares. Most of the rotational bands are regular energy sequences. These gaps in lifetime and transition data underscore the need for targeted high-resolution spectroscopy to establish a complete empirical foundation for quantitative modelling and reliable systematics.

## POLICES OF THE TABLE 3

Level Energies: The listed level energies are taken form the first reference given for a band. In some cases, the energy values are extracted from the ENSDF/XUNDL database. However, the original reference alone has been quoted. The two signature partners are not listed separately.

## EXPLANATION OF TABLE 3

$^{A}_{Z}X$: Denotes the specific nuclide, represented by
- X: Chemical symbol
- A: Mass number
- Z: Atomic number

A single blank row separates the entries for each band. The number in the first column indicates the band number.

$E_{level}$: Level energy in keV. Energies in parentheses are tentative. An 'X' label indicates that the excitation energy is unknown, typically due to missing information on linking transitions to lower levels. Level energies are primarily taken from original references, or, if not available, sourced from ENSDF or XUNDL.

$I^{\pi}$: I: Spin of each band member
- π: Parity (+ or −), If given in parentheses, the spin and/or parity assignments are tentative.

Eγ(M1): Gamma transition energy (keV) for the M1 (I → I−1) transition, values listed in parentheses are tentative.

Eγ(E2): Gamma transition energy (keV) for the E2 (I → I−2) transition, values listed in parentheses are tentative.

ENSDF: Evaluated Nuclear Structure Data File (ENSDF) database ([www.nndc.bnl.gov](www.nndc.bnl.gov)).

XUNDL: Experimental Unevaluated Nuclear Data List (XUNDL) database

| | |
|---|---|
| B(M1)/B(E2): | The ratio of reduced transition probabilities, in units of $\mu_N/(eb)^2$, with uncertainties in the last digits shown in parentheses. Where only plots are provided in the original literature, numerical values are extracted from the plots and appropriately rounded. If neither values nor plots are available, these ratios are deduced, if possible, in this work using the rotational model and experimental gamma-ray energies/intensities. The mixing ratio from the rotational model is used for these calculations. |
| $\|g_K-g_R\|$: | Values are given with uncertainties in parentheses. If only plots are available in the references, values are digitized from the plots. |
| Keywords: | Keywords follow key numbers as assigned in the Nuclear Science References (NSR) database at Brookhaven National Laboratory, USA. Data are cited from the first listed keyword (shown in bold). Additional information from other keywords appears in the "Configuration and Comments" column. |
| Configuration: | The quasiparticle configuration for each band is listed.<br>• $\pi$ stands for protons, $\nu$ for neutrons.<br>• Nilsson quantum numbers label the orbitals, though notations can differ among authors. Explicit Nilsson configurations and author-specific notations are listed in Table 3. |
| Signature Splitting: | In rotational bands with $\Delta I = 1$ between members, each state belongs to a different signature. When odd-even staggering is observed in the band energies, it is referred to as "signature splitting," which primarily arises due to Coriolis coupling. |
| Signature Inversion: | When the expected favored signature branch becomes unfavored at higher spins i.e., the branch lower in energy becomes higher it is called "signature inversion." |
| Regular Band: | A band in which excitation energy changes smoothly with spin (though not necessarily as I(I+1)). Most 2qp bands are regular. |
| Irregular Band: | A band in which the energy variation with spin is abrupt. |

# TABLE 3: Two-quasiparticle Rotational Bands

$^{156}_{67}Ho_{89}$

| S.No. | $E_{level}$ keV | $I^\pi$ | $E_\gamma$(M1) keV | $E_\gamma$(E2) keV | $|g_K-g_R|$ | B(M1)/B(E2) $(\mu_N/eb)^2$ | Keywords | Configuration and Comments |
|---|---|---|---|---|---|---|---|---|
| 1 | 0 | 4⁻ | | | | | **1999KAZV** | 1. π5/2[402]⊗ν3/2[521] |
| | | | | | | | 1989AL27 | 2. Nuclear reaction: |
| | | | | | | | | $^{156}$Er ε decay $^{156}$Ho |
| | | | | | | | | 3. π7/2[413]⊗ν3/2[521] if K=5 |
| | | | | | | | | and spin = 4⁽⁺⁾,5⁺ (1989Al27) |
| | | | | | | | | 4. Half-life of bandhead is 56(1) min |
| 2 | 52.37 | 1⁻ | | | | | **1999KAZV** | 1. π5/2[402]⊗ν3/2[521] |
| | | | | | | | 1975AL26 | 2. Half-life of bandhead is 9.5(15) |
| | | | | | | | | 3. π7/2[523]⊗ν3/2[521] and spin=2⁺ |
| | | | | | | | | (1975Al26) |
| 3 | 82.23 | 2⁻ | | | | | **1999KAZV** | 1. π7/2[404]⊗ν3/2[521] |
| | | | | | | | 1975AL26 | 2. π7/2[523]⊗ν5/2[523] and |
| | | | | | | | | spin =1⁽⁺⁾ (1975AL26) |
| | | | | | | | | 3. Half-life of bandhead is 1.38(12) ns |
| | | | | | | | | 4. B(M1) (W.u.) = 0.036(7) and |
| | | | | | | | | B(E2) (W.u.) =23(10) (Adopted data set |
| 4 | 117.58 | 1⁺ | | | | | **1999KAZV** | 1. π7/2[523]⊗ν5/2[523] |
| | | | | | | | 1978SC10 | 2. π1/2[411]⊗ν3/2[521] when spin= 1⁻ |
| | | | | | | | | and E = 87.5 |
| | | | | | | | | 3. Nuclear reaction: |
| | | | | | | | | $^{156}$Er (EC) $^{156}$Ho |
| | | | | | | | | 4. Half-life of bandhead is 58.5(35) ns |
| | | | | | | | | 5. B(E1) (W.u.) =1.9×10⁻⁵(5) and |
| | | | | | | | | B(E2) (W.u.) = 0.21(6) (Adopted data) |
| 5 | (A<150)? | (9⁻) | | | | | **1982LO05** | 1. πh₁₁/₂⊗νi₁₃/₂ |
| | (A+134.27) | (11⁻) | | 134.27 | | | | 2. Nuclear reaction: |
| | (A+344.57) | (12⁻) | 210.30 | | | | | $^{146}$N ($^{14}$N, 4nγ) $^{156}$Ho |
| | (A+474.04) | (13⁻) | 129.50 | 339.74 | | | | E($^{14}$N) = 72 MeV |
| | (A+738.1) | (14⁻) | 264.05 | 393.49 | | | | 3. Regular band |
| | (A+950.4) | (15⁻) | 212.31 | 476.47 | | | | 4. Signature splitting with signature |
| | (A+1249.8) | (16⁻) | 299.31 | 511.72 | | | | inversion at I = (19⁻) |
| | (A+1533.3) | (17⁻) | 283.45 | 582.86 | | | | 5. Level energy is adopted from ENSDF |
| | (A+1859.7) | (18⁻) | 326.54 | 609.86 | | | | |
| | (A+2204.1) | (19⁻) | 343 | 670.88 | | | | |
| | (A+2552.2) | (20⁻) | 347.98 | 692.50 | | | | |
| | (A+2947.7) | (21⁻) | 395.85 | 743.23 | | | | |
| | (A+3315.8) | (22⁻) | | 763.64 | | | | |
| 6 | (A+573.8) | (13⁻) | | | | | **1982LO05** | 1. Competing configurations are: |
| | (A+1105.3) | (15⁻) | | 532.57 | | | | πh₁₁/₂⊗νf₇/₂ |
| | (A+1711.1) | (17⁻) | | 605.8 | | | | πh₁₁/₂⊗νh₉/₂ |
| | (A+2386) | (19⁻) | | 656 | | | | 2. Regular band |
| | (A+2992) | (21⁻) | | 675 | | | | 3. Level energy is adopted from ENSDF |

| S.No. | E$_{level}$ keV | I$^\pi$ | E$_\gamma$(M1) keV | E$_\gamma$(E2) keV | \|g$_K$-g$_R$\| | B(M1)/B(E2) ($\mu_N$/eb)$^2$ | Keywords | Configuration and Comments |
|---|---|---|---|---|---|---|---|---|
| 7 | 0+y | (8$^-$) | | | | | **1998CU01** | 1. At lower spin configuration is |
| | 91.0+y | (9$^-$) | 91 | | | | | $\pi$7/2[523]⊗$\nu$1/2[660] |
| | 185.5+y | (10$^-$) | 94 | | | | | At higher spin configuration is |
| | 320.0+y | (11$^-$) | 134 | 229 | | | | $\nu$1/2[660]⊗ $\nu$1/2[660]⊗ $\nu$3/2[651]⊗ |
| | 530.2+y | (12$^-$) | | 211 | | | | $\pi$7/2[523] |
| | 660.1+y | (13$^-$) | 130 | 340 | | | | 2. Nuclear reaction: |
| | 923.8+y | (14$^-$) | 264 | 394 | | | | $^{148}$Nd($^{14}$N,6n$\gamma$)$^{156}$Ho |
| | 1135.7+y | (15$^-$) | 212 | 476 | | | | E($^{14}$N) =96MeV |
| | 1436.3+y | (16$^-$) | 300 | 512 | | | | 3. Band crossing at ℏ$\omega$≈0.40MeV with an |
| | 1719.2+y | (17$^-$) | 284 | 583 | | | | aligned angular momentum gain Δi=5.0ℏ |
| | 2045.9+y | (18$^-$) | 327 | 610 | | | | 4. Regular band |
| | 2390.0+y | (19$^-$) | 344 | 671 | | | | 5. Signature splitting with double signature |
| | 2738.7+y | (20$^-$) | 348 | 693 | | | | inversion at I = (12$^-$) and I = (19$^-$) |
| | 3133.6+y | (21$^-$) | 394 | 744 | | | | 7. Level energy is adopted from ENSDF |
| | 3496.3+y | (22$^-$) | 382 | 758 | | | | |
| | 3930.5+y | (23$^-$) | 433 | 797 | | | | |
| | 4297.2+y | (24$^-$) | 366 | 802 | | | | |
| | 4750.8+y | (25$^-$) | 454 | 820 | | | | |
| | 5123.2+y | (26$^-$) | 373 | 826 | | | | |
| | 5587.5+y | (27$^-$) | 464 | 836 | | | | |
| | 5978.3+y | (28$^-$) | 390 | 856 | | | | |
| | 6464.5+y | (29$^-$) | | 877 | | | | |
| | 6884.3+y | (30$^-$) | | 906 | | | | |
| | 7402.5+y | (31$^-$) | | 938 | | | | |
| | 7854.4+y | (32$^-$) | | 970 | | | | |
| | 8411+y | (33$^-$) | | 1009 | | | | |
| | 8890+y | (34$^-$) | | 1036 | | | | |
| | 9489+y | (35$^-$) | | 1078 | | | | |
| | 9994+y | (36$^-$) | | 1104 | | | | |
| | | | | | | | | |
| 8 | 660.5+y | (10$^+$) | | 383 | | | **1998CU01** | 1. $\pi$7/2[404]⊗$\nu$1/2[660] |
| | 849.1+y | (11$^+$) | 189 | 529 | | | | 2. Regular band |
| | 1044.0+y | (12$^+$) | 196 | 382 | | | | 3. Small signature splitting with signature |
| | 1267.0+y | (13$^+$) | 223 | 418 | | | | inversion at I = (20$^+$) |
| | 1505.7+y | (14$^+$) | 239 | 462 | | | | 4. Undergo a smooth increase in aligned |
| | 1769.6+y | (15$^+$) | 263 | 502 | | | | angular momentum from 0.25 to 0.4 |
| | 2037.2+y | (16$^+$) | 267 | 532 | | | | MeV |
| | 2337.6+y | (17$^+$) | 300 | 568 | | | | 5. Increase arise from AD crossing |
| | 2640.1+y | (18$^+$) | 303 | 602 | | | | A= 1/2[660] ($\alpha$= +1/2) |
| | 2969.6+y | (19$^+$) | 329 | 631 | | | | D = 3/2[651] ($\alpha$= -1/2) |
| | 3315.9+y | (20$^+$) | 346 | 676 | | | | 6. Level energy is adopted from ENSDF |
| | 3668.8+y | (21$^+$) | 352 | 699 | | | | |
| | 4066.8+y | (22$^+$) | 396 | 752 | | | | |
| | 4439.8+y | (23$^+$) | 372 | 772 | | | | |
| | 4922.8+y | (24$^+$) | | 856 | | | | |
| | 5301.8+y | (25$^+$) | | 862 | | | | |
| | 5853.8+y | (26$^+$) | | 931 | | | | |
| | 6214.8+y | (27$^+$) | | 913 | | | | |
| | 6809.8+y | (28$^+$) | | 956 | | | | |
| | 7066.8+y | (29$^+$) | | 852 | | | | |
| | 7701+y | (30$^+$) | | 891 | | | | |
| | 8611+y | (32$^+$) | | 910 | | | | |

| S.No. | E$_{level}$ keV | I$^\pi$ | E$_\gamma$(M1) keV | E$_\gamma$(E2) keV | \|g$_K$-g$_R$\| | B(M1)/B(E2) ($\mu_N$/eb)$^2$ | Keywords | Configuration and Comments |
|---|---|---|---|---|---|---|---|---|
| 9 | 0+y | (12) | | | | | **1998CU01** | 1. At lower spin configuration is |
| | 227.9+y | (13) | 228 | | | | | $\pi$7/2[523]⊗$\nu$11/2[505] |
| | 484.1+y | (14) | 256 | 484 | | | | At higher spin configuration is |
| | 761.7+y | (15) | 278 | 534 | | | | $\nu$11/2[505]⊗ $\nu$1/2[660]⊗ $\nu$1/2[660]⊗ |
| | 1056.7+y | (16) | 295 | 572 | | | | $\pi$7/2[523] |
| | 1367.1+y | (17) | 310 | 606 | | | | 2. Regular band |
| | 1691.0+y | (18) | 324 | 634 | | | | 3. At rotational frequency of ℏω ≈ 0.3 MeV |
| | 2023.9+y | (19) | | 657 | | | | undergo a backbend at I = 21 |
| | 2365.2+y | (20) | | 674 | | | | 4. Level energy is adopted from ENSDF |
| | 2700.1+y | (21) | | 676 | | | | |
| | 2984.3+y | (22) | | 619 | | | | |
| | 3234.3+y | (23) | 250 | 534 | | | | |
| | 3501.4+y | (24) | 267 | 517 | | | | |
| | 3789.2+y | (25) | 288 | 555 | | | | |
| | 4103.7+y | (26) | 315 | 602 | | | | |
| | 4443.3+y | (27) | 340 | 654 | | | | |
| | 4804.0+y | (28) | 361 | 700 | | | | |
| | 5191.3+y | (29) | | 748 | | | | |
| | 5598.0+y | (30) | | 794 | | | | |
| | 6030.3+y | (31) | | 839 | | | | |
| | 6477.0+y | (32) | | 879 | | | | |
| | 6951.3+y | (33) | | 921 | | | | |
| | 7436+y | (34) | | 959 | | | | |
| | 7945+y | (35) | | 994 | | | | |
| | 8467+y | (36) | | 1031 | | | | |
| | 9016+y | (37) | | 1071 | | | | |
| 10 | 0+z | (11) | | | | | **1998CU01** | 1. $\pi$7/2[404]⊗$\nu$1/2[660] |
| | 161.9+z | (12) | 162 | | | | | 2. gain in aligned angular momentum occurs |
| | 324.1+z | (13) | 163 | 324 | | | | around 0.35 MeV which is consistent with |
| | 541.5+z | (14) | 218 | 379 | | | | the BC crossing |
| | 758.1+z | (15) | 217 | 434 | | | | B=1/2[660] (α= -1/2) |
| | 1023.9+z | (16) | 266 | 482 | | | | C = 3/2[651] (α= +1/2) |
| | 1285.9+z | (17) | 262 | 528 | | | | 3. Regular band |
| | 1592.1+z | (18) | 306 | 568 | | | | 4. Signature splitting with signature |
| | 1891.6+z | (19) | 299 | 606 | | | | inversion at I = (23) |
| | 2228.8+z | (20) | 337 | 637 | | | | 5. Level energy is adopted from ENSDF |
| | 2557.7+z | (21) | 329 | 666 | | | | |
| | 2902.8+z | (22) | 345 | 674 | | | | |
| | 3285.7+z | (23) | | 728 | | | | |
| | 3581.8+z | (24) | | 679 | | | | |
| | 4039.7+z | (25) | | 754 | | | | |
| | 4828.7+z | (27) | | 789 | | | | |
| 11 | 0+v | (24) | | | | | **1998CU01** | 1. One possible configuration is |
| | 624+v | (26) | | 624 | | | | $\pi$1/2[541]⊗$\nu$1/2[660], becoming |
| | 1304+v | (28) | | 680 | | | | $\pi$1/2[541]⊗ $\nu$1/2[660]⊗ $\nu$1/2[660]⊗ |
| | 2068+v | (30) | | 764 | | | | $\nu$ 3/2[651] |
| | 2922+v | (32) | | 854 | | | | 2. Regular band |
| | 3843+v | (34) | | 921 | | | | 3. Decoupled band |
| | 4814+v | (36) | | 971 | | | | 4. At a rotational frequency of ℏω ≈0.43 |
| | 5812+v | (38) | | 1007 | | | | MeV a smooth increase in aligned angular |
| | | | | | | | | momentum is observed |
| | | | | | | | | 5. Level energy is adopted from ENSDF |

$^{158}_{67}Ho_{91}$

| S.No. | E$_{level}$ keV | I$^\pi$ | E$_\gamma$(M1) keV | E$_\gamma$(E2) keV | \|g$_K$-g$_R$\| | B(M1)/B(E2) ($\mu_N$/eb)$^2$ | Keywords | Configuration and Comments |
|---|---|---|---|---|---|---|---|---|
| 1 | 0.0 | 5$^+$ | | | | | **1999LU03** | 1. $\pi 7/2[523] \otimes \nu 3/2[521]$ |
| | 102.8 | 6$^+$ | 102.7 | | | | 1969EK01 | 2. Nuclear reaction: |
| | 225.0 | 7$^+$ | 121.8 | 224.9 | | | | $^{152}$Sm($^{11}$B,5n$\gamma$)$^{158}$Ho, E = 60 MeV |
| | 368.0 | 8$^+$ | 142.9 | 265.5 | | | | 3. Half-life of bandhead is 11.3(4) min |
| | 531.5 | 9$^+$ | 163.4 | 306.4 | | | | (1969EK01) |
| | 715.8 | 10$^+$ | 183.9 | 348.1 | | | | 4. Regular band |
| | 917.2 | 11$^+$ | 201.4 | 183.9 | | | | 5. Signature splitting pronounced at higher |
| | 1111.4 | 12$^+$ | 194.3 | 395.3 | | | | spins |
| | 1362.0 | 13$^+$ | 250.9 | 194.3 | | | | 6. The value of g$_R$ = 0.48(9) |
| | 1557.2 | 14$^+$ | 195.7 | 445.4 | | | | 7. Level energy is adopted from ENSDF |
| | 1868.3 | 15$^+$ | | 506.3 | | | | |
| | 2013.7 | 16$^+$ | | 465.5 | | | | |
| | 2433.8 | 17$^+$ | | 565.5 | | | | |
| 2 | 67.25 | 2$^-$ | | | | | **1986SO02** | 1. $\pi 7/2[404] \otimes \nu 3/2[521]$ |
| | | | | | | | 1969EK01 | 2. Half-life of bandhead is 28(2) min |
| | | | | | | | | 3. J measured by atomic-beam magnetic resonance (1969EK01) |
| 3 | 118.0 or | 2$^+$ | | | | | **1982VY06** | 1. $\pi 7/2[523] \otimes \nu 3/2[521]$ |
| | 134.4 | | | | | | | 2. Nuclear reaction: |
| | | | | | | | | $^{158}$Er $\epsilon$ decay $^{158}$Ho |
| 4 | 139.2 | 1$^-$ | | | | | **1986SO02** | 1. Competing configurations |
| | 160.3 or | 2$^-$ | | | | | 1978SC10 | $\pi 7/2[523] \otimes \nu 5/2[642]$ |
| | 182.6 | | | | | | | $\pi 1/2[411] \otimes \nu 3/2[521]$ |
| | | | | | | | | $\pi 5/2[402] \otimes \nu 3/2[521]$ |
| | | | | | | | | 2. Half-life of bandhead is 1.85(10) ns |
| 5 | 143.5 | (2$^-$) | | | | | **1968AB18** | 1. $\pi 7/2[523] \otimes \nu 3/2[651]$ |
| | | | | | | | | 2. Nuclear reaction: |
| | | | | | | | | $^{158}$Ho IT decay (28 min) |
| 6 | (156.9) | (5$^-$) | | | | | **1986SO02** | 1. Competing configurations |
| | 227.8 | (6$^-$) | | | | | | $\pi 7/2[523] \otimes \nu 3/2[402]$ |
| | 328.4 | (7$^-$) | | | | | | $\pi 7/2[404] \otimes \nu 3/2[521]$ |
| | 443.5 | (8$^-$) | | | | | | $\pi 7/2[523] \otimes \nu 5/2[642]$ |
| | 602.8 | (9$^-$) | | | | | | 2. Half-life of bandhead is 29(3) ns |
| | 768.9 | (10$^-$) | | | | | | 3. Small signature splitting in $\Delta E\gamma$ vs I |
| | 982.7 | (11$^-$) | | | | | | 4. Signature inversion at I = (14$^-$) |
| | 1204.7 | (12$^-$) | | | | | | |
| | 1460.5 | (13$^-$) | | | | | | |
| | 1737.9 | (14$^-$) | | | | | | |
| | 2024.3 | (15$^-$) | | | | | | |
| | (2354.9) | (16$^-$) | | | | | | |

| S.No. | $E_{level}$ keV | $I^\pi$ | $E_\gamma$(M1) keV | $E_\gamma$(E2) keV | $\|g_K-g_R\|$ | B(M1)/B(E2) $(\mu_N/eb)^2$ | Keywords | Configuration and Comments |
|---|---|---|---|---|---|---|---|---|
| 7 | 180.0 | $9^+$ | | | | | **1999LU03** | 1. $\pi7/2[523]\otimes v11/2[505]$ |
| | 405.0 | $10^+$ | 225.0 | | | | 1986SO02 | 2. Half-life of bandhead is 21 min |
| | 652.0 | $11^+$ | 247.3 | 472.1 | | | | 3. Regular band |
| | 918.0 | $12^+$ | 265.5 | 513.4 | | | | 4. The value of $g_R = 0.55(6)$ |
| | 1200.0 | $13^+$ | 282.1 | 548.4 | | | | 5. Level energy is adopted from ENSDF |
| | 1497.0 | $14^+$ | 296.1 | 578.9 | | | | |
| | 1806.0 | $15^+$ | 309.3 | 605.6 | | | | |
| | 2121.8 | $16^+$ | 315.9 | 624.8 | | | | |
| | 2444.0 | $17^+$ | | 638.5 | | | | |
| | 2765.0 | $18^+$ | | 642.9 | | | | |
| | | | | | | | | |
| 8 | 315 | $1^+$ | | | | | **1972HA41** | 1. $\pi7/2[404]\otimes v5/2[642]$ |
| | | | | | | | | 2. Nuclear reaction: $^{158}$Ho IT decay (28 min) |
| | | | | | | | | |
| 9 | 378 | $1^-$ | | | | | **1972HA41** | 1. $\pi7/2[404]\otimes v5/2[642]$ |
| | | | | | | | | |
| 10 | 408 | $(2)^+$ | | | | | **1972HA41** | 1. $\pi7/2[404]\otimes v3/2[651]$ |
| | | | | | | | | |
| 11 | 279.2 | $7^+$ | | | | | **1999LU03** | 1. $\pi7/2[404]\otimes v3/2[651]$, $K^\pi = 5^+$ |
| | 359.3 | $8^+$ | 80.2 | | | | 1996YUZY | 2. This band expanded to I =$43^+$ |
| | 454.5 | $9^+$ | 95.4 | 174.9 | | | | (1996YUZY) |
| | 582.4 | $10^+$ | 127.4 | 222.9 | | | | 3. Regular band |
| | 735.9 | $11^+$ | 153.3 | 281.6 | | | | 4. Small Signature splitting in $\Delta E\gamma$ vs I |
| | 918.8 | $12^+$ | 182.5 | 336.3 | | | | 5. The value of $g_R = 0.23(8)$ |
| | 1120.1 | $13^+$ | 201.0 | 384.5 | | | | 6. Level energy is adopted from XUNDL |
| | 1349.4 | $14^+$ | 229.3 | 430.7 | | | | |
| | 1590.0 | $15^+$ | 240.6 | 470.1 | | | | |
| | 1860.3 | $16^+$ | 270.7 | 510.6 | | | | |
| | 2132.6 | $17^+$ | 272.5 | 542.4 | | | | |
| | 2431.1 | $18^+$ | 299.3 | 571.9 | | | | |
| | 2728.5 | $19^+$ | | 595.9 | | | | |
| | 3052.1 | $20^+$ | | 620.0 | | | | |
| | | | | | | | | |
| 12 | 207.9 | $8^-$ | | | | | **1999LU03** | 1. $\pi7/2[523]\otimes v3/2[651]$, $K^\pi = 5^-$ |
| | 278.6 | $9^-$ | 70.8 | | | | 1996YUZY | 2. This band expanded to I =$42^-$ |
| | 379.2 | $10^-$ | 100.3 | 171.3 | | | | (1996YUZY) |
| | 494.1 | $11^-$ | 115.1 | 215.7 | | | | 3. Regular band |
| | 653.4 | $12^-$ | 159.4 | 274.1 | | | | 4. Small Signature splitting in $\Delta E\gamma$ vs I |
| | 819.4 | $13^-$ | 165.7 | 325.3 | | | | 5. The value of $g_R = 0.39(6)$ |
| | 1033.0 | $14^-$ | 213.7 | 379.6 | | | | 6. Level energy is adopted from XUNDL |
| | 1254.7 | $15^-$ | 221.8 | 435.2 | | | | |
| | 1510.0 | $16^-$ | 255.4 | 477.3 | | | | |
| | 1787.8 | $17^-$ | 277.3 | 532.9 | | | | |
| | 2074.0 | $18^-$ | 258.6 | 564.6 | | | | |
| | 2403.7 | $19^-$ | 329.8 | 615.8 | | | | |
| | 2710.6 | $20^-$ | 307.2 | 636.4 | | | | |
| | 3085.3 | $21^-$ | 374.5 | 681.4 | | | | |
| | 3404.7 | $22^-$ | 319.0 | 694.6 | | | | |
| | 3812.8 | $23^-$ | 408.0 | 727.4 | | | | |
| | 4143.0 | $24^-$ | 330.0 | 738.5 | | | | |
| | 4584.8 | $25^-$ | | 772.0 | | | | |

$^{160}_{67}Ho_{93}$

| S.No. | E_level keV | I^π | E_γ(M1) keV | E_γ(E2) keV | \|g_K-g_R\| | B(M1)/B(E2) (μ_N/eb)^2 | Keywords | Configuration and Comments |
|---|---|---|---|---|---|---|---|---|
| 1 | 0 | 5+ | | | | | **2004ES01** | 1. π7/2[523]⊗ν3/2[521] |
| | 107 | 6+ | 107 | | | | 1996DR03 | 2. Nuclear reaction: |
| | 232 | 7+ | 125 | 233 | | | 1970LE22 | $^{158}$Gd($^{7}$Li,5nγ)$^{160}$Ho |
| | 375 | 8+ | 143 | 269 | | | | 3. Half-life of bandhead is 25.6 min |
| | 536 | 9+ | 161 | 304 | | | | (1990GO22) |
| | 707 | 10+ | 171 | 332 | | | | 3. Regular band |
| | 919 | 11+ | 211 | 383 | | | | 4. Signature splitting |
| | 1117 | 12+ | 198 | 410 | | | | 5. Signature inversion at I = 20+ |
| | 1345 | 13+ | 228 | 426 | | | | |
| | 1546 | 14+ | 201 | 428 | | | | |
| | 1820 | 15+ | | 475 | | | | |
| | 2037 | 16+ | 217 | 491 | | | | |
| | 2330 | 17+ | | 510 | | | | |
| | 2592 | 18+ | 262 | 555 | | | | |
| | 2887 | 19+ | | 557 | | | | |
| | 3214 | 20+ | | 622 | | | | |
| | 3524 | 21+ | | 637 | | | | |
| | 3877 | 22+ | | 663 | | | | |
| | 4205 | (23+) | | (681) | | | | |
| | 4572 | (24+) | | (695) | | | | |
| 2 | 59.98 | 2- | | | | | **1990GO22** | 1. Competing configurations |
| | | | | | | | 1969EK01 | π7/2[404]⊗ν3/2[521] |
| | | | | | | | | π1/2[411]⊗ν3/2[521] |
| | | | | | | | | 2. Nuclear reaction: |
| | | | | | | | | $^{160}$Er ε decay $^{160}$Ho |
| | | | | | | | | 3. Half-life of bandhead is 5.02 h |
| 3 | 67.11 | 1+ | | | | | **1990GO22** | 1. π7/2[523]⊗ν5/2[523] |
| | | | | | | | | 2. Half-life of bandhead is 28(2) ns |
| | | | | | | | | (2006KAZX) |
| 4 | 118 | 6- | | | | | **2004ES01** | 1. π7/2[523]⊗ν5/2[642] |
| | 170 | 7- | | | | | 1996DR03 | 2. Half-life of bandhead is 56(8) ns |
| | 242 | 8- | 73 | 126 | | | 1990SA19 | (1990SA19) |
| | 336 | 9- | 93 | 166 | | | | 3. Regular band |
| | 451 | 10- | 115 | 209 | | | | 4. Signature splitting |
| | 586 | 11- | 135 | 250 | | | | |
| | 746 | 12- | 160 | 293 | | | | |
| | 924 | 13- | 178 | 338 | | | | |
| | 1128 | 14- | 203 | 382 | | | | |
| | 1353 | 15- | 225 | 428 | | | | |
| | 1595 | 16- | 242 | 467 | | | | |
| | 1869 | 17- | 274 | 515 | | | | |
| | 2142 | 18- | 272 | 546 | | | | |
| | 2465 | 19- | 323 | 596 | | | | |
| | 2761 | 20- | 295 | 619 | | | | |
| | 3133 | 21- | 371 | 668 | | | | |
| | 3446 | 22- | 312 | 684 | | | | |
| | 3860 | 23- | 414 | 727 | | | | |
| | 4185 | 24- | | 738 | | | | |
| | 4630 | 25- | | 771 | | | | |
| | 4966 | 26- | | 781 | | | | |
| | 5784 | 28- | | 817 | | | | |

| S.No. | E_level keV | $I^\pi$ | $E_\gamma$(M1) keV | $E_\gamma$(E2) keV | $|g_K-g_R|$ | B(M1)/B(E2) $(\mu_N/eb)^2$ | Keywords | Configuration and Comments |
|---|---|---|---|---|---|---|---|---|
| 5 | 176 | 9+ | | | | | **2004ES01** | 1. $\pi7/2[523]\otimes\nu11/2[505]$ |
| | 396 | 10+ | 220 | | | | 1996DR03 | 2. Half-life of bandhead is 3s |
| | 635 | 11+ | 240 | 459 | | | 1990ANZZ | 3. Regular band |
| | 893 | 12+ | 257 | 497 | | | | 4. Small signature splitting in $\Delta E\gamma$ vs I |
| | 1167 | 13+ | 274 | 532 | | | | |
| | 1456 | 14+ | 288 | 563 | | | | |
| | 1756 | 15+ | 300 | 589 | | | | |
| | 2067 | 16+ | 310 | 610 | | | | |
| | 2379 | 17+ | | 624 | | | | |
| | 2694 | 18+ | | 627 | | | | |
| | 2998 | 19+ | | 620 | | | | |
| | 3301 | 20+ | | 607 | | | | |
| | 3599 | 21+ | | 601 | | | | |
| | 3902 | 22+ | | (601) | | | | |
| | 4194 | 23+ | | (595) | | | | |
| | 4512 | 24+ | | (609) | | | | |
| 6 | 112 | 6- | | | | | **2004ES01** | 1. $\pi7/2[404]\otimes\nu5/2[523]$ |
| | 238 | 7- | 126 | | | | 1996DR03 | 2. Regular band |
| | 402 | 8- | 163 | 289 | | | | 3. Signature splitting |
| | 578 | 9- | 177 | 340 | | | | |
| | 795 | 10- | 216 | 393 | | | | |
| | 1019 | 11- | 225 | 441 | | | | |
| | 1284 | 12- | 264 | 489 | | | | |
| | 1552 | 13- | 268 | 533 | | | | |
| | 1856 | 14- | 303 | 572 | | | | |
| | 2161 | 15- | 305 | 609 | | | | |
| | 2494 | 16- | 333 | 638 | | | | |
| | 2821 | 17- | | 660 | | | | |

$^{162}_{67}Ho_{95}$

| S.No. | E_level keV | $I^\pi$ | $E_\gamma$(M1) keV | $E_\gamma$(E2) keV | $|g_K-g_R|$ | B(M1)/B(E2) $(\mu_N/eb)^2$ | Keywords | Configuration and Comments |
|---|---|---|---|---|---|---|---|---|
| 1 | 0.0 | 1+ | | | | | **2005LI63** | 1. $\pi7/2[523]\otimes\nu5/2[523]$ |
| | 38.3 | 2+ | 38.3 | | | | 1972WAYO | 2. Nuclear reaction: |
| | 96.1 | 3+ | 57.8 | | | | | $^{160}Gd(^7Li,5n\gamma)^{162}Ho$, E = 49 MeV |
| | 171.1 | 4+ | 75.6 | | | | | 3. Half-life of bandhead is 15.0(10) min |
| | 270.0 | 5+ | 98.2 | | | | | (1965ST08) |
| | 385.9 | 6+ | 115.9 | 214.3 | | | | 4. Regular band |
| | 521.5 | 7+ | 135.5 | 251.4 | | | | 5. Small signature splitting in $\Delta E\gamma$ vs I |
| | 672.8 | 8+ | 151.0 | 287.0 | | | | 6. Level energy is adopted from ENSDF |
| | 846.4 | 9+ | 173.5 | 324.7 | | | | |
| | 1030.8 | 10+ | 184.5 | 358.2 | | | | |
| | 1244.2 | 11+ | 213.5 | 397.5 | | | | |
| | 1457.2 | 12+ | 213.0 | 426.7 | | | | |
| | 1709.1 | 13+ | 252.0 | 464.5 | | | | |
| | 1948.4 | 14+ | 239.0 | 491.5 | | | | |
| | (2234.6) | 15+ | | (525.5) | | | | |

| S.No. | $E_{level}$ keV | $I^\pi$ | $E_\gamma$(M1) keV | $E_\gamma$(E2) keV | $\|g_K-g_R\|$ | B(M1)/B(E2) $(\mu_N/eb)^2$ | Keywords | **Configuration and Comments** |
|---|---|---|---|---|---|---|---|---|
| 2 | 106 | $6^-$ | | | | | **2005LI63** | 1. $\pi 7/2[523] \otimes \nu 5/2[642]$ |
| | 176 | $7^-$ | 70.5 | | | | 2004ES01 | 2. Half-life of bandhead is 68 min |
| | 266 | $8^-$ | 90.4 | 160.5 | | | 1970LE22 | (1970LE22) |
| | 377 | $9^-$ | 110.5 | 201.0 | | | | 3. Regular band |
| | 507 | $10^-$ | 130.5 | 240.9 | | | | 4. Signature splitting more pronounced at |
| | 657 | $11^-$ | 150.5 | 280.7 | | | | Higher spins. |
| | 828 | $12^-$ | 170.0 | 320.4 | | | | 5. Level energies taken from 2004ES01 |
| | 1018 | $13^-$ | 189.9 | 360.0 | | | | |
| | 1225 | $14^-$ | 207.5 | 397.7 | | | | |
| | 1456 | $15^-$ | 230.3 | 438.0 | | | | |
| | 1697 | $16^-$ | 241.2 | 472.0 | | | | |
| | 1970 | $17^-$ | 273.0 | 514.4 | | | | |
| | 2240 | $18^-$ | 270.5 | 543.5 | | | | |
| | 2558 | $19^-$ | 317.5 | 589.0 | | | | |
| | 2852 | $20^-$ | 293.5 | 611.5 | | | | |
| | 3216 | $21^-$ | 364.0 | 675.5 | | | | |
| | 3528 | $22^-$ | 312.5 | 676.5 | | | | |
| | 3938 | $23^-$ | 410.0 | 722.0 | | | | |
| | 4264 | $24^-$ | 326 | 736.5 | | | | |
| | 4716 | $25^-$ | 452 | 778 | | | | |
| | 5052 | $26^-$ | 336 | 788 | | | | |
| | 5536 | $27^-$ | 484 | 820 | | | | |
| | 5881 | $28^-$ | 345 | 829 | | | | |
| 3 | 179.8 | $1^-$ | | | | | **2005LI63** | 1. $\pi 7/2[523] \otimes \nu 5/2[642]$ |
| | 389.7 | $(6^-)$ | | | | | 1978SC10 | 2. Half-life of bandhead is 8.7(2) ns |
| | 476.0 | $(7^-)$ | 86.5 | | | | | 3. Regular band |
| | 563.0 | $(8^-)$ | 87.2 | 173.5 | | | | 4. Signature splitting |
| | 687.4 | $(9^-)$ | 124.5 | 211.5 | | | | 5. Level energy is adopted from ENSDF |
| | 811.2 | $(10^-)$ | 124.0 | 248.0 | | | | |
| | 978.6 | $(11^-)$ | 167.4 | 291.3 | | | | |
| | 1144.0 | $(12^-)$ | 165.3 | 332.7 | | | | |
| | 1358.6 | $(13^-)$ | 214.5 | 380.1 | | | | |
| | 1566.7 | $(14^-)$ | 208.2 | 422.7 | | | | |
| | 1827.6 | $(15^-)$ | 261.0 | 469.0 | | | | |
| | 2078.2 | $(16^-)$ | 250.5 | 511.5 | | | | |
| | 2380.4 | $(17^-)$ | 302.0 | 552.9 | | | | |
| | 2672.5 | $(18^-)$ | 292.0 | 594.5 | | | | |
| | 3006.0 | $(19^-)$ | 333.5 | 625.5 | | | | |
| | 3338.0 | $(20^-)$ | | 665.5 | | | | |
| 4 | 184.8 | $(6^+)$ | | | | | **2005LI63** | 1. $\pi 7/2[523] \otimes \nu 5/2[523]$ |
| | 301.2 | $(7^+)$ | 116.5 | | | | | 2. Regular band |
| | 437.1 | $(8^+)$ | 136.0 | 252.5 | | | | 3. Small signature splitting |
| | 592.3 | $(9^+)$ | 155.5 | 291.1 | | | | 4. Level energy is adopted from ENSDF |
| | 765.8 | $(10^+)$ | 173.5 | 328.5 | | | | |
| | 940.1 | $(11^+)$ | 174.0 | 348.0 | | | | |
| | 1146.7 | $(12^+)$ | 206.5 | 381.0 | | | | |
| | 1355.0 | $(13^+)$ | 208.4 | 414.7 | | | | |
| | (1598.7) | $(14^+)$ | | (452.0) | | | | |
| | 1834.5 | $(15^+)$ | | 479.5 | | | | |

| S.No. | $E_{level}$ keV | $I^\pi$ | $E_\gamma$(M1) keV | $E_\gamma$(E2) keV | $|g_K-g_R|$ | B(M1)/B(E2) $(\mu_N/eb)^2$ | Keywords | Configuration and Comments |
|---|---|---|---|---|---|---|---|---|
| 5 | 0+x | (5+) | | | | | **2005LI63** | 1. $\pi7/2[523]\otimes\nu3/2[521]$ |
| | 101.0+x | (6+) | 100.9 | | | | | 2. Regular band |
| | 219.4+x | (7+) | 118.5 | 219.5 | | | | 3. Small signature splitting |
| | 365.2+x | (8+) | 146.0 | 264.1 | | | | 4. Level energy is adopted from ENSDF |
| | 521.0+x | (9+) | 155.6 | 301.5 | | | | |
| | 709.6+x | (10+) | 188.7 | 344.6 | | | | |
| | 899.5+x | (11+) | 189.8 | 378.3 | | | | |
| | 1129.3+x | (12+) | 229.5 | 420.0 | | | | |
| 6. | 0+y | (9+) | | | | | **2005LI63** | 1. $\pi7/2[523]\otimes\nu11/2[505]$ |
| | 217.2+y | (10+) | 217.4 | | | | | 2. Regular band |
| | 454.7+y | (11+) | 237.5 | 454.5 | | | | 3. Level energy is adopted from ENSDF |
| | 710.3+y | (12+) | 255.6 | 493.2 | | | | |
| | 984.0+y | (13+) | 273.5 | 529.2 | | | | |
| | 1272.6+y | (14+) | 288.5 | 562.5 | | | | |
| | 1581.5+y | (15+) | 309.0 | 597.5 | | | | |

$^{164}_{67}Ho_{97}$

| S.No. | $E_{level}$ keV | $I^\pi$ | $E_\gamma$(M1) keV | $E_\gamma$(E2) keV | $|g_K-g_R|$ | B(M1)/B(E2) $(\mu_N/eb)^2$ | Keywords | Configuration and Comments |
|---|---|---|---|---|---|---|---|---|
| 1 | 0.0 | 1+ | | | | | **2004HO19** | 1. $\pi7/2[523]\otimes\nu5/2[523]$ |
| | 37.3 | 2+ | 37.3 | | | | 1972WAYR | 2. Nuclear reaction is $^{160}$Gd($^{11}$B,$\alpha$3n$\gamma$)$^{164}$Ho, E=61MeV |
| | 94.0 | 3+ | 56.7 | | | | | |
| | 167.7 | (4+) | 73.7 | | | | | 3. Half-life of bandhead is 29(1) min (1972WAYR) |
| | 262.2 | (5+) | 94.5 | | | | | |
| | 370.7 | (6+) | 108.5 | | | | | 4. Regular band |
| | 503.4 | (7+) | 132.5 | 241.0 | | | | 5. Signature splitting more pronounced at higher spins |
| | 644.9 | (8+) | 141.3 | 274.0 | | | | |
| | 814.7 | (9+) | 169.8 | 311.5 | | | | 6. Level energy is adopted from ENSDF except (1+) to (6+) |
| | 985.9 | (10+) | 171.2 | 340.8 | | | | |
| | 1192.7 | (11+) | 206.7 | 378.2 | | | | 7. The value of $\hbar\omega_c \geq 0.35$MeV follows the systematics for the N = 96 isotones for which $\hbar\omega_c$ increases when Z decreases |
| | 1387.9 | (12+) | 195.1 | 402.0 | | | | |
| | 1631.7 | (13+) | 243.8 | 439.1 | | | | |
| | 1865.3 | (14+) | | 477.4 | | | | |
| | 2113.5 | (15+) | | 481.8 | | | | |

| S.No. | E_level keV | I^π | E_γ(M1) keV | E_γ(E2) keV | \|g_K-g_R\| | B(M1)/B(E2) (μ_N/eb)^2 | Keywords | Configuration and Comments |
|---|---|---|---|---|---|---|---|---|
| 2 | 139.9 | 6⁻ | | | | | **2004HO19** | 1. π7/2[523]⊗ν5/2[642] |
| | 204.0 | 7⁻ | 63.9 | | | | 1972WAYR | 2. Half-life of bandhead is 37.5(+15-5) |
| | 294.0 | (8⁻) | 90.1 | | | | | min(1972WAYR) and 36.3(4) min |
| | 406.8 | (9⁻) | 112.7 | 202.8 | | | | (2004HO19) |
| | 540.6 | (10⁻) | 133.7 | 264.7 | | 1.1 | | 3. Regular band |
| | 694.5 | (11⁻) | 153.9 | 287.6 | | 0.8 | | 4. Signature splitting more pronounced at |
| | 867.3 | (12⁻) | 172.9 | 326.7 | | 0.85 | | higher spins |
| | 1058.5 | (13⁻) | 191.2 | 363.9 | | 0.90 | | 5. Level energy is adopted from ENSDF |
| | 1265.6 | (14⁻) | 207.1 | 398.5 | | 0.6 | | except 6⁻ |
| | 1492.2 | (15⁻) | 226.8 | 433.6 | | 0.65 | | |
| | 1728.5 | (16⁻) | 236.3 | 462.9 | | 0.4 | | |
| | 1990.4 | (17⁻) | 261.9 | 498.1 | | 0.5 | | |
| | 2249.9 | (18⁻) | 259.2 | 521.3 | | 0.7 | | |
| | 2548.4 | (19⁻) | 298.5 | 558.4 | | 0.5 | | |
| | 2826.0 | (20⁻) | 277.7 | 575.9 | | | | |
| | 3162.3 | (21⁻) | 336.1 | 614.1 | | | | |
| | 3452.2 | (22⁻) | | 6622 | | | | |
| | 3827.0 | (23⁻) | | 664.7 | | | | |
| | 4125.7 | (24⁻) | | 673.5 | | | | |
| | 4541.8 | (25⁻) | | 714.8 | | | | |
| | | | | | | | | |
| 3 | 188.6 | 3⁺ | | | | | **1972WAYR** | 1. π7/2[523]⊗ν1/2[521] |
| | 297.4 | 4⁺ | | | | | | 2. Nuclear reaction is |
| | 429.9 | 5⁺ | | | | | | ¹⁶⁴Dy(d,2nγ)¹⁶⁴Ho, E=13.5MeV |
| | | | | | | | | 3. Regular band |
| | | | | | | | | 4. Level energy is adopted from ENSDF |
| | | | | | | | | |
| 4 | 191.1 | (6⁺) | | | | | **2004HO19** | 1. π7/2[523]⊗ν5/2[523] |
| | 317.4 | (7⁺) | 126.3 | | | | 1970JO11 | 2. Regular band |
| | 461.2 | (8⁺) | 143.5 | | | | | |
| | 621.9 | (9⁺) | 160.7 | 304.5 | | | | |
| | 799.4 | (10⁺) | 177.5 | 338.1 | | | | |
| | 993.1 | (11⁺) | 193.6 | 371.2 | | | | |
| | 1202.4 | (12⁺) | 209.4 | 403.1 | | | | |
| | 1427.0 | (13⁺) | 224.8 | 433.7 | | | | |
| | 1666.3 | (14⁺) | 239.2 | 464.0 | | | | |
| | 1919.3 | (15⁺) | 253.1 | 492.2 | | | | |
| | 2184.1 | (16⁺) | | 517.8 | | | | |
| | 2462.8 | (17⁺) | | 543.5 | | | | |

| S.No. | E$_{level}$ keV | I$^\pi$ | E$_\gamma$(M1) keV | E$_\gamma$(E2) keV | \|g$_K$-g$_R$\| | B(M1)/B(E2) ($\mu_N$/eb)$^2$ | Keywords | Configuration and Comments |
|---|---|---|---|---|---|---|---|---|
| 5 | 193.3 | (1$^-$) | | | | | **2004HO19** | 1. $\pi$7/2[523]$\otimes$v5/2[642] |
|   | 208.0 | (2$^-$) | | | | | 1970JO11 | 2. Regular band |
|   | 236.1 | (3$^-$) | | | | | | 3. Signature splitting more pronounced at |
|   | 273.8 | (4$^-$) | 37.7 | | | | | higher spins |
|   | 331.2 | (5$^-$) | | | | | | 4. Level energy is adopted from ENSDF |
|   | 398.5 | (6$^-$) | 67.3 | | | | | except from (1$^-$) to (7$^-$) |
|   | 488.3 | (7$^-$) | 89.8 | | | | | |
|   | 586.9 | (8$^-$) | 99.0 | | | | | |
|   | 711.9 | (9$^-$) | 125.1 | 223.8 | | | | |
|   | 843.1 | (10$^-$) | 131.1 | 256.2 | | 1.4 | | |
|   | 1006.7 | (11$^-$) | 163.6 | 131.1 | | 1.45 | | |
|   | 1171.3 | (12$^-$) | 164.5 | 328.1 | | 1.6 | | |
|   | 1377.2 | (13$^-$) | 205.8 | 370.6 | | 1.4 | | |
|   | 1577.2 | (14$^-$) | 200.0 | 405.9 | | 1.6 | | |
|   | 1825.7 | (15$^-$) | 248.5 | 448.6 | | 1.5 | | |
|   | 2062.4 | (16$^-$) | 236.7 | 485.2 | | | | |
|   | 2350.1 | (17$^-$) | 287.6 | 524.5 | | | | |
|   | 2623.9 | (18$^-$) | 273.9 | 561.4 | | | | |
|   | 2941.0 | (19$^-$) | 317.0 | 590.9 | | | | |
|   | 3250.4 | (20$^-$) | | 626.5 | | | | |
| 6 | 342.7 | 5$^+$ | | | | | **2004HO19** | 1. $\pi$7/2[523]$\otimes$v3/2[521] |
|   | 452.4 | 6$^+$ | 109.6 | | | | 1972WAYR | 2. Regular band |
|   | 579.6 | (7$^+$) | 127.3 | | | | | 3. Level energy is adopted from ENSDF |
|   | 723.1 | (8$^+$) | 143.3 | 270.6 | | | | except 5$^+$ |
|   | 881.5 | (9$^+$) | 158.2 | 302.0 | | | | |
|   | 1049.9 | (10$^+$) | 168.4 | 326.9 | | | | |
|   | 1234.8 | (11$^+$) | 185.0 | 353.3 | | | | |
|   | 1420.2 | (12$^+$) | | 370.3 | | | | |
| 7 | 486 | (2$^+$) | | | | | **1970JO11** | 1. $\pi$7/2[523]$\otimes$v3/2[521] |
|   | 558 | (3$^+$) | | | | | | 2. Nuclear reaction: |
|   | 650 | (4$^+$) | | | | | | $^{165}$Ho(d, t)$^{164}$Ho, E=12 MeV |
|   | 777 | (5$^+$) | | | | | | 3. Regular band |
| 8 | E | 3 | | | | | **1972WAYR** | 1. $\pi$1/2[411]$\otimes$v5/2[523] |
|   | E+90.0 | 4 | | | | | | 2. Regular band |
|   | E+202.8 | 5 | | | | | | |
|   | E+336.6 | 6 | | | | | | |
|   | E+487.8 | 7 | | | | | | |

$^{166}_{67}Ho_{99}$

| S.No. | E$_{level}$ keV | I$^\pi$ | E$_\gamma$(M1) keV | E$_\gamma$(E2) keV | \|g$_K$-g$_R$\| | B(M1)/B(E2) ($\mu_N$/eb)$^2$ | Keywords | Configuration and Comments |
|---|---|---|---|---|---|---|---|---|
| 1 | 190.9038(20) | 3$^+$ | | | | | **2000PR03** | 1. π7/2[523]⊗ν1/2[521] |
| | 260.6653(23) | 4$^+$ | 69.7604(14) | | | | 1965ST06 | 2. Nuclear reaction: |
| | 348.2617(26) | 5$^+$ | 87.5946(16) | 157.344(8) | | | | $^{165}$Ho(d,p)$^{166}$Ho, E = 17 MeV |
| | 453.773(4) | 6$^+$ | 105.517(4) | 193.107(6) | | | | 3. Half-life of bandhead is 185(15) μs |
| | 577.216(6) | 7$^+$ | 123.437(5) | 229.00(7) | | | | (1965BJ03) |
| | | | | | | | | 4. Regular band |
| 2 | 263.7895(23) | 5$^+$ | | | | | **2000PR03** | 1. Competing configurations |
| | 379.549(4) | 6$^+$ | 115.759(3) | | | | 1992KV01 | π3/2[411]⊗ν7/2[633] + |
| | 514.363(7) | 7$^+$ | 134.815(6) | 250.49(9) | | | | π7/2[523]⊗ν3/2[521] |
| | | | | | | | | 2. Regular band |
| 3 | 295.088(9) | 6$^+$ | | | | | **2000PR03** | 1. π7/2[523]⊗ν5/2[512] |
| | 423.654(10) | 7$^+$ | 128.566(5) | | | | 1967MO05 | 2. Regular band |
| 4 | 595.841(50) | 1$^-$ | | | | | **2000PR03** | 1. π1/2[411]⊗ν1/2[521] |
| | 628.435(12) | 2$^-$ | | | | | 1989SH18 | 2. Regular band |
| | 683.810(5) | 3$^-$ | | | | | | 3. Band is extended to 5$^-$ in 1989SH18 |
| | 742.08(8) | 4$^-$ | 266.53(5) | 113.644(4) | | | | diff energies but compressed in |
| | | | | | | | | 2000PR03 |
| 5 | 371.9878(25) | 4$^+$ | | | | | **2000PR03** | 1. π7/2[523]⊗ν1/2[521] |
| | 470.8433(27) | 5$^+$ | 98.8572(15) | 279.79(10) | | | 1967MO05 | 2. Regular band |
| | 588.104(4) | 6$^+$ | 117.264(3) | 216.160(5) | | | | |
| | 723.256(18) | 7$^+$ | 135.15(2) | | | | | |
| 6 | 373.158(6) | 1$^-$ | | | | | **2000PR03** | 1. π3/2[411]⊗ν1/2[521] |
| | 416.016(5) | 2$^-$ | (43.08) | | | | 1989SH18 | 2. Regular band |
| | 475.736(7) | 3$^-$ | (59.62) | 102.55(4) | | | | 3. Small signature splitting |
| | 562.859(7) | 4$^-$ | 87.193(15) | 146.808(8) | | | | |
| | 658.086(10) | 5$^-$ | | 182.302(16) | | | | |
| | 788.610(12) | 6$^-$ | 130.641(16) | 225.722(9) | | | | |
| 7 | 426.090(5) | 1$^+$ | | | | | **2000PR03** | 1. π7/2[523]⊗ν5/2[512] |
| | 464.558(6) | 2$^+$ | 38.493(6) | | | | 1992KV01 | 2. Competing configurations |
| | 522.045(5) | 3$^+$ | 57.517(8) | 95.953(2) | | | | π7/2[523]⊗ν5/2[512] |
| | 598.511(6) | 4$^+$ | 76.4663(14) | 134.00(3) | | | | π7/2[523]⊗ν5/2[523](1992KV01) |
| | 693.701(6) | 5$^+$ | 95.190(3) | 171.67(3) | | | | 3. Regular band |
| | 807.074(10) | 6$^+$ | 113.373(3) | | | | | |
| 8 | 430.040(4) | 2$^+$ | | | | | **2000PR03** | 1. Competing configurations |
| | 481.854(4) | 3$^+$ | 51.8155(7) | | | | 1992KV01 | π3/2[411]⊗ν7/2[633] + |
| | 547.934(5) | 4$^+$ | 66.103(7) | | | | | π7/2[523]⊗ν3/2[521] |
| | 634.329(4) | 5$^+$ | 86.359(11) | 152.45(3) | | | | 2. Regular band |
| | 732.549(14) | 6$^+$ | 98.200(15) | | | | | |
| 9 | 658.99(3) | 0$^-$ | | | | | **2000PR03** | 1. π1/2[411]⊗ν1/2[521] |
| | 774.516(15) | 1$^-$ | 115.51(3) | | | | 1989SH18 | 2. Irregular band |
| | 725.586(12) | 2$^-$ | | | | | | 3. Signature splitting |
| | 881.089(19) | 3$^-$ | 155.42(3) | | | | | 4. Diff energies in 1989SH18 |
| | 868.27(15) | 4$^-$ | | | | | | |

| S.No. | E$_{level}$ keV | I$^\pi$ | E$_\gamma$(M1) keV | E$_\gamma$(E2) keV | \|g$_K$-g$_R$\| | B(M1)/B(E2) ($\mu_N$/eb)$^2$ | Keywords | Configuration and Comments |
|---|---|---|---|---|---|---|---|---|
| 10 | 558.579(4) | 4$^+$ | | | | | **2000PR03** | 1. $\pi$7/2[523]⊗v1/2[510] |
| | 654.802(11) | 5$^+$ | 96.265(20) | | | | 1970BO29 | 2. Regular band |
| | 771.77(8) | 6$^+$ | 117.264(3) | | | | | |
| | | | | | | | | |
| 11 | 638.229(9) | 2$^-$ | | | | | **2000PR03** | 1. $\pi$3/2[411]⊗v1/2[521] |
| | 704.947(9) | 3$^-$ | | | | | 1989SH18 | 2. Regular band |
| | 792.98(2) | 4$^-$ | | 154.71(3) | | | | |
| | | | | | | | | |
| 12 | 567.654(6) | 1$^+$ | | | | | **2000PR03** | 1. $\pi$7/2[523]⊗v5/2[523] |
| | 605.109(6) | 2$^+$ | | | | | 1992KV01 | 2. Competing configurations |
| | 662.235(7) | 3$^+$ | 57.19(1) | 94.529(11) | | | | $\pi$7/2[523]⊗v5/2[512] |
| | 736.495(8) | 4$^+$ | 74.261(16) | 131.41(3) | | | | $\pi$7/2[523]⊗v5/2[523](1992KV01) |
| | 832.264(9) | 5$^+$ | 95.767(3) | 170.09(3) | | | | 3. Regular band |
| | 942.605(13) | 6$^+$ | 110.327(12) | 206.15(2) | | | | 4. Signature splitting |
| | | | | | | | | |
| 13 | 592.460(9) | 3$^+$ | | | | | **2000PR03** | 1. $\pi$1/2[411]⊗v7/2[633] |
| | 671.750(12) | 4$^+$ | | | | | 1970BO29 | 2. Regular band |
| | 769.549(16) | 5$^+$ | | | | | | |
| | 884.055(14) | 6$^+$ | 114.50(3) | 212.30(6) | | | | |
| | | | | | | | | |
| 14 | 599.4 | 4$^-$ | | | | | **1988BA79** | 1. $\pi$3/2[411]⊗v5/2[512] |
| | | | | | | | | |
| 15 | 719.44(4) | 4$^+$ | | | | | **2000PR03** | 1. $\pi$1/2[411]⊗v7/2[633] |
| | 806.68(18) | 5$^+$ | | | | | 1970BO29 | 2. Competing configurations |
| | 911.40(4) | 6$^+$ | 191.961(11) | | | | | $\pi$1/2[411]⊗v7/2[633] |
| | | | | | | | | $\pi$7/2[523]⊗v1/2[510](1970BO29) |
| | | | | | | | | 3. Regular band |
| | | | | | | | | |
| 16 | 740.6 | 1$^-$ | | | | | **1988BA79** | 1. $\pi$3/2[411]⊗v5/2[512] |
| | | | | | | | | |
| 17 | 771.5 | 4$^-$ | | | | | **1988BA79** | 1. $\pi$1/2[541]⊗v7/2[633] |
| | | | | | | | | |
| 18 | 815.072(10) | 3$^+$ | | | | | **2000PR03** | 1. $\pi$7/2[523]⊗v1/2[510] |
| | 890.988(12) | 4$^+$ | 75.985(8) | | | | 1970BO29 | 2. Regular band |
| | 985.15(4) | 5$^+$ | | 170.09(3) | | | | |
| | 1098.61(21) | 6$^+$ | | | | | | |
| | | | | | | | | |
| 19 | 1150 | 1$^+$ | | | | | **1982DE37** | 1. $\pi$5/2[413]⊗v7/2[633] |
| | 1184 | 2$^+$ | | | | | | 2. Nuclear reaction: |
| | 1235 | 3$^+$ | | | | | | $^{167}$Er(t,$\alpha$)$^{166}$Ho, E=17 MeV |
| | 1303 | 4$^+$ | | | | | | 2. Regular band |
| | 1387 | 5$^+$ | | | | | | |
| | | | | | | | | |
| 20 | 1272 | 6$^+$ | | | | | **1982DE37** | 1. $\pi$5/2[413]⊗v7/2[633] |
| | 1417 | 7$^+$ | | | | | | 2. Regular band |
| | | | | | | | | |
| 21 | 1560 | 6$^-$ | | | | | **1982DE37** | 1. $\pi$5/2[532]⊗v7/2[633] |
| | 1692 | 7$^-$ | | | | | | 2. Regular band |
| | 1834 | 8$^-$ | | | | | | |
| | 1998 | 9$^-$ | | | | | | |

| S.No. | E_level keV | $I^\pi$ | $E_\gamma$(M1) keV | $E_\gamma$(E2) keV | $\|g_K - g_R\|$ | B(M1)/B(E2) $(\mu_N/eb)^2$ | Keywords | Configuration and Comments |
|---|---|---|---|---|---|---|---|---|
| 22 | 0 | $0^-$ | | | | | **2000PR03** | 1. $\pi 7/2[523] \otimes \nu 7/2[633]$ |
| | 82.4709(19) | $1^-$ | 82.470(2) | | | | 1992KV01 | 2. Competing configurations |
| | 54.2388(7) | $2^-$ | | 54.2392(7) | | | | $\pi 7/2[523] \otimes \nu 7/2[633]$ |
| | 171.0726(12) | $3^-$ | 116.835(10) | 88.60(3) | | | | $\pi 7/2[523] \otimes \nu 3/2[651]$-$Q_{22}$ |
| | 180.4686(28) | $4^-$ | (9.39) | 126.228(3) | | | | (1992KV01) |
| | 329.777(4) | $5^-$ | 149.307(3) | 158.702(9) | | | | 3. Half-life of bandhead is 26.74(5) h |
| | 377.808(4) | $6^-$ | 48.0315(7) | 197.339(8) | | | | (1996DA04) |
| | 557.690(5) | $7^-$ | 179.882(4) | 227.88(7) | | | | 3. Irregular band |
| | | | | | | | | 4. Signature splitting |
| 23 | 5.971(12) | $7^-$ | | | | | **2000PR03** | 1. $\pi 7/2[523] \otimes \nu 7/2[633]$ |
| | 137.731(13) | $8^-$ | 131.759(5) | | | | 1992KV01 | 2. Competing configurations |
| | 286.96(10) | $9^-$ | | 280.99(10) | | | | $\pi 7/2[523] \otimes \nu 7/2[633]$ |
| | | | | | | | | $\pi 7/2[523] \otimes \nu 3/2[651]$-$Q_{22}$ |
| | | | | | | | | (1992KV01) |
| | | | | | | | | 3. Half-life of bandhead is $1.20 \times 10^3$ |
| | | | | | | | | (15) y (1965FA01) |
| | | | | | | | | 4. Regular band |
| 24 | 760.375(7) | $3^-$ | | | | | **2000PR03** | 1. Competing configurations |
| | 837.734(8) | $4^-$ | | | | | | $\pi 1/2[541] \otimes \nu 7/2[633]$ |
| | 935.047(17) | $5^-$ | 97.253(20) | 174.77(4) | | | | $\pi 1/2[411] \otimes \nu 5/2[512]$ |
| | | | | | | | | 2. Regular band |
| 25 | 722.00(15) | $6^+$ | | | | | **2000PR03** | 1. $\pi 7/2[523] \otimes \nu 5/2[523]$ |
| | 848.49(21) | $7^+$ | | | | | | 2. Regular band |
| 26 | 905.60(1) | $2^+$ | | | | | **2000PR03** | 1. $\pi 7/2[523] \otimes \nu 3/2[521]$ |
| | 961.23(16) | $3^+$ | | | | | | 2. Tentative band |
| | 1030.47(23) | $4^+$ | | | | | | 3. Regular band |
| 27 | 925.51(3) | $5^+$ | | | | | **2000PR03** | 1. $\pi 7/2[523] \otimes \nu 3/2[521]$ |
| | 1038.43(20) | $6^+$ | | | | | | 2. Tentative band |
| | | | | | | | | 3. Regular band |

$^{168}_{67}Ho_{101}$

| S.No. | E_level keV | $I^\pi$ | $E_\gamma$(M1) keV | $E_\gamma$(E2) keV | $\|g_K - g_R\|$ | B(M1)/B(E2) $(\mu_N/eb)^2$ | Keywords | Configuration and Comments |
|---|---|---|---|---|---|---|---|---|
| 1 | 0.0 | $3^+$ | | | | | **1971HA42** | 1. $\pi 7/2[523] \otimes \nu 1/2[521]$ |
| | | | | | | | | 2. Nuclear reaction: |
| | | | | | | | | $^{170}$Er($\gamma$, np)$^{168}$Ho, E = 70 MeV |
| | | | | | | | | 3. Half-life of bandhead is 3.3(5) min |
| 2 | (59) | $(6^+)$ | | | | | **1990CH37** | 1. $\pi 7/2[523] \otimes \nu 5/2[512]$ |
| | | | | | | | | 2. Nuclear reaction: |
| | | | | | | | | $^{168}$Ho IT decay |
| | | | | | | | | 3. Half-life of bandhead is 132(4) s |
| 3 | (143.5) | $(1)^-$ | | | | | **1991SH20** | 1. $\pi 3/2[411] \otimes \nu 1/2[521]$ |
| | | | | | | | 1990CH37 | 2. Nuclear reaction: |
| | | | | | | | | $^{168}$Dy $\beta^-$ decay |
| | | | | | | | | 2. Half-life of bandhead is >4 µs |
| | | | | | | | | (1990CH37) |

| S.No. | E_level keV | $I^\pi$ | $E_\gamma$ (M1) keV | $E_\gamma$ (E2) keV | $|g_K - g_R|$ | B(M1)/B(E2) $(\mu_N/eb)^2$ | Keywords | Configuration and Comments |
|---|---|---|---|---|---|---|---|---|
| 4 | (187.3) | (1)⁻ | | | | | **1991SH20** | 1. Competing configurations |
| | | | | | | | 1990CH37 | π1/2[411]⊗ν1/2[521] |
| | | | | | | | | π3/2[411]⊗ν1/2[521] |
| | | | | | | | | |
| 5 | 192.5 | 1⁺ | | | | | **1991SH20** | 1. π7/2[523]⊗ν5/2[512] |
| | | | | | | | 1990CH37 | |

$^{170}_{67}Ho_{103}$

| S.No. | E_level keV | $I^\pi$ | $E_\gamma$ (M1) keV | $E_\gamma$ (E2) keV | $|g_K - g_R|$ | B(M1)/B(E2) $(\mu_N/eb)^2$ | Keywords | Configuration and Comments |
|---|---|---|---|---|---|---|---|---|
| 1 | (0.0) | (6⁺) | | | | | **1978KA16** | 1. π7/2[523]⊗ν5/2[512] |
| | | | | | | | | 2. Nuclear reaction: |
| | | | | | | | | ¹⁷⁰Er(n,p)¹⁷⁰Ho, E = 14-15 MeV |
| | | | | | | | | 3. Half-life of bandhead is 2.76(5) min |
| | | | | | | | | |
| 2 | (120(70)) | 1⁽⁺⁾ | | | | | **1978TU04** | 1. π7/2[523]⊗ν5/2[512] |
| | | | | | | | | 2. Nuclear reaction: |
| | | | | | | | | ¹⁷⁰Er(n,p)¹⁷⁰Ho, E = 14-15 MeV |
| | | | | | | | | 2. Half-life of bandhead is 43(2)s |

$^{156}_{69}Tm_{87}$

| S.No. | E_level keV | $I^\pi$ | $E_\gamma$ (M1) keV | $E_\gamma$ (E2) keV | $|g_K - g_R|$ | B(M1)/B(E2) $(\mu_N/eb)^2$ | Keywords | Configuration and Comments |
|---|---|---|---|---|---|---|---|---|
| 1 | 0.0 | 2⁻ | | | | | **1983ML01** | 1. Competing configurations |
| | | | | | | | | $\pi d_{3/2}(h_{11/2})^4_{0+}\otimes(f_{7/2})^5_{7/2}$ |
| | | | | | | | | π7/2[404]⊗ν3/2[532] |
| | | | | | | | | π5/2[402]⊗ν1/2[530] |
| | | | | | | | | 2. Nuclear reaction: |
| | | | | | | | | ¹⁵⁶Yb ε decay |
| | | | | | | | | 3. Half-life of bandhead is 82(5) s |
| | | | | | | | | |
| 2 | 115.2(2) | 1⁺ | | | | | **1983ML01** | 1. Competing configurations |
| | | | | | | | | $h_{11/2}\otimes f_{7/2}$ |
| | | | | | | | | π5/2[532]⊗ν3/2[532] |
| | | | | | | | | π3/2[541]⊗ν1/2[530] |
| | | | | | | | | |
| 3 | (203) | (3⁻) | | | | | **1993GAZU** | 1. Competing configurations |
| | | | | | | | | $\pi d_{3/2}\otimes\nu f_{7/2}$ |
| | | | | | | | | $\pi s_{1/2}\otimes\nu f_{7/2}$ |
| | | | | | | | | |
| 4 | (419) | (5⁻) | | | | | **1993GAZU** | 1. Competing configurations |
| | | | | | | | | $\pi d_{3/2}\otimes\nu f_{7/2}$ |
| | | | | | | | | $\pi s_{1/2}\otimes\nu f_{7/2}$ |
| | | | | | | | | |
| 5 | 203.6+x | (11⁻) | | | | | **1995SU12** | 1. π7/2[523]⊗ν1/2[660] |
| | 771.2+x | (13⁻) | | 568 | | | | 2. Nuclear reaction: |
| | 1366.0+x | (15⁻) | | 595 | | | | ¹⁴⁴Sm(¹⁹F,2p5n)¹⁵⁶Tm |
| | 1725.7+x | (16⁻) | 359 | | | | | 3. Regular band |
| | 2056.6+x | (17⁻) | 330 | | | | | 4. Level energy is adopted from ENSDF |
| | 2335.6+x | (18⁻) | | 610 | | | | |
| | 2535.0+x | (19⁻) | 199 | 479 | | | | |
| | 3234+x | (21⁻) | | 699 | | | | |
| | 3407+x | (22⁻) | 173 | | | | | |
| | 3978+x | (23⁻) | | 744 | | | | |
| | 4773+x | (25⁻) | | 795 | | | | |

$^{158}_{69}Tm_{89}$

| S.No. | E$_{level}$ keV | I$^\pi$ | E$_\gamma$(M1) keV | E$_\gamma$(E2) keV | \|g$_K$-g$_R$\| | B(M1)/B(E2) ($\mu_N$/eb)$^2$ | Keywords | Configuration and Comments |
|---|---|---|---|---|---|---|---|---|
| 1 | 0.0 | 2$^-$ | | | | | **1975AG01** | 1. $\pi 7/2[404] \otimes \nu 3/2[521]$ |
| | | | | | | | | 2. Nuclear reaction: |
| | | | | | | | | $^{158}$Tm $\varepsilon$ decay |
| | | | | | | | | 3. Half-life of bandhead is 4.02(10) min |
| | | | | | | | | |
| 2 | A+130 | 11$^-$ | | | | | **1975AG01** | 1. $\pi h_{11/2} \otimes \nu i_{13/2}$, K$^\pi$ = 4$^-$ |
| | A+384 | 12$^-$ | | | | | 1986DR06 | 2. Regular band |
| | A+513 | 13$^-$ | | | | | | 3. Signature splitting |
| | A+812 | 14$^-$ | | | | | | |
| | A+1030 | 15$^-$ | | | | | | |
| | A+1357 | 16$^-$ | | | | | | |
| | A+1641 | 17$^-$ | | | | | | |
| | A+1996 | 18$^-$ | | | | | | |
| | A+2324 | 19$^-$ | | | | | | |
| | A+2705 | 20$^-$ | | | | | | |
| | A+3058 | 21$^-$ | | | | | | |
| | A+3466 | 22$^-$ | | | | | | |
| | A+3843 | 23$^-$ | | | | | | |
| | A+4264 | (24$^-$) | | | | | | |

$^{160}_{69}Tm_{91}$

| S.No. | E$_{level}$ keV | I$^\pi$ | E$_\gamma$(M1) keV | E$_\gamma$(E2) keV | \|g$_K$-g$_R$\| | B(M1)/B(E2) ($\mu_N$/eb)$^2$ | Keywords | Configuration and Comments |
|---|---|---|---|---|---|---|---|---|
| 1 | 0.0 | 1$^-$ | | | | | **1978AD03** | 1. $\pi 1/2[411] \otimes \nu 3/2[521]$ |
| | 42.02 | 2$^-$ | 42.02 | | | | 1971EK01 | 2. Nuclear reaction: |
| | | | | | | | | $^{160}$Yb $\varepsilon$ decay |
| | | | | | | | | 3. Half-life of bandhead is 9 min |
| | | | | | | | | (1971EK01) |
| | | | | | | | | |
| 2 | 244.8 | (7$^+$) | | | | | **2008SU08** | 1. $\pi h_{11/2} \otimes h_{9/2}$, K$^\pi$ = (5$^+$) |
| | 342.7 | (8$^+$) | 97.9 | | | | 2005LA32 | $\pi 7/2[523] \otimes \nu 3/2[521]$ (1989AN08) |
| | 484.8 | (9$^+$) | 142.0 | | | | 1989AN08 | 2. Nuclear reaction: |
| | 655.4 | (10$^+$) | 170.6 | 312.8 | | | | $^{146}$Nd($^{19}$F,5n$\gamma$)$^{160}$Tm, E = 102 MeV |
| | 865.7 | (11$^+$) | 210.3 | 380.8 | | | | 3. Half-life of bandhead is 74.5(15) s |
| | 1094.5 | (12$^+$) | 228.6 | 439.2 | | | | (1983SI20) |
| | 1359.1 | (13$^+$) | 264.3 | 493.5 | | | | 4. Regular band |
| | 1632.5 | (14$^+$) | 273.2 | 538.1 | | | | 5. Small signature splitting in $\Delta E_\gamma$ vs I |
| | 1938.9 | (15$^+$) | 306.3 | 579.9 | | | | 6. Signature inversion at I = (19$^+$) and |
| | 2243.2 | (16$^+$) | 304.6 | 610.7 | | | | I = (24$^+$) |
| | 2570.9 | (17$^+$) | 327.7 | 631.4 | | | | 7. Band extended utpo 37$^+$ with |
| | 2814.6 | (18$^+$) | 243.6 | 571.7 | | | | confirmed spin and parity but |
| | 3052.0 | (19$^+$) | 237.7 | 480.6 | | | | uncertainty in energy (2005LA32) |
| | 3314.6 | (20$^+$) | 262.9 | 499.6 | | | | different energies |
| | 3597.3 | (21$^+$) | 282.7 | 545.4 | | | | 8. Band extended upto (31$^+$) (1989AN08) |
| | 3911.8 | (22$^+$) | 314.4 | 597.1 | | | | 9. Level energy is adopted from XUNDL |
| | 4250.0 | (23$^+$) | 338.5 | 652.8 | | | | |
| | 4611.0 | (24$^+$) | 361.3 | 699.0 | | | | |
| | 5006.2 | (25$^+$) | | 756.2 | | | | |
| | 5410.5 | (26$^+$) | | 799.5 | | | | |
| | 5848.0 | (27$^+$) | | 841.8 | | | | |

| S.No. | E$_{level}$ keV | I$^\pi$ | E$_\gamma$ (M1) keV | E$_\gamma$ (E2) keV | \|g$_K$-g$_R$\| | B(M1)/B(E2) ($\mu_N$/eb)$^2$ | Keywords | Configuration and Comments |
|---|---|---|---|---|---|---|---|---|
| 3 | (140.32) | (2$^+$) | | | | | **1978AD03** | 1. $\pi$1/2[411]$\otimes$v5/2[642] |
| 4 | (174.40) | (2$^+$) | | | | | **1978AD03** | 1. $\pi$7/2[523]$\otimes$v3/2[521] |
| | | | | | | | | 2. Half-life of bandhead is 17(1) ns |
| 5 | 215.78 | 1$^+$ | | | | | **1978AD03** | 1. $\pi$7/2[523]$\otimes$v3/2[521] |
| | | | | | | | | 2. Half-life of bandhead is 0.65(15) ns |
| 6 | 444.3 | (9$^-$) | | | | | **2008SU08** | 1. $\pi$h$_{11/2}\otimes$ vi$_{13/2}$ |
| | 523.5 | (10$^-$) | 79.3 | | | | 2005LA32 | 2. Side band (1989AN08) |
| | 606.8 | (11$^-$) | 83.3 | 162.7 | | | 1989AN08 | 3. Regular band |
| | 783.5 | (12$^-$) | 176.6 | 259.6 | | | | 4. signature splitting |
| | 936.4 | (13$^-$) | 152.6 | 330.0 | | | | 5. Signature inversion at I = (19$^-$) |
| | 1182.0 | (14$^-$) | 245.7 | 398.4 | | | | 6. Band extended upto 46$^-$ with |
| | 1405.9 | (15$^-$) | 223.8 | 469.6 | | | | confirmed spin and parity but |
| | 1695.8 | (16$^-$) | 289.8 | 513.9 | | | | uncertainty in energy (2005LA32) |
| | 1985.6 | (17$^-$) | 289.8 | 579.8 | | | | different energies |
| | 2302.6 | (18$^-$) | 316.8 | 606.7 | | | | 8. Band extended upto (34$^-$) (1989AN08) |
| | 2647.5 | (19$^-$) | 344.3 | 662.2 | | | | Diff energies |
| | 2977.5 | (20$^-$) | 329.8 | 675.0 | | | | 9. Level energy is adopted from XUNDL |
| | 3358.4 | (21$^-$) | 380.6 | 711.0 | | | | |
| | 3688.5 | (22$^-$) | 330.1 | 711.2 | | | | |
| | 4081.8 | (23$^-$) | 393.5 | 723.3 | | | | |
| | 4411.5 | (24$^-$) | 329.7 | 722.8 | | | | |
| | 4812.8 | (25$^-$) | 401.0 | 731.2 | | | | |
| | 5155.3 | (26$^-$) | 342.5 | 743.9 | | | | |
| | 5582.2 | (27$^-$) | 425.0 | 769.4 | | | | |
| | 5946.9 | (28$^-$) | 364.4 | 791.8 | | | | |
| | 6410.6 | (29$^-$) | | 828.4 | | | | |
| | 6798.7 | (30$^-$) | | 851.8 | | | | |
| | 7304.0 | (31$^-$) | | 893.4 | | | | |
| 7 | x | (8$^+$) | | | | | **2008SU08** | 1. $\pi$d$_{3/2}\otimes$ vi$_{13/2}$ |
| | 234.4+x | (10$^+$) | | 234.4 | | | | 2. Regular band |
| | 599.0+x | (12$^+$) | | 364.6 | | | | 3. Level energy is adopted from XUNDL |
| | 1067.8+x | (14$^+$) | | 468.8 | | | | |
| | 1624.4+x | (16$^+$) | | 556.6 | | | | |
| | 2268.7+x | (18$^+$) | | 644.3 | | | | |
| | 2973.4+x | (20$^+$) | | 704.7 | | | | |

$^{162}_{69}Tm_{93}$

| S.No. | E_level keV | I^π | E_γ(M1) keV | E_γ(E2) keV | \|g_K-g_R\| | B(M1)/B(E2) (μ_N/eb)^2 | Keywords | Configuration and Comments |
|---|---|---|---|---|---|---|---|---|
| 1 | 0.0 | 1^- | | | | | **1974DE47** | 1. π1/2[411]⊗ν3/2[521] |
| | 44.66 | 2^- | 44.66 | | | | | 2. Nuclear reaction; $^{162}$Tm IT decay |
| | | | | | | | | 3. Half-life of bandhead is 21.55(30) min |
| 2 | 66.90 | 2^- | | | | | **1974DE47** | 1. π7/2[404]⊗ν3/2[521] |
| 3 | 0.0+x | 5^+ | | | | | **1998ES06** | 1. π7/2[523]⊗ν3/2[521] |
| | 96.4+x | 6^+ | 96.00(40) | | | | 1974DE47 | 2. Nuclear reaction; $^{130}$Te($^{37}$Cl,5nγ)$^{162}$Tm, E=166MeV |
| | 203.1+x | 7^+ | 106.60(30) | | | | | 3. Half-life of bandhead is 24.3(17) s (1974DE47) |
| | 323.4+x | 8^+ | 120.30(30) | 111.93(16) | | | | 4. Regular band |
| | 450.2+x | 9^+ | 126.87(16) | 135.35(17) | | | | 5. Signature splitting from 10 to 18 then 35 to 45 |
| | 596.0+x | 10^+ | 145.48(16) | 272.34(18) | | | | |
| | 787.2+x | 11^+ | 192.15(22) | 337.03(17) | | | | |
| | 979.5+x | 12^+ | 192.28(22) | 384.01(17) | | | | |
| | 1227.3+x | 13^+ | 247.49(17) | 439.36(17) | | | | |
| | 1459.3+x | 14^+ | 232.03(17) | 479.71(17) | | | | |
| | 1741.6+x | 15^+ | 282.27(17) | 514.82(16) | | | | |
| | 1999.7+x | 16^+ | 258.11(18) | 540.21(17) | | | | |
| | 2273.7+x | 17^+ | 274.00(18) | 532.17(17) | | | | |
| | 2536.3+x | 18^+ | 262.45(18) | 622.08(17) | | | | |
| | 2800.0+x | 19^+ | 263.70(17) | 544 | | | | |
| | 3076.2+x | 20^+ | 276.26(17) | 540.21(17) | | | | |
| | 3360.5+x | 21^+ | 284.25(16) | 560.37(16) | | | | |
| | 3662.2+x | 22^+ | 301.75(16) | 586.14(16) | | | | |
| | 3974.6+x | 23^+ | 312.41(17) | 614.27(17) | | | | |
| | 4309.9+x | 24^+ | 335.25(17) | 647.38(17) | | | | |
| | 4655.5+x | 25^+ | 345.61(17) | 681.20(17) | | | | |
| | 5025.8+x | 26^+ | 370.26(17) | 716.86(17) | | | | |
| | 5406.5+x | 27^+ | 380.43(18) | 750.92(17) | | | | |
| | 5808.3+x | 28^+ | 401.48(18) | 782.20(17) | | | | |
| | 6221.3+x | 29^+ | 412.96(18) | 814.84(17) | | | | |
| | 6650.8+x | 30^+ | 429.45(25) | 842.63(17) | | | | |
| | 7094.8+x | 31^+ | 444.70(30) | 873.24(18) | | | | |
| | 7552.3+x | 32^+ | (458) | 901.16(18) | | | | |
| | 8024.1+x | 33^+ | | 929.17(20) | | | | |
| | 8513.0+x | 34^+ | | 961.08(19) | | | | |
| | 9008.6+x | 35^+ | | 984.36(23) | | | | |
| | 9534.5+x | 36^+ | | 1021.33(22) | | | | |
| | 10046.3+x | 37^+ | | 1038.34(25) | | | | |
| | 10616.8+x | 38^+ | | 1082.05(27) | | | | |
| | 11136.5+x | 39^+ | | 1090.21(31) | | | | |
| | 11756.1+x | 40^+ | | 1139.28(35) | | | | |
| | 12274.5+x | 41^+ | | 1138.16(38) | | | | |
| | 12935.6+x | 42^+ | | 1179.39(48) | | | | |
| | 13451.6+x | 43^+ | | 1176.53(43) | | | | |
| | 14129.1+x | 44^+ | | (1194) | | | | |
| | 14650.3+x | 45^+ | | 1198.88(40) | | | | |
| | 15865.3+x | 47^+ | | (1215) | | | | |

| S.No. | E_level keV | I^π | E_γ (M1) keV | E_γ (E2) keV | \|g_K-g_R\| | B(M1)/B(E2) (μ_N/eb)^2 | Keywords | Configuration and Comments |
|---|---|---|---|---|---|---|---|---|
| 4 | 163.35 | 1^+ | | | | | **1982AD03** | 1. π7/2[523]⊗ν5/2[523] |
| | 290.30 | 2^+ | | | | | | 2. Nuclear reaction: |
| | | | | | | | | $^{162}$Yb ε+β^+ decay |
| | | | | | | | | 3. Half-life of bandhead is 1.1(1) ns |
| | | | | | | | | (1978SC10) |
| | | | | | | | | |
| 5 | 163.8 | 6^- | | | | | **1998ES06** | 1. π7/2[523]⊗ν5/2[642] |
| | 189.8 | 7^- | | | | | 1986DR06 | 2. Regular band |
| | 237.8 | 8^- | | 73.87(16) | | | | 3. Signature splitting more pronounced at |
| | 305.2 | 9^- | 68.14(16) | 115.55(19) | | | | at higher spins |
| | 401.1 | 10^- | 95.88(16) | 162.93(16) | | | | 4. Signature inversion at I=17^- (ΔE_γ vs I) |
| | 515.0 | 11^- | 113.95(15) | 209.32(16) | | | | 5. Decoupled band |
| | 671.5 | 12^- | 156.64(15) | 270.28(16) | | | | 6. Aligned band |
| | 839.0 | 13^- | 167.55(15) | 323.79(16) | | | | |
| | 1051.9 | 14^- | 212.90(15) | 380.17(16) | | | | |
| | 1274.6 | 15^- | 222.71(15) | 436.03(30) | | | | |
| | 1530.8 | 16^- | 256.21(15) | 479.70(16) | | | | |
| | 1809.7 | 17^- | 278.49(16) | 534.55(16) | | | | |
| | 2096.3 | 18^- | 286.55(16) | 562.12(16) | | | | |
| | 2425.0 | 19^- | 328.37(16) | 615.72(16) | | | | |
| | 2731.4 | 20^- | 306.34(16) | 635.09(16) | | | | |
| | 3101.9 | 21^- | 370.31(16) | 677.06(16) | | | | |
| | 3418.6 | 22^- | 316.74(16) | 687.37(16) | | | | |
| | 3817.3 | 23^- | 398.43(16) | 715.60(16) | | | | |
| | 4140.3 | 24^- | 323.07(17) | 721.25(16) | | | | |
| | 4554.9 | 25^- | 414.62(16) | 737.53(16) | | | | |
| | 4890.8 | 26^- | 335.48(19) | 750.45(16) | | | | |
| | 5324.9 | 27^- | 434.07(48) | 769.97(16) | | | | |
| | 5684.9 | 28^- | 360.00(17) | 794.50(16) | | | | |
| | 6150.8 | 29^- | 465.91(18) | 826.00(17) | | | | |
| | 6539.2 | 30^- | 388.38(34) | 853.82(16) | | | | |
| | 7045.7 | 31^- | 506.44(48) | 894.96(17) | | | | |
| | 7458.6 | 32^- | 412.87(48) | 919.48(18) | | | | |
| | 8015.0 | 33^- | 556.37(48) | 969.57(19) | | | | |
| | 8442.5 | 34^- | 427.35(45) | 98401(21) | | | | |
| | 9058.3 | 35^- | | 1042.74(24) | | | | |
| | 9488.2 | 36^- | (430) | 1045.60(50) | | | | |
| | 10171.1 | 37^- | | 1113.00(48) | | | | |
| | 10592.3 | 38^- | | 1104.20(50) | | | | |
| | 11346.5 | 39^- | | 1175.00(48) | | | | |
| | 11744.6 | 40^- | | 1152.20(50) | | | | |
| | 12576.0 | 41^- | | 1230.00(48) | | | | |
| | 12896.7 | 42^- | | 1152.20(50) | | | | |

| S.No. | $E_{level}$ keV | $I^\pi$ | $E_\gamma$(M1) keV | $E_\gamma$(E2) keV | $|g_K-g_R|$ | B(M1)/B(E2) $(\mu_N/eb)^2$ | Keywords | Configuration and Comments |
|---|---|---|---|---|---|---|---|---|
| 6 | 151.5 | $6^+$ | | | | | **1998ES06** | 1. $\pi 7/2[404] \otimes \nu 5/2[642]$ |
| | 233.0 | $7^+$ | 81.50(16) | | | | 1989AN08 | 2. Regular band |
| | 294.0 | $8^+$ | | | | | | 3. Signature splitting more pronounced at |
| | 393.4 | $9^+$ | 99.55(16) | 160.55(17) | | | | at higher spins |
| | 526.5 | $10^+$ | 133.05(16) | 232.68(16) | | | | 4. Signature inversion at I=$21^+$ and |
| | 690.5 | $11^+$ | 164.00(16) | 297.10(16) | | | | I=$32^+$ ($\Delta E_\gamma$ vs I) |
| | 883.4 | $12^+$ | 192.95(16) | 356.95(16) | | | | |
| | 1100.5 | $13^+$ | 217.05(16) | 410.10(16) | | | | |
| | 1344.3 | $14^+$ | 243.80(16) | 460.54(16) | | | | |
| | 1605.2 | $15^+$ | 260.95(17) | 504.45(16) | | | | |
| | 1890.7 | $16^+$ | 285.54(18) | 546.51(16) | | | | |
| | 2188.0 | $17^+$ | 297.35(19) | 583.08(16) | | | | |
| | 2507.4 | $18^+$ | 319.40(18) | 616.87(16) | | | | |
| | 2834.1 | $19^+$ | 326.69(19) | 646.21(16) | | | | |
| | 3178.9 | $20^+$ | 344.37(22) | 670.91(17) | | | | |
| | 3540.8 | $21^+$ | | 706.35(17) | | | | |
| | 3890.0 | $22^+$ | | 711.72(17) | | | | |
| | 4259.6 | $23^+$ | | 718.25(16) | | | | |
| | 4613.0 | $24^+$ | | 722.93(19) | | | | |
| | 5013.7 | $25^+$ | | 754.05(20) | | | | |
| | 5369.6 | $26^+$ | | 756.61(18) | | | | |
| | 5812.7 | $27^+$ | | 798.94(22 | | | | |
| | 6187.8 | $28^+$ | | 818.34(20) | | | | |
| | 6658.7 | $29^+$ | | 846.09(23) | | | | |
| | 7069.0 | $30^+$ | | 881.01(23) | | | | |
| | 7528.7 | $31^+$ | | 869.93(27) | | | | |
| | 8008.5 | $32^+$ | | 939.54(22) | | | | |
| | 8418.1 | $33^+$ | | 889.40(47) | | | | |
| | | | | | | | | |
| 7 | 373.2 | $6^-$ | | | | | **1998ES06** | 1. $\pi 1/2[541] \otimes \nu 5/2[642]$ |
| | 567.0 | $8^-$ | | 193.23(18) | | | 1989AN08 | 2. Regular band |
| | 717.9 | $9^-$ | | | | | | 3. Signature splitting |
| | 815.0 | $10^-$ | | 248.21(17) | | | | 4. Signature inversion at I = $18^-$ |
| | 1010.6 | $11^-$ | | 293.01(28) | | | | |
| | 1141.0 | $12^-$ | | 325.88(17) | | | | |
| | 1375.5 | $13^-$ | 234.49(30) | 364.64(17) | | | | |
| | 1553.9 | $14^-$ | | 413.48(19) | | | | |
| | 1816.5 | $15^-$ | 263.03(28) | 440.70(19) | | | | |
| | 2053.8 | $16^-$ | | 500.23(17) | | | | |
| | 2334.0 | $17^-$ | 280.21(20) | 517.26(18) | | | | |
| | 2636.0 | $18^-$ | | 582.14(18) | | | | |
| | 2923.7 | $19^-$ | | 590.14(17) | | | | |
| | 3297.5 | $20^-$ | | 634.47(36) | | | | |
| | 3583.4 | $21^-$ | | 659.68(18) | | | | |
| | 4008.2 | $22^-$ | | 710.64(28) | | | | |
| | 4308.7 | $23^-$ | | 725.16(19) | | | | |
| | 4752.2 | $24^-$ | | 744.18(41) | | | | |
| | 5085.5 | $25^-$ | | 776.83(20) | | | | |
| | 5905.2 | $27^-$ | | 819.68(23) | | | | |
| | 6753.0 | $29^-$ | | 847.82(20) | | | | |

| S.No. | E$_{level}$ keV | I$^\pi$ | E$_\gamma$(M1) keV | E$_\gamma$(E2) keV | \|g$_K$-g$_R$\| | B(M1)/B(E2) ($\mu_N$/eb)$^2$ | Keywords | Configuration and Comments |
|---|---|---|---|---|---|---|---|---|
| 8 | 107.7 | 6$^+$ | | | | | **1998ES06** | 1. $\pi$1/2[411]$\otimes\nu$5/2[642] |
| | 211.4 | 7$^+$ | 103.60(40) | | | | 1989AN08 | 2. Competing configurations are: |
| | 314.9 | 8$^+$ | 103.40(40) | 207.26(18) | | | | $\pi$1/2[541]$\otimes\nu$5/2[642] |
| | 413.8 | 9$^+$ | 98.86(31) | 202.29(20) | | | | $\pi$1/2[411]$\otimes\nu$5/2[642] (1989AN08) |
| | 556.0 | 10$^+$ | 142.28(27) | 241.17(16) | | | | 3. Regular band |
| | 729.1 | 11$^+$ | 173 | 314.72(24) | | | | 4. Signature splitting more pronounced at |
| | 911.4 | 12$^+$ | 182.32(34) | 355.20(16) | | | | at higher spins |
| | 1150.7 | 13$^+$ | 239.80(40) | 421.35(16) | | | | |
| | 1369.2 | 14$^+$ | | 457.86(16) | | | | |
| | 1664.4 | 15$^+$ | 296 | 513.87(17) | | | | |
| | 1914.2 | 16$^+$ | | 545.07(16) | | | | |
| | 2255.4 | 17$^+$ | | 591.32(17) | | | | |
| | 2524.2 | 18$^+$ | | 610.36(16) | | | | |
| | 2910.1 | 19$^+$ | | 654.36(18) | | | | |
| | 3183.7 | 20$^+$ | | 658.50(55) | | | | |
| | 3609.4 | 21$^+$ | | 699.60(18) | | | | |
| | 3861.8 | 22$^+$ | | 678.34(18) | | | | |
| | 4340.5 | 23$^+$ | | 731.01(23) | | | | |
| | 4519.9 | 24$^+$ | | 658.50(55) | | | | |
| | 5100.5 | 25$^+$ | | 759.94(21) | | | | |
| | 5234.1 | 26$^+$ | | 714.70(50) | | | | |
| | | | | | | | | |
| 9 | 3517.6 | 21$^+$ | | | | | **1998ES06** | 1. $\pi$7/2[523]$\otimes\nu$5/2[523] |
| | 3821.5 | 22$^+$ | 303.48(18) | | | | | 2. Regular band |
| | 4146.5 | 23$^+$ | 324.47(23) | 628.06(19) | | | | 3. Small signature splitting in $\Delta$E$_\gamma$ vs I |
| | 4491.1 | 24$^+$ | 344.62(18) | 669.93(22) | | | | |
| | 4855.3 | 25$^+$ | 364.17(19) | 709.21(22) | | | | |
| | 5249.1 | 26$^+$ | 393.89(24) | 758.65(25) | | | | |
| | 5642.7 | 27$^+$ | 393.35(22) | 786.89(21) | | | | |
| | 6061.2 | 28$^+$ | 418.58(29) | 811.74(28) | | | | |
| | 6482.0 | 29$^+$ | 420.79(26) | 839.57(23) | | | | |
| | 7371.5 | 31$^+$ | | 889.40(47) | | | | |
| | | | | | | | | |
| 10 | 3878.7 | 22$^-$ | | | | | **1998ES06** | 1. $\pi$7/2[404]$\otimes\nu$3/2[521] |
| | 4196.5 | 23$^-$ | | | | | | 2. Regular band |
| | 4510.4 | 24$^-$ | | 631.70(18) | | | | 3. Small signature splitting |
| | 4868.1 | 25$^-$ | | 671.30(17) | | | | |
| | 5215.5 | 26$^-$ | | 705.71(19) | | | | |
| | 5602.7 | 27$^-$ | | 734.52(18) | | | | |
| | 5976.6 | 28$^-$ | | 761.12(17) | | | | |
| | 6395.0 | 29$^-$ | | 792.24(18) | | | | |
| | 6799.3 | 30$^-$ | | 822.65(18) | | | | |
| | 7256.2 | 31$^-$ | | 861.20(17) | | | | |
| | 7694.7 | 32$^-$ | | 895.49(19) | | | | |
| | 8194.3 | 33$^-$ | | 938.10(19) | | | | |
| | 8665.3 | 34$^-$ | | 970.32(22) | | | | |
| | 9209.3 | 35$^-$ | | 1014.98(20) | | | | |
| | 9704.3 | 36$^-$ | | 1038.98(24) | | | | |
| | 10298.9 | 37$^-$ | | 1089.64(29) | | | | |

| S.No. | E$_{level}$ keV | I$^\pi$ | E$_\gamma$ (M1) keV | E$_\gamma$ (E2) keV | |g$_K$-g$_R$| | B (M1)/B (E2) ($\mu_N$/eb)$^2$ | Keywords | Configuration and Comments |
|---|---|---|---|---|---|---|---|---|
| 11 | x | 6$^+$ | | | | | **1998ES06** | 1. π5/2[402]⊗ν5/2[642] |
| | x+97 | 7$^+$ | (97) | | | | | 2. Regular band |
| | x+200 | 8$^+$ | 102.60(48) | | | | | 3. Signature splitting |
| | x+326 | 9$^+$ | 126.32(17) | (228) | | | | 4. signature inversion at I = 10$^+$ |
| | x+490 | 10$^+$ | 164.00(16) | 290.57(28) | | | | 5. B(M1)/B(E2) values read from plot |
| | x+674 | 11$^+$ | 183.88(17) | 348.03(19) | | | | |
| | x+900 | 12$^+$ | 225.83(17) | 409.81(20) | | | | |
| | x+1137 | 13$^+$ | 236.53(18) | 462.34(18) | | | | |
| | x+1411 | 14$^+$ | 274.00(18) | 510.48(18) | | | | |
| | x+1692 | 15$^+$ | 281.07(18) | 555.84(18) | | | | |
| | x+2001 | 16$^+$ | 309.10(30) | 590.48(19) | | | | |
| | x+2314 | 17$^+$ | 313.50(30) | 622.45(20) | | | | |
| | x+2651 | 18$^+$ | | 649.50(34) | | | | |
| | x+2965 | 19$^+$ | | 651.20(23) | | | | |

$^{164}_{69}Tm_{95}$

| S.No. | E$_{level}$ keV | I$^\pi$ | E$_\gamma$ (M1) keV | E$_\gamma$ (E2) keV | |g$_K$-g$_R$| | B (M1)/B (E2) ($\mu_N$/eb)$^2$ | Keywords | Configuration and Comments |
|---|---|---|---|---|---|---|---|---|
| 1 | 0.0 | 1$^+$ | | | | | **1995RE05** | 1. π7/2[523]⊗ν5/2[523] |
| | 37.5 | 2$^+$ | | | | | 1971DE22 | 2. Nuclear reaction: |
| | | | | | | | | $^{150}$Nd($^{19}$F,5nγ), E=85 MeV |
| | | | | | | | | 3. Half-life of bandhead is 2.0(5) min |
| | | | | | | | | (1971DE22) |
| 2 | 0<x<E(6$^-$) | (3$^-$) | | | | | **1971DE22** | 1. π1/2[411]⊗ν5/2[523] |
| | | | | | | | | 2. Nuclear reaction: |
| | | | | | | | | $^{164}$Tm IT decay (5.1 min) |
| 3 | 0.0+x | 6$^-$ | | | | | **1995RE05** | 1. π7/2[404]⊗ν5/2[523] |
| | 157.8+x | 7$^-$ | 158.2 | | | | 1971DE22 | πg$_{7/2}$⊗νh$_{9/2}$ |
| | 338.4+x | 8$^-$ | 180.6 | 338.3 | | | | 2. Half-life of bandhead is 5.1(1) min |
| | 535.6+x | 9$^-$ | 197.6 | 377.8 | | | | (1971DE22) |
| | 747.9+x | 10$^-$ | 212.6 | 409.3 | | | | 3. Regular band |
| | 974.4+x | 11$^-$ | 226.8 | 438.7 | | | | 4. Typical uncertainties on the γ-ray |
| | 1211.1+x | 12$^-$ | 236.6 | 462.8 | | | | energies are 0.3keV for the strongest |
| | 1455.2+x | 13$^-$ | 243.8 | 481.3 | | | | transitions and upto 0.7keV for the |
| | 1702.6+x | 14$^-$ | 247.6 | 491.3 | | | | weakest transitions |
| | 1951.4+x | 15$^-$ | 248.8 | 496.3 | | | | 5. νi$_{13/2}$ AB crossing occurs at |
| | 2199.0+x | 16$^-$ | | 496.4 | | | | ℏω~0.25MeV |
| | 2447.8+x | 17$^-$ | | 496.4 | | | | A = νi$_{13/2}$ 5/2[642] (α = +½) |
| | 2704.9+x | 18$^-$ | | 505.9 | | | | B = νi$_{13/2}$ 5/2[642] (α = -½) |
| | 2970.0+x | 19$^-$ | | 522.3 | | | | 6. Level energy is adopted from ENSDF |
| | 3250.4+x | 20$^-$ | | 545.5 | | | | |
| | 3541.1+x | 21$^-$ | | 571.0 | | | | |
| | 3855.4+x | 22$^-$ | | 605.0 | | | | |
| | 4176.6+x | 23$^-$ | | 635.5 | | | | |
| | 4525.4+x | 24$^-$ | | 670.0 | | | | |

| S.No. | E<sub>level</sub> keV | $I^\pi$ | $E_\gamma$(M1) keV | $E_\gamma$(E2) keV | $|g_K-g_R|$ | B(M1)/B(E2) $(\mu_N/eb)^2$ | Keywords | Configuration and Comments |
|---|---|---|---|---|---|---|---|---|
| 4 | 110.1 | $2^+$ | | | | | **1995RE05** | 1. $\pi1/2[411]\otimes\nu5/2[642]$ |
| | 107.9 | $3^+$ | | | | | 1987DR07 | $\pi d_{3/2}\otimes\nu i_{13/2}$ |
| | 158.1 | $4^+$ | | 48 | | | | 2. Regular band |
| | 172.9 | $5^+$ | | 65 | | | | 3. signature splitting |
| | 253.4 | $6^+$ | 80.2 | 95.3 | | | | 4. Signature inversion at I = $11^+$ |
| | 301.5 | $7^+$ | 48 | 128.6 | | | | 5. Competing configurations are |
| | 410.2 | $8^+$ | 109.0 | 156.6 | | | | $\pi1/2[411]\otimes\nu5/2[642]$, K= $2^+$ |
| | 520.2 | $9^+$ | 110.4 | 218.7 | | | | $\pi7/2[404]\otimes\nu5/2[642]$, K= $1^+$ |
| | 656.9 | $10^+$ | | 246.6 | | | | observes at 97.17 Mev (1987DR07) |
| | 830.3 | $11^+$ | 173.1 | 310.3 | | | | 6. Level energy is adopted from ENSDF |
| | 1000.5 | $12^+$ | | 343.4 | | | | 7. Typical uncertainties on the γ-ray |
| | 1231.7 | $13^+$ | 230.6 | 401.6 | | | | energies are 0.3keV for the strongest |
| | 1438.3 | $14^+$ | | 437.8 | | | | transitions and upto 0.7keV for the |
| | 1720.1 | $15^+$ | 281.7 | 488.5 | | | | weakest transitions |
| | 1963.9 | $16^+$ | | 525.4 | | | | |
| | 2287.9 | $17^+$ | 323.7 | 568.0 | | | | |
| | 2566.9 | $18^+$ | | 603.0 | | | | |
| | 2926.8 | $19^+$ | | 639.9 | | | | |
| | 3231.7 | $20^+$ | | 664.8 | | | | |
| | 3649.8 | $21^+$ | | 723.0 | | | | |
| | 3945.2 | $22^+$ | | 713.5 | | | | |
| | 4377.2 | $23^+$ | | 727.4 | | | | |
| | 4598.7 | $24^+$ | | 653.5 | | | | |
| 5 | 270.7+y | $(5^+)$ | | | | | **1995RE05** | 1. $\pi7/2[404]\otimes\nu5/2[642]$, K = $1^+$ |
| | 354.6+y | $(6^+)$ | 83.9 | | | | 1987DR07 | $\pi g_{7/2}\otimes\nu i_{13/2}$ |
| | 461.7+y | $(7^+)$ | 107.1 | 191.1 | | | | 2. Regular band |
| | 587.9+y | $(8^+)$ | 125.9 | 233.3 | | | | 3. Small signature splitting in $\Delta E_\gamma$ vs I |
| | 736.9+y | $(9^+)$ | 148.8 | 275.4 | | | | 4. Signature inversion at both I = $(12^+)$ |
| | 903.0+y | $(10^+)$ | | 315.1 | | | | And I = $(18^+)$ |
| | 1091.2+y | $(11^+)$ | | 354.3 | | | | 5. Competing configurations are |
| | 1301.2+y | $(12^+)$ | | 398.2 | | | | $\pi1/2[411]\otimes\nu5/2[642]$, K= $2^+$ |
| | 1525.0+y | $(13^+)$ | | 433.8 | | | | $\pi7/2[404]\otimes\nu5/2[642]$, K= $1^+$ |
| | 1774.5+y | $(14^+)$ | | 473.3 | | | | observes at 97.17 Mev (1987DR07) |
| | 2031.4+y | $(15^+)$ | | 506.4 | | | | 6. Level energy is adopted from ENSDF |
| | 2316.0+y | $(16^+)$ | | 541.5 | | | | 7. Typical uncertainties on the γ-ray |
| | 2610.3+y | $(17^+)$ | | 578.9 | | | | energies are 0.3keV for the strongest |
| | 2911.4+y | $(18^+)$ | | 595.4 | | | | transitions and upto 0.7keV for the |
| | 3264.1+y | $(19^+)$ | | 653.8 | | | | weakest transitions |

| S.No. | $E_{level}$ keV | $I^\pi$ | $E_\gamma$(M1) keV | $E_\gamma$(E2) keV | $|g_K-g_R|$ | B(M1)/B(E2) $(\mu_N/eb)^2$ | Keywords | Configuration and Comments |
|---|---|---|---|---|---|---|---|---|
| 6 | 124.3+x | $6^-$ | | | | | **1995RE05** | 1. $\pi 7/2[523] \otimes \nu 5/2[642]$ |
| | 182.3+x | $7^-$ | 58 | | | | 1986DR06 | $\pi h_{11/2} \otimes \nu i_{13/2}$, $\alpha=+1/2$ |
| | 257.4+x | $8^-$ | 75.5 | 133.1 | | | | 2. Regular band |
| | 353.4+x | $9^-$ | 95.9 | 170.7 | | | | 3. Signature splitting more pronounced at |
| | 470.1+x | $10^-$ | 116.8 | 212.9 | | | | at higher spins |
| | 607.8+x | $11^-$ | 138.5 | 254.1 | | | | 4. Typical uncertainties on the γ-ray |
| | 769.9+x | $12^-$ | 162.7 | 299.4 | | | | energies are 0.3keV for the strongest |
| | 952.7+x | $13^-$ | 183.1 | 344.8 | | | | transitions and upto 0.7keV for the |
| | 1159.9+x | $14^-$ | 207.4 | 389.9 | | | | weakest transitions |
| | 1389.5+x | $15^-$ | 229.8 | 436.6 | | | | 5. Side band (1986DR06) |
| | 1636.4+x | $16^-$ | 247.1 | 476.4 | | | | 6. Level energy is adopted from ENSDF |
| | 1914.4+x | $17^-$ | 278.1 | 524.9 | | | | 7. Aligned band |
| | 2193.7+x | $18^-$ | 279.5 | 557.1 | | | | |
| | 2520.5+x | $19^-$ | 326.7 | 605.9 | | | | |
| | 2523.8+x | $20^-$ | 303.0 | 630.1 | | | | |
| | 3198.7+x | $21^-$ | 374.7 | 678.3 | | | | |
| | 3519.4+x | $22^-$ | 320.5 | 695.6 | | | | |
| | 3939.2+x | $23^-$ | 419.7 | 740.6 | | | | |
| | 4272.5+x | $24^-$ | | 753.1 | | | | |
| | 4732.0+x | $25^-$ | | 792.8 | | | | |
| | 5068.1+x | $26^-$ | | 795.6 | | | | |
| | 5575.4+x | $27^-$ | | 843.4 | | | | |
| | 5888.5+x | $28^-$ | | 820.4 | | | | |
| | | | | | | | | |
| 7 | 140.7+x | $6^+$ | | | | | **1995RE05** | 1. $\pi 7/2[404] \otimes \nu 5/2[642]$ |
| | 185.0+x | $7^+$ | 44 | | | | 1987DR07 | $\pi g_{7/2} \otimes \nu i_{13/2}$ |
| | 264.1+x | $8^+$ | 79.2 | 123.5 | | | | 2. Regular band |
| | 376.3+x | $9^+$ | 112.3 | 191.0 | | | | 3. Signature splitting more pronounced at |
| | 517.2+x | $10^+$ | 141.0 | 253.1 | | | | at higher spins |
| | 684.7+x | $11^+$ | 167.5 | 308.3 | | | | 4. Typical uncertainties on the γ-ray |
| | 875.5+x | $12^+$ | 190.6 | 358.3 | | | | energies are 0.3keV for the strongest |
| | 1090.8+x | $13^+$ | 214.9 | 406.2 | | | | transitions and upto 0.7keV for the |
| | 1325.5+x | $14^+$ | 234.7 | 450.1 | | | | weakest transitions |
| | 1583.3+x | $15^+$ | 257.7 | 492.4 | | | | 5. Aligned band |
| | 1856.1+x | $16^+$ | 272.6 | 530.7 | | | | 6. Level energy is adopted from ENSDF |
| | 2153.2+x | $17^+$ | 297.7 | 569.9 | | | | 7. Different energies (1987DR07) |
| | 2459.6+x | $18^+$ | 307.4 | 603.2 | | | | |
| | 2793.1+x | $19^+$ | 333.7 | 639.9 | | | | |
| | 3129.6+x | $20^+$ | 336.4 | 670.0 | | | | |
| | 3495.0+x | $21^+$ | | 701.9 | | | | |
| | 3854.6+x | $22^+$ | | 725.0 | | | | |
| | 4249.7+x | $23^+$ | | 754.7 | | | | |
| | 4633.2+x | $24^+$ | | 778.6 | | | | |
| | 5020.8+x | $25^+$ | | 771.1 | | | | |

| S.No. | $E_{level}$ keV | $I^\pi$ | $E_\gamma$(M1) keV | $E_\gamma$(E2) keV | $|g_K-g_R|$ | B(M1)/B(E2) $(\mu_N/eb)^2$ | Keywords | Configuration and Comments |
|---|---|---|---|---|---|---|---|---|
| 8 | 381.2 | $6^-$ | | | | | **1995RE05** | 1. $\pi 1/2[541]\otimes\nu 5/2[642]$, $\alpha=+1/2$, $K^\pi=2^-$ |
| | 463.1 | $7^-$ | | | | | 1987DR07 | $\pi h_{9/2}\otimes\nu i_{13/2}$ |
| | 547.7 | $8^-$ | 84.9 | 166.9 | | | | 2. Regular band |
| | 685.2 | $9^-$ | | | | | | 3. Signature splitting |
| | 785.7 | $10^-$ | 99.7 | 237.8 | | | | 4. Signature inversion at I = $20^-$ |
| | 975.2 | $11^-$ | 189.1 | 290.3 | | | | 5. Typical uncertainties on the γ-ray |
| | 1104.9 | $12^-$ | 129.7 | 319.4 | | | | energies are 0.3keV for the strongest |
| | 1343.7 | $13^-$ | 238.1 | 368.4 | | | | transitions and upto 0.7keV for the |
| | 1511.9 | $14^-$ | | 407.4 | | | | weakest transitions |
| | 1787.8 | $15^-$ | 275.6 | 443.8 | | | | 6. Aligned band |
| | 2005.3 | $16^-$ | | 493.6 | | | | 7. Level energy is adopted from ENSDF |
| | 2307.6 | $17^-$ | 302.7 | 519.3 | | | | 8. Side band and bandhead is $4^-$ or $5^-$ or $6^-$ |
| | 2580.9 | $18^-$ | | 575.6 | | | | (1987DR07) |
| | 2895.9 | $19^-$ | | 588.3 | | | | |
| | 3229.8 | $20^-$ | | 648.9 | | | | |
| | 3548.4 | $21^-$ | | 652.5 | | | | |
| | 3943.1 | $22^-$ | | 713.3 | | | | |
| | 4262.0 | $23^-$ | | 713.6 | | | | |
| | 4650.5 | $24^-$ | | 707.4 | | | | |
| | 5016.3 | $25^-$ | | 754.3 | | | | |
| | | | | | | | | |
| 9 | 542.3 | $8^-$ | | | | | **1995RE05** | 1. $\pi 7/2[523]\otimes\nu 5/2[642]$, $K^\pi=1^-$ |
| | 657.3 | $9^-$ | 115.2 | | | | | $\pi h_{11/2}\otimes\nu i_{13/2}$ |
| | 776.6 | $10^-$ | 119.3 | 234.2 | | | | 2. Regular band |
| | 937.1 | $11^-$ | 160.6 | 280.1 | | | | 3. Small signature splitting |
| | 1107.8 | $12^-$ | 170.7 | 331.2 | | | | 4. Signature inversion at I = $18^-$ |
| | 1319.6 | $13^-$ | 211.8 | 383.0 | | | | 5. Typical uncertainties on the γ-ray |
| | 1541.0 | $14^-$ | 221.3 | 433.3 | | | | energies are 0.3keV for the strongest |
| | 1802.2 | $15^-$ | 261.0 | 483.1 | | | | transitions and upto 0.7keV for the |
| | 2070.4 | $16^-$ | 268.1 | 529.6 | | | | weakest transitions |
| | 2369.3 | $17^-$ | 298.7 | 567.1 | | | | 6. Level energy is adopted from ENSDF |
| | 2682.4 | $18^-$ | 312.6 | 612.0 | | | | |
| | 3002.5 | $19^-$ | 319.7 | 633.6 | | | | |
| | 3349.9 | $20^-$ | | 667.5 | | | | |
| | 3634.7 | $21^-$ | | 632.2 | | | | |
| | 4085.2 | $22^-$ | | 735.3 | | | | |
| | | | | | | | | |
| 10 | 168.3+x | $(6^+)$ | | | | | **1995RE05** | 1. $\pi 7/2[523]\otimes\nu 3/2[521]$ |
| | 302.6+x | $(7^+)$ | 134.5 | | | | | $\pi h_{11/2}\otimes\nu f_{7/2}$ |
| | 443.3+x | $(8^+)$ | 140.9 | 274.8 | | | | 2. Regular band |
| | 602.5+x | $(9^+)$ | 159.0 | 299.9 | | | | 3. Small signature splitting |
| | 773.5+x | $(10^+)$ | 170.9 | 330.2 | | | | 4. Typical uncertainties on the γ-ray |
| | 964.0+x | $(11^+)$ | 190.5 | 361.6 | | | | energies are 0.3keV for the strongest |
| | 1170.1+x | $(12^+)$ | 206.2 | 396.6 | | | | transitions and upto 0.7keV for the |
| | 1409.2+x | $(13^+)$ | | 445.2 | | | | weakest transitions |
| | 1642.9+x | $(14^+)$ | | 472.8 | | | | 5. Level energy is adopted from ENSDF |

$^{166}_{69}Tm_{97}$

| S.No. | $E_{level}$ keV | $I^\pi$ | $E_\gamma$ (M1) keV | $E_\gamma$ (E2) keV | $|g_K-g_R|$ | B(M1)/B(E2) $(\mu_N/eb)^2$ | Keywords | Configuration and Comments |
|---|---|---|---|---|---|---|---|---|
| 1 | x+0.0 | (3+) | | | | | **2002CA46** | 1. $\pi1/2[411]\otimes\nu5/2[642]$ $K^\pi=2^+,3^+$ |
| | x+33.6 | (4+) | | | | | 1962WA27 | 2. Nuclear reaction: |
| | x+74.9 | (5+) | | 74.9 | | | | $^{160}$Gd($^{11}$B,5n$\gamma$)$^{166}$Tm, E=61 MeV |
| | x+152.2 | (6+) | 77.3 | 18.4 | | 0.02 | | 3. Half-life of bandhead is 7.70(3) h |
| | x+226.5 | (7+) | 74.5 | 151.6 | | 0.00 | | (1962WA27) |
| | x+341.9 | (8+) | 115.2 | 189.7 | | 0.03 | | 4. Regular band |
| | x+460.30 | (9+) | | 233.7 | | | | 5. Signature splitting more |
| | x+605.20 | (10+) | 144.9 | 263.5 | | 0.03 | | pronounced at higher spins |
| | x+772.73 | (11+) | 167.4 | 312.5 | | 0.01 | | 6. Signature inversion at I = (11+) |
| | x+946.15 | (12+) | 173.5 | 341.0 | | 0.02 | | and (16+) |
| | x+1157.08 | (13+) | 210.9 | 384.4 | | 0.05 | | 8. Level energy is adopted from |
| | x+1368.06 | (14+) | 211.2 | 421.9 | | 0.01 | | ENSDF except from (3+) to (8+) |
| | x+1609.95 | (15+) | 241.9 | 452.9 | | 0.02 | | 9. Level energy of x<16 KeV from |
| | x+1865.88 | (16+) | | 497.8 | | | | ENSDF |
| | x+2120.36 | (17+) | 254.5 | 510.4 | | 0.19 | | 10. Level energies matches with $K^\pi$ |
| | x+2423.33 | (18+) | | 557.4 | | | | =2+(1962WA27) |
| | x+2690.07 | (19+) | 266.6 | 569.7 | | 0.1 | | |
| | x+3031.59 | (20+) | | 608.2 | | | | |
| | x+3308.58 | (21+) | 277.1 | 618.5 | | 0.04 | | |
| | x+3686.72 | (22+) | | 655.1 | | | | |
| | x+3975.85 | (23+) | | 667.3 | | | | |
| | x+4391.02 | (24+) | | 704.3 | | | | |
| | x+4697.4 | (25+) | | 721.5 | | | | |
| | x+5150.8 | (26+) | | 759.8 | | | | |
| | x+5480.3 | (27+) | | 782.9 | | | | |
| | x+5972.7 | (28+) | | 821.9 | | | | |
| | x+6329.6 | (29+) | | 849.3 | | | | |
| | x+6861.0 | (30+) | | 888.3 | | | | |
| | x+7247.9 | (31+) | | 918.3 | | | | |
| | x+7816.0 | (32+) | | 955.0 | | | | |
| | x+8234.4 | (33+) | | 986.5 | | | | |
| | (x+8845.1) | (34+) | | (1029.1) | | | | |
| 2 | 82.29 | 1+ | | | | | **1995MA07** | 1. $\pi7/2[523]\otimes\nu5/2[523]$ |
| | | | | | | | | 2. Nuclear reaction: |
| | | | | | | | | $^{165}$Ho($\alpha$,3n$\gamma$)$^{166}$Tm, |
| | | | | | | | | E=32.6 to 47.9 MeV |
| 3 | 109.44 | 3- | | | | | **1995MA07** | 1. $\pi1/2[411]\otimes\nu5/2[523]$ |
| | 131.74 | 4- | | | | | | 2. Regular band |
| | 207.54 | 5- | 75.793(4) | 98.10(5) | | | | 3. Signature Splitting |
| | 288.12 | 6- | 80.584(4) | 156.409(7) | | | | 4. The value of $g_R$ for bandhead is |
| | 423.68 | 7- | 135.554(3) | 216.139(12) | | | | 0.33(3) calculated in paper by |
| | 524.62 | 8- | 100.939(3) | 236.484(10) | | | | taking $g_R^{ee}$= 0.3485(75) |
| | 733.68 | 9- | 209.081(13) | 309.977(16) | | | | (1989RA17) |
| | 850.02 | 10- | (116.3) | 325.423(12) | | | | |
| | 1130.47 | 11- | 280.446(8) | 396.786(42) | | | | |
| | 1268.61 | 12- | | 418.603(22) | | | | |
| | 1612.07 | 13- | | 481.60 | | | | |
| | 1774.58 | 14- | | 505.97 | | | | |

| S.No. | $E_{level}$ keV | $I^\pi$ | $E_\gamma$(M1) keV | $E_\gamma$(E2) keV | $|g_K-g_R|$ | B(M1)/B(E2) $(\mu_N/eb)^2$ | Keywords | Configuration and Comments |
|---|---|---|---|---|---|---|---|---|
| 4 | x+109.34 | (6⁻) | | | | | **2002CA46** | 1. $\pi 7/2[404]\otimes \nu 5/2[523]$ |
| | x+256.52 | (7⁻) | 147.5 | | | | 1995MA07 | 2. Half-life of bandhead is 340(25) ms |
| | x+423.92 | (8⁻) | 167.2 | 314.7 | | 0.75 | | 3. Regular band |
| | x+609.06 | (9⁻) | 185.4 | 352.6 | | 0.5 | | 4. Level energy is adopted from ENSDF |
| | x+811.92 | (10⁻) | 202.5 | 388.0 | | 0.34 | | 5. Level energy of x<25 KeV from ENSDF |
| | x+1030.41 | (11⁻) | 218.2 | 421.4 | | 0.3 | | |
| | x+1263.64 | (12⁻) | 233.2 | 451.7 | | 0.32 | | |
| | x+1510.09 | (13⁻) | 246.4 | 479.7 | | 0.24 | | |
| | x+1768.45 | (14⁻) | 258.5 | 504.8 | | 0.18 | | |
| | x+2037.1 | (15⁻) | 268.6 | 527.0 | | 0.2 | | |
| | x+2315.4 | (16⁻) | 278.4 | 547.0 | | | | |
| | x+2601.7 | (17⁻) | 286.5 | 564.4 | | | | |
| | x+2893.2 | (18⁻) | 291.4 | 577.9 | | | | |
| | x+3200.1 | (19⁻) | | 598.4 | | | | |
| | | | | | | | | |
| 5 | x+171.6 | (7⁺) | | | | | 1995MA07 | 1. $\pi 7/2[404]\otimes \nu 7/2[633]$ |
| | | | | | | | | 2. The value of $g_R$ for bandhead is 0.146(15) |
| | | | | | | | | 3. Energy includes x<20 KeV |
| | | | | | | | | |
| 6 | x+231.05 | (6⁻) | | | | | **2002CA46** | 1. $\pi 7/2[523]\otimes \nu 5/2[642]$ |
| | x+287.54 | (7⁻) | 56.4 | | | | 1995MA07 | 2. Half-life of bandhead is 36(2) ns |
| | x+367.45 | (8⁻) | 79.9 | 136.6 | | 1.7 | | 3. Aligned band |
| | x+469.13 | (9⁻) | 101.6 | 181.6 | | 1.1 | | 4. Regular band |
| | x+592.48 | (10⁻) | 123.2 | 225.1 | | 0.8 | | 5. Signature splitting more pronounced at |
| | x+737.66 | (11⁻) | 145.1 | 268.6 | | 0.75 | |    higher spins |
| | x+904.47 | (12⁻) | 166.8 | 312.0 | | 0.6 | | 6. The value of $g_R$ for bandhead is |
| | x+1092.35 | (13⁻) | 187.8 | 354.7 | | 0.6 | |    0.23(2) calculated in paper by |
| | x+1299.71 | (14⁻) | 207.3 | 395.3 | | 0.5 | |    taking $g_R^{ee}$ = 0.3485(75) (1995MA07) |
| | x+1528.34 | (15⁻) | 228.6 | 436.0 | | 0.6 | | 7. Level energy of x<25 KeV from ENSDF |
| | x+1770.39 | (16⁻) | 242.1 | 470.7 | | 0.5 | | |
| | x+2038.61 | (17⁻) | 268.2 | 510.2 | | 0.45 | | |
| | x+2307.60 | (18⁻) | 269.0 | 537.3 | | 0.75 | | |
| | x+2614.30 | (19⁻) | 306.8 | 575.6 | | 0.5 | | |
| | x+2902.81 | (20⁻) | 288.6 | 595.2 | | 0.75 | | |
| | x+3246.19 | (21⁻) | 343.6 | 631.8 | | 0.6 | | |
| | x+3546.76 | (22⁻) | 300.7 | 643.8 | | 0.75 | | |
| | x+3923.98 | (23⁻) | 377.2 | 677.8 | | 0.4 | | |
| | x+4232.6 | (24⁻) | 306.9 | 685.8 | | 0.75 | | |
| | x+4642.9 | (25⁻) | 410.7 | 718.9 | | 0.5 | | |
| | x+4957.7 | (26⁻) | | 725.1 | | | | |
| | x+5407.2 | (27⁻) | | 764.3 | | | | |
| | x+5725.6 | (28⁻) | | 767.9 | | | | |
| | x+6227.2 | (29⁻) | | 820.0 | | | | |
| | x+6542.6 | (30⁻) | | 817.0 | | | | |
| | x+7111.2 | (31⁻) | | 884.0 | | | | |
| | x+7414.5 | (32⁻) | | 871.9 | | | | |
| | x+8065.8 | (33⁻) | | 954.6 | | | | |
| | x+8345.3 | (34⁻) | | 930.8 | | | | |
| | x+9338.8 | (36⁻) | | 993.5 | | | | |

| S.No. | E$_{level}$ keV | I$^\pi$ | E$_\gamma$(M1) keV | E$_\gamma$(E2) keV | \|g$_K$-g$_R$\| | B(M1)/B(E2) ($\mu_N$/eb)$^2$ | Keywords | Configuration and Comments |
|---|---|---|---|---|---|---|---|---|
| 7 | y+85.97 | (4$^-$) | | | | | **1995MA07** | 1. $\pi$7/2[404]$\otimes$v1/2[521] |
| | y+184.1 | (5$^-$) | 98.10(5) | | | | | 2. Energy Y could be x+149.2 KeV |
| | y+302.4 | (6$^-$) | 118.284(4) | | | | | 3. The value of g$_R$ for bandhead is |
| | y+433.6 | (7$^-$) | 131.215(4) | 249.52(7) | | | | 0.247(12) |
| | y+591.7 | (8$^-$) | 158.148(14) | 289.36(7) | | | | 4. Regular band |
| | y+760.3 | (9$^-$) | 168.609(5) | 326.89(8) | | | | 5. Small signature splitting |
| | y+963.0 | (10$^-$) | 202.649(8) | 371.2 | | | | |
| | y+1171.6 | (11$^-$) | 208.659(15) | 411.21(11) | | | | |
| | y+1419.7 | (12$^-$) | 248.077(31) | 456.91(16) | | | | |
| | y+1669.2 | (13$^-$) | 249.52(7) | 497.77(3) | | | | |
| | | | | | | | | |
| 8 | x+171.56 | (6$^+$) | | | | | **2002CA46** | 1. $\pi$7/2[404]$\otimes$v5/2[642] |
| | x+211.44 | (7$^+$) | 39.9 | | | | 1995MA07 | 2. Aligned band |
| | x+298.12 | (8$^+$) | 86.7 | 126.5 | | 0.52 | | 3. Regular band |
| | x+417.45 | (9$^+$) | 119.3 | 206.0 | | 0.22 | | 4. Signature splitting more |
| | x+563.37 | (10$^+$) | 145.8 | 265.3 | | 0.16 | | pronounced at higher spins |
| | x+733.18 | (11$^+$) | 169.7 | 315.8 | | 0.12 | | 5. The value of g$_R$ for bandhead is |
| | x+922.13 | (12$^+$) | 188.9 | 358.8 | | 0.1 | | 0.104(23) calculated in paper by |
| | x+1132.36 | (13$^+$) | 210.2 | 399.2 | | 0.1 | | taking $g_R^{ee}$= 0.3485(75) |
| | x+1350.26 | (14$^+$) | 217.8 | 428.2 | | 0.08 | | (1995MA07) |
| | x+1599.57 | (15$^+$) | 249.3 | 467.2 | | 0.04 | | 6. Level energy is adopted from |
| | x+1836.42 | (16$^+$) | 236.8 | 486.2 | | 0.06 | | ENSDF |
| | x+2131.85 | (17$^+$) | 295.6 | 532.2 | | 0.22 | | |
| | x+2381.20 | (18$^+$) | 249.3 | 544.8 | | 0.04 | | |
| | x+2713.74 | (19$^+$) | 332.6 | 581.9 | | 0.1 | | |
| | x+2978.39 | (20$^+$) | | 597.2 | | | | |
| | x+3345.13 | (21$^+$) | 366.8 | 631.4 | | 0.12 | | |
| | x+3623.41 | (22$^+$) | | 645.0 | | | | |
| | x+4024.64 | (23$^+$) | 401.3 | 679.5 | | 0.1 | | |
| | x+4316.91 | (24$^+$) | | 693.5 | | | | |
| | x+4755.7 | (25$^+$) | | 731.1 | | | | |
| | x+5064.6 | (26$^+$) | | 747.7 | | | | |
| | x+5544.8 | (27$^+$) | | 789.1 | | | | |
| | x+5873.6 | (28$^+$) | | 809.0 | | | | |
| | x+6396.3 | (29$^+$) | | 851.5 | | | | |
| | x+6748.9 | (30$^+$) | | 875.3 | | | | |
| | x+7313.9 | (31$^+$) | | 917.5 | | | | |
| | x+7692.5 | (32$^+$) | | 943.6 | | | | |
| | x+8297.4 | (33$^+$) | | 983.5 | | | | |
| | x+8692.2 | (34$^+$) | | 999.7 | | | | |
| | | | | | | | | |
| 9 | 334.25 | (5$^-$) | | | | | **1995MA07** | 1. $\pi$7/2[404]$\otimes$v3/2[521] |
| | 474.89 | (6$^-$) | 140.641(12) | | | | | 2. The value of g$_R$ for bandhead is |
| | 633.21 | (7$^-$) | 158.329(15) | 298.89(9) | | | | 0.25(3) |
| | 808.72 | (8$^-$) | 175.514(9) | 333.783(34) | | | | 3. Regular band |
| | 1000.74 | (9$^-$) | 192.023(11) | 367.52(5) | | | | |
| | 1208.74 | (10$^-$) | 208.0 | | | | | |

| S.No. | E$_{level}$ keV | I$^\pi$ | E$_\gamma$(M1) keV | E$_\gamma$(E2) keV | \|g$_K$-g$_R$\| | B(M1)/B(E2) ($\mu_N$/eb)$^2$ | Keywords | Configuration and Comments |
|---|---|---|---|---|---|---|---|---|
| 10 | x+266.29 | (6$^+$) | | | | | **2002CA46** | 1. $\pi 7/2[523] \otimes \nu 5/2[523]$ |
| | x+389.05 | (7$^+$) | 122.7 | | | | 1995MA07 | 2. Regular band |
| | x+529.72 | (8$^+$) | 140.6 | 263.5 | | 1.75 | | 3. The value of g$_R$ for bandhead is |
| | x+687.86 | (9$^+$) | 158.1 | 298.8 | | 1.1 | | 0.350(35) calculated in paper by |
| | x+863.5 | (10$^+$) | 175.4 | 333.9 | | 1.75 | | taking $g_R^{ee}$= 0.3485(75) (1995MA07) |
| | x+1055.5 | (11$^+$) | 191.9 | 367.7 | | 1.1 | | 4. Level energy is adopted from ENSDF |
| | x+1263.3 | (12$^+$) | 207.7 | 399.9 | | 1.1 | | |
| | x+1486.7 | (13$^+$) | 223.5 | 431.2 | | 0.9 | | |
| | x+1723.9 | (14$^+$) | 237.1 | 460.5 | | 1.0 | | |
| | x+1976.4 | (15$^+$) | 252.6 | 489.9 | | 0.75 | | |
| | x+2245.5 | (16$^+$) | 269.2 | 521.4 | | 0.55 | | |
| | x+2520.9 | (17$^+$) | | 544.5 | | | | |
| | x+2839.3 | (18$^+$) | | 593.8 | | | | |
| 11 | x+423.21 | (7$^-$) | | | | | **2002CA46** | 1. $\pi 1/2[541] \otimes \nu 5/2[642]$, K=2$^-$,3$^-$ |
| | x+507.83 | (8$^-$) | | | | | | 2. Regular band |
| | x+634.34 | (9$^-$) | 126.5 | 211.4 | | 0.12 | | 3. Signature splitting |
| | x+736.29 | (10$^-$) | 101.8 | 228.5 | | 0.16 | | 4. Signature inversion at I = (24$^-$) |
| | x+915.99 | (11$^-$) | 179.7 | 281.7 | | 0.15 | | 5. Level energy is adopted from ENSDF |
| | x+1043.03 | (12$^-$) | 127.0 | 306.8 | | 0.16 | | |
| | x+1279.73 | (13$^-$) | 236.7 | 363.7 | | 0.16 | | |
| | x+1433.82 | (14$^-$) | 154.0 | 390.8 | | 0.14 | | |
| | x+1722.83 | (15$^-$) | 289.0 | 443.1 | | 0.13 | | |
| | x+1908.53 | (16$^-$) | 185.4 | 474.7 | | 0.12 | | |
| | x+2237.44 | (17$^-$) | 329.1 | 514.6 | | 0.15 | | |
| | x+2463.43 | (18$^-$) | | 554.9 | | | | |
| | x+2814.92 | (19$^-$) | 351.5 | 577.5 | | 0.18 | | |
| | x+3092.54 | (20$^-$) | | 629.1 | | | | |
| | x+3449.1 | (21$^-$) | 356.5 | 634.2 | | 0.19 | | |
| | x+3788.1 | (22$^-$) | | 695.6 | | | | |
| | x+4136.1 | (23$^-$) | | 687.0 | | | | |
| | x+4542.0 | (24$^-$) | | 753.9 | | | | |
| | x+4874.1 | (25$^-$) | | 738.0 | | | | |
| | x+5346.2 | (26$^-$) | | 804.2 | | | | |
| | x+5662.5 | (27$^-$) | | 788.4 | | | | |
| | x+6192.6 | (28$^-$) | | 846.4 | | | | |
| | x+6503.1 | (29$^-$) | | 840.6 | | | | |
| | x+7062.9 | (30$^-$) | | 870.3 | | | | |
| | x+7398.7 | (31$^-$) | | 895.6 | | | | |
| | x+7969.7 | (32$^-$) | | 906.7 | | | | |
| | x+8352.6 | (33$^-$) | | 953.9 | | | | |

| S.No. | E$_{level}$ keV | I$^\pi$ | E$_\gamma$(M1) keV | E$_\gamma$(E2) keV | \|g$_K$-g$_R$\| | B(M1)/B(E2) ($\mu_N$/eb)$^2$ | Keywords | Configuration and Comments |
|---|---|---|---|---|---|---|---|---|
| 12 | x+131.69 | (5$^+$) | | | | | 2002CA46 | 1. $\pi$1/2[541]⊗v5/2[523], K=2$^+$,3$^+$ |
| | x+207.32 | (6$^+$) | 75.7 | | | | | 2. Regular band |
| | x+287.89 | (7$^+$) | 80.6 | 156.3 | | 0.25 | | 3. Signature splitting |
| | x+423.48 | (8$^+$) | 135.6 | 216.1 | | 0.06 | | 4. Level energy is adopted from ENSDF |
| | x+524.54 | (9$^+$) | 100.8 | 236.6 | | 0.15 | | |
| | x+733.49 | (10$^+$) | 209.0 | 310.0 | | 0.14 | | |
| | x+849.96 | (11$^+$) | | 325.5 | | | | |
| | x+1130.28 | (12$^+$) | 280.4 | 396.8 | | 0.08 | | |
| | x+1268.67 | (13$^+$) | | 418.6 | | | | |
| | x+1603.88 | (14$^+$) | 343.6 | 473.6 | | 0.04 | | |
| | x+1774.78 | (15$^+$) | | 506.1 | | | | |
| | x+2122.31 | (16$^+$) | | 518.2 | | | | |
| | x+2357.18 | (17$^+$) | | 582.4 | | | | |
| | x+2696.21 | (18$^+$) | | 573.9 | | | | |
| | x+2987.29 | (19$^+$) | | 630.1 | | | | |
| | x+3328.11 | (20$^+$) | | 631.9 | | | | |
| | x+3640.6 | (21$^+$) | | 653.4 | | | | |
| | x+4018.4 | (22$^+$) | | 690.3 | | | | |
| | x+4359.1 | (23$^+$) | | 718.5 | | | | |
| | x+4762.6 | (24$^+$) | | 744.2 | | | | |
| | x+5111.0 | (25$^+$) | | 751.9 | | | | |
| | x+5559.2 | (26$^+$) | | 796.6 | | | | |
| | x+5923.5 | (27$^+$) | | 812.5 | | | | |
| | x+6407.2 | (28$^+$) | | 848.0 | | | | |
| | x+6788.7 | (29$^+$) | | 865.2 | | | | |
| | x+7304.5 | (30$^+$) | | 897.3 | | | | |
| | | | | | | | | |
| 13 | 212.8 | (5$^+$) | | | | | 2002CA46 | 1. $\pi$1/2[541]⊗v3/2[521], K=1$^+$,2$^+$ |
| | 293.73 | (6$^+$) | 81.1 | | | | | 2. Regular band |
| | 383.13 | (7$^+$) | 89.6 | 170.1 | | | | 3. Signature splitting |
| | 504.79 | (8$^+$) | 121.4 | 211.1 | | | | 4. Level energy is adopted from ENSDF |
| | 634.48 | (9$^+$) | 129.5 | 251.4 | | | | |
| | 799.30 | (10$^+$) | 164.8 | 294.5 | | | | |
| | 965.93 | (11$^+$) | 166.5 | 331.5 | | | | |
| | 1173.02 | (12$^+$) | 206.9 | 373.7 | | | | |
| | 1379.26 | (13$^+$) | 206.0 | 413.4 | | | | |
| | 1634.69 | (14$^+$) | | 461.9 | | | | |
| | 1858.45 | (15$^+$) | | 479.2 | | | | |
| | 2181.59 | (16$^+$) | | 546.9 | | | | |
| | 2399.16 | (17$^+$) | | 540.7 | | | | |
| | 2785.3 | (18$^+$) | | 603.7 | | | | |
| | 3016.45 | (19$^+$) | | 617.2 | | | | |
| | 3457.6 | (20$^+$) | | 672.3 | | | | |
| | 3732.4 | (21$^+$) | | 715.9 | | | | |
| | 4481.9 | (23$^+$) | | 749.5 | | | | |

| S.No. | E$_{level}$ keV | I$^\pi$ | E$_\gamma$(M1) keV | E$_\gamma$(E2) keV | \|g$_K$-g$_R$\| | B(M1)/B(E2) ($\mu_N$/eb)$^2$ | Keywords | **Configuration and Comments** |
|---|---|---|---|---|---|---|---|---|
| 14 | 281.4 | (6$^+$) | | | | | **2002CA46** | 1. $\pi$7/2[404]$\otimes\nu$5/2[642], K = 1$^+$ |
| | 401.64 | (7$^+$) | 120.2 | | | | | 2. Regular band |
| | 488.59 | (8$^+$) | 86.8 | 207.2 | | 0.65 | | 3. Signature splitting |
| | 649.49 | (9$^+$) | 160.9 | 247.9 | | 0.25 | | 4. Level energy is adopted from ENSDF |
| | 778.48 | (10$^+$) | 128.7 | 289.9 | | 0.3 | | |
| | 982.17 | (11$^+$) | 203.8 | 332.7 | | 0.1 | | |
| | 1156.45 | (12$^+$) | 174.4 | 377.9 | | 0.14 | | |
| | 1397.05 | (13$^+$) | 240.4 | 414.9 | | 0.25 | | |
| | 1612.04 | (14$^+$) | | 455.6 | | | | |
| | 1900.7 | (15$^+$) | | 503.6 | | | | |
| | 2153.10 | (16$^+$) | | 541.0 | | | | |
| | 2479.0 | (17$^+$) | | 578.3 | | | | |
| | 2751.10 | (18$^+$) | | 598.0 | | | | |
| | 3133.8 | (19$^+$) | | 654.8 | | | | |
| | 3374.7 | (20$^+$) | | 623.6 | | | | |
| | 3804.2 | (21$^+$) | | 670.4 | | | | |
| 15 | (376.9) | (5$^-$) | | | | | **2002CA46** | 1. $\pi$7/2[523]$\otimes\nu$5/2[642], K = 1$^-$ |
| | (453.8) | (6$^-$) | 76.9 | | | | | 2. Regular band |
| | 539.76 | (7$^-$) | 86.0 | | | | | 3. Level energy is adopted from ENSDF |
| | 637.74 | (8$^-$) | 98.0 | | | | | |
| | 756.05 | (9$^-$) | 118.2 | 216.1 | | 1.55 | | |
| | 887.29 | (10$^-$) | 131.1 | 249.7 | | 0.85 | | |
| | 1045.40 | (11$^-$) | 158.1 | 289.3 | | 1.6 | | |
| | 1213.99 | (12$^-$) | 168.5 | 326.8 | | 1.65 | | |
| | 1416.61 | (13$^-$) | 202.5 | 371.3 | | 1.2 | | |
| | 1625.27 | (14$^-$) | 208.6 | 411.3 | | 1.3 | | |
| | 1837.38 | (15$^-$) | 248.1 | 456.8 | | 1.0 | | |
| | 2123.08 | (16$^-$) | 249.8 | 497.8 | | 1.4 | | |
| | 2411.96 | (17$^-$) | 289.6 | 538.5 | | 0.8 | | |
| | 2702.68 | (18$^-$) | 290.8 | 579.6 | | 1.0 | | |
| | 3024.4 | (19$^-$) | 321.7 | 612.3 | | 0.85 | | |
| | 3354.5 | (20$^-$) | 330.1 | 651.9 | | 0.8 | | |
| | 3699.8 | (21$^-$) | 345.3 | 675.4 | | 0.75 | | |
| | 4058.7 | (22$^-$) | 359.0 | 704.2 | | | | |
| | 4420.9 | (23$^-$) | | 721.1 | | | | |
| 16 | 423.64 | 6$^+$ | | | | | **1995MA07** | 1. $\pi$1/2[411]$\otimes\nu$5/2[642], K=3$^+$ |
| | 507.79 | 7$^+$ | | | | | | 2. The value of g$_R$ for bandhead is |
| | 634.37 | 8$^+$ | 126.577(4) | (210.7) | | | | 0.20(3) calculated in paper by |
| | 736.30 | 9$^+$ | 101.929(5) | 228.533(15) | | | | taking $g_R^{ee}$= 0.3485(75) |
| | 915.96 | 10$^+$ | 179.664(7) | 281.597(13) | | | | 3. Regular band |
| | 1042.99 | 11$^+$ | 127.030(4) | 306.685(9) | | | | 4. Signature splitting |
| | 1279.69 | 12$^+$ | 236.688(15) | 363.756(45) | | | | |
| | 1433.77 | 13$^+$ | 154.18(4) | 390.773(28) | | | | |
| | 1722.64 | 14$^+$ | 288.9 | 442.95 | | | | |
| | 1908.43 | 15$^+$ | | 474.664(31) | | | | |
| | 2463.69 | 17$^+$ | | 555.26(9) | | | | |

$^{168}_{69}Tm_{99}$

| S.No. | E<sub>level</sub> keV | I<sup>π</sup> | E<sub>γ</sub>(M1) keV | E<sub>γ</sub>(E2) keV | \|g<sub>K</sub>-g<sub>R</sub>\| | B(M1)/B(E2) (μ<sub>N</sub>/eb)² | Keywords | Configuration and Comments |
|---|---|---|---|---|---|---|---|---|
| 1 | 0.0 | 3⁺ | | | | | **2007CAZW** | 1. π1/2[411]⊗ν7/2[633] |
| | 64.1 | 4⁺ | 64.2 | | | | 1995SI20 | 2. Nuclear reaction: |
| | 144.7 | 5⁺ | 80.7 | 144.7 | | | | ¹⁶⁴Dy(¹¹B,α3nγ)¹⁶⁸Tm, E = 65 MeV |
| | 243.4 | 6⁺ | 98.7 | 179.3 | | | | 3. Half-life of bandhead is 93 d |
| | 358.0 | 7⁺ | 114.6 | 213.3 | | | | 4. Regular band |
| | 493.7 | 8⁺ | 135.7 | 250.4 | | | | 5. Signature splitting more pronounced at |
| | 644.5 | 9⁺ | 150.7 | 286.5 | | | | higher spins |
| | 816.9 | 10⁺ | 172.4 | 323.3 | | | | 6. Level energy is adopted from ENSDF |
| | 1008.5 | 11⁺ | 191.7 | 363.7 | | | | |
| | 1212.2 | 12⁺ | 203.6 | 395.5 | | | | |
| | 1451.0 | 13⁺ | | 442.5 | | | | |
| | 1677.8 | 14⁺ | | 465.6 | | | | |
| | 1967.9 | 15⁺ | | 516.9 | | | | |
| | 2212.2 | 16⁺ | | 534.4 | | | | |
| | 2548.2 | 17⁺ | | 580.3 | | | | |
| | 2812.6 | 18⁺ | | 600.4 | | | | |
| | 3186.3 | 19⁺ | | 638.1 | | | | |
| | 3475.5 | 20⁺ | | 662.9 | | | | |
| | 3881.4 | 21⁺ | | 695.1 | | | | |
| | (4190) | 22⁺ | | (714) | | | | |
| 2 | 41 | (2⁻) | | | | | **2007CAZW** | 1. π1/2[411]⊗ν1/2[521], K<sup>π</sup> = 1⁻ |
| | 178.9 | (4⁻) | | 137.8 | | | 1995SI20 | 2. Regular band |
| | 392.6 | (6⁻) | | 213.6 | | | | 3. Level energy is adopted from ENSDF |
| | 676.2 | (8⁻) | | 283.6 | | | | |
| | 1021.4 | (10⁻) | | 345.2 | | | | |
| | 1422.0 | (12⁻) | | 400.6 | | | | |
| | 1873.7 | (14⁻) | | 451.7 | | | | |
| | 2377.0 | (16⁻) | | 503.3 | | | | |
| | 2934.8 | (18⁻) | | 557.8 | | | | |
| | 3549.0 | (20⁻) | | 614.2 | | | | |
| | 4221.3 | (22⁻) | | 672.3 | | | | |
| | 4952.1 | (24⁻) | | 730.8 | | | | |
| 3 | 0.0+x | (2⁺) | | | | | **2007CAZW** | 1. π7/2[404]⊗ν7/2[633], K<sup>π</sup> = 0⁺ |
| | 121.2+x | (4⁺) | | 121.2 | | | 1987SO08 | 2. Irregular band |
| | 291.8+x | (5⁺) | 170.6 | | | | | 3. Signature splitting |
| | 315.1+x | (6⁺) | | 193.9 | | | | 4. Level energy is adopted from ENSDF |
| | 508.7+x | (7⁺) | 193.4 | | | | | |
| | 582.5+x | (8⁺) | 73.7 | 267.6 | | | | |
| | 795.8+x | (9⁺) | 213.2 | 287.1 | | | | |
| | 922.4+x | (10⁺) | 126.7 | 340.0 | | | | |
| | 1143.4+x | (11⁺) | 221.1 | 347.5 | | | | |
| | 1333.1+x | (12⁺) | | 410.7 | | | | |
| 4 | 148.4 | (4⁺) | | | | | 1995SI20 | 1. π1/2[411]⊗ν7/2[633] |
| | 228.9 | (5⁺) | 80.5 | | | | | 2. Nuclear Reaction: |
| | 327.7 | (6⁺) | 98.7 | 179.3 | | | | ¹⁶⁸Er(d,2nγ)¹⁶⁸Tm, E=10-16 MeV and |
| | 442.0 | (7⁺) | 114.4 | (213.3) | | | | ¹⁶⁵Ho(α,nγ)¹⁶⁸Tm, E=18 MeV |
| | 577.8 | (8⁺) | (135.5) | (250.2) | | | | 3. Regular band |

| S.No. | E$_{level}$ keV | I$^\pi$ | E$_\gamma$(M1) keV | E$_\gamma$(E2) keV | \|g$_K$-g$_R$\| | B(M1)/B(E2) ($\mu_N$/eb)$^2$ | Keywords | Configuration and Comments |
|---|---|---|---|---|---|---|---|---|
| 5 | 168 | 0$^-$ | | | | | **1973PR06** | 1. $\pi 1/2[411]\otimes\nu 1/2[521]$ |
| | 230 | 1$^-$ | | | | | | 2. Nuclear reaction: |
| | 235 | 2$^-$ | | | | | | $^{167}$Er($\alpha$,t)$^{168}$Tm, E=25 MeV |
| | 348 | 3$^-$ | | | | | | 3. Irregular band |
| | 365 | 4$^-$ | | | | | | 4. Signature splitting |
| 6 | 203.7 | (3$^-$) | | | | | **1995SI20** | 1. $\pi 1/2[541]\otimes\nu 7/2[633]$ |
| | 193.3 | (4$^-$) | | | | | | 2. Regular band |
| | 247.0 | (5$^-$) | | | | | | |
| | 314.1 | (6$^-$) | 67.1 | 120.8 | | | | |
| | 391.7 | (7$^-$) | | (144.7) | | | | |
| | 484.2 | (8$^-$) | | 170.1 | | | | |
| | 755.6 | (10$^-$) | | 271.4 | | | | |
| | 1097.3 | (12$^-$) | | 341.7 | | | | |
| 7 | 312 | 7$^+$ | | | | | **1973PR06** | 1. $\pi 7/2[404]\otimes\nu 7/2[633]$ |
| 8 | 323 | 3$^-$ | | | | | **1971JO18** | 1. $\pi 1/2[411]\otimes\nu 5/2[512]$, K$^\pi$=2$^-$ |
| | | | | | | | | 2. Nuclear reaction: |
| | | | | | | | | $^{169}$Tm(d,t)$^{168}$Tm, E= 12 MeV |
| 9 | 337 | 4$^-$ | | | | | **1973PR06** | 1. $\pi 1/2[541]\otimes\nu 7/2[633]$ |
| | 448 | 5$^-$ | | | | | | 2. Regular band |
| | 564 | 6$^-$ | | | | | | |
| | 698 | 7$^-$ | | | | | | |
| 10 | 495 | 3$^-$ | | | | | **1971JO18** | 1. $\pi 1/2[411]\otimes\nu 5/2[512]$ |
| | 593 | 4$^-$ | | | | | | |
| 11 | 611 | (1$^-$) | | | | | **1973KO06** | 1. Tentative configuration |
| | 661 | (2$^-$) | | | | | | $\pi 1/2[411]\otimes\nu 3/2[521]$ |
| | 723 | (3$^-$) | | | | | | 2. Nuclear reaction: |
| | (851) | (4$^-$) | | | | | | $^{169}$Tm(d,t)$^{168}$Tm, E= 17 MeV |
| 12 | 699 | (2$^-$) | | | | | **1973KO06** | 1. Tentative configuration |
| | X | (3$^-$) | | | | | | $\pi 1/2[411]\otimes\nu 3/2[521]$ |
| | 915 | (4$^-$) | | | | | | |
| 13 | 731 | 6$^+$ | | | | | **1973PR06** | 1. Competing configurations are |
| | | | | | | | | $\pi 5/2[402]\otimes\nu 7/2[633]$ |
| | | | | | | | | $\pi 7/2[523]\otimes\nu 5/2[512]$ |
| 14 | (766) | (2$^+$) | | | | | **1973KO06** | 1. Tentative configuration |
| | | | | | | | | $\pi 1/2[411]\otimes\nu 5/2[642]$ |
| 15 | 789 | (1$^-$) | | | | | **1973KO06** | 1. Tentative configuration |
| | 833 | (2$^-$) | | | | | | $\pi 1/2[411]\otimes\nu 1/2[510]$, K$^\pi$=0$^-$ |
| 16 | 815 | 1$^+$ | | | | | **1973PR06** | 1. $\pi 5/2[402]\otimes\nu 7/2[633]$ |
| | 847 | 2$^+$ | | | | | | 2. Regular band |
| | 890 | 3$^+$ | | | | | | |

| S.No. | E$_{level}$ keV | I$^\pi$ | E$_\gamma$(M1) keV | E$_\gamma$(E2) keV | \|g$_K$-g$_R$\| | B(M1)/B(E2) ($\mu_N$/eb)$^2$ | Keywords | Configuration and Comments |
|---|---|---|---|---|---|---|---|---|
| 17 | (860) | (3$^-$) | | | | | **1973KO06** | 1. Tentative configuration |
| | 963 | (4$^-$) | | | | | | $\pi$1/2[411]$\otimes\nu$5/2[523] |
| | 1095 | (5$^-$) | | | | | | 2. Regular band |
| | | | | | | | | |
| 18 | 882 | (2$^-$) | | | | | **1973KO06** | 1. Tentative configuration |
| | | | | | | | | $\pi$1/2[411]$\otimes\nu$1/2[510], K$^\pi$=1$^-$ |
| | | | | | | | | |
| 19 | 899 | (2$^-$) | | | | | **1973KO06** | 1. Tentative configuration |
| | (950) | (3$^-$) | | | | | | $\pi$1/2[411]$\otimes\nu$5/2[523] |
| | (1050) | (4$^-$) | | | | | | 2. Regular band |
| | 1162 | (5$^-$) | | | | | | |
| | | | | | | | | |
| 20 | 1057 | 0$^+$ | | | | | **1973PR06** | 1. $\pi$1/2[411]$\otimes\nu$1/2[400] |
| | 1127 | 1$^+$ | | | | | | 2. Regular band |
| | 1133 | 2$^+$ | | | | | | |
| | 1239 | 3$^+$ | | | | | | |
| | | | | | | | | |
| 21 | 1116 | 2$^+$ | | | | | **1973PR06** | 1. $\pi$1/2[411]$\otimes\nu$3/2[402] |
| | 1183 | 3$^+$ | | | | | | |
| | | | | | | | | |
| 22 | 1348 | 1$^+$ | | | | | **1973PR06** | 1. $\pi$1/2[411]$\otimes\nu$1/2[400] |
| | 1382 | 2$^+$ | | | | | | |
| | | | | | | | | |
| 23 | 1389 | 4$^-$ | | | | | **1973PR06** | 1. $\pi$1/2[530]$\otimes\nu$7/2[633] |
| | 1482 | 5$^-$ | | | | | | 2. Regular band |
| | 1590 | 6$^-$ | | | | | | |
| | | | | | | | | |
| 24 | 1427 | 1$^+$ | | | | | **1973PR06** | 1. $\pi$1/2[411]$\otimes\nu$3/2[402] |
| | 1466 | 2$^+$ | | | | | | 2. Regular band |
| | 1507 | 3$^+$ | | | | | | |
| | | | | | | | | |
| 25 | 1439 | 3$^-$ | | | | | **1973PR06** | 1. $\pi$1/2[530]$\otimes\nu$7/2[633] |
| | 1540 | 4$^-$ | | | | | | 2. Regular band |
| | 1628 | 5$^-$ | | | | | | |
| | | | | | | | | |
| 26 | 198.9 | 4$^-$ | | | | | **2007CAZW** | 1. $\pi$h$_{9/2}\otimes\nu$i$_{13/2}$ |
| | 242.4 | 5$^-$ | | | | | | 2. Regular band |
| | 307.4 | 6$^-$ | 65.0 | 108.5 | | | | 3. Signature splitting more pronounced at higher spins |
| | 392.7 | 7$^-$ | 85.2 | | | | | |
| | 495.7 | 8$^-$ | 103.1 | 188.4 | | | | 4. Level energy is adopted from ENSDF |
| | 620.0 | 9$^-$ | 124.3 | 227.4 | | | | |
| | 759.3 | 10$^-$ | 139.1 | 263.7 | | | | |
| | 925.5 | 11$^-$ | 166.3 | 305.4 | | | | |
| | 1098.7 | 12$^-$ | 173.2 | 339.4 | | | | |
| | 1310.3 | 13$^-$ | 211.6 | 384.7 | | | | |
| | 1514.5 | 14$^-$ | 204.2 | 415.9 | | | | |
| | 1774.9 | 15$^-$ | 260.5 | 464.6 | | | | |
| | 2007.0 | 16$^-$ | 232.1 | 495.2 | | | | |
| | 2318.3 | 17$^-$ | 311.3 | 543.3 | | | | |
| | 2575.0 | 18$^-$ | 256.7 | 568.0 | | | | |
| | 2936.5 | 19$^-$ | | 618.2 | | | | |
| | 3217.0 | 20$^-$ | | 642.0 | | | | |
| | 3625.4 | 21$^-$ | | 688.9 | | | | |
| | 3929.3 | 22$^-$ | | 712.3 | | | | |
| | 4379.7 | 23$^-$ | | 754.3 | | | | |
| | 4705.1 | 24$^-$ | | 775.8 | | | | |

| S.No. | E$_{level}$ keV | I$^\pi$ | E$_\gamma$(M1) keV | E$_\gamma$(E2) keV | \|g$_K$-g$_R$\| | B(M1)/B(E2) ($\mu_N$/eb)$^2$ | Keywords | Configuration and Comments |
|---|---|---|---|---|---|---|---|---|
| 27 | 182.6+y | (7$^-$) | | | | | 2007CAZW | 1. $\pi7/2[523]\otimes v i_{13/2}$ |
| | 292.9+y | (8$^-$) | 110.2 | | | | | 2. Regular band |
| | 420.9+y | (9$^-$) | 127.8 | 238.4 | | | | 3. Signature splitting more pronounced at |
| | 567.5+y | (10$^-$) | 146.6 | 274.6 | | | | higher spins |
| | 733.6+y | (11$^-$) | 166.1 | 312.7 | | | | 4. Level energy is adopted from ENSDF |
| | 919.5+y | (12$^-$) | 185.9 | 351.9 | | | | |
| | 1125.2+y | (13$^-$) | 205.8 | 391.7 | | | | |
| | 1350.4+y | (14$^-$) | 225.2 | 430.8 | | | | |
| | 1595.4+y | (15$^-$) | 245.0 | 470.2 | | | | |
| | 1857.3+y | (16$^-$) | 262.0 | 506.9 | | | | |
| | 2140.9+y | (17$^-$) | 283.5 | 545.4 | | | | |
| | 2434.6+y | (18$^-$) | 293.6 | 577.5 | | | | |
| | 2757.0+y | (19$^-$) | | 616.1 | | | | |
| | 3076.1+y | (20$^-$) | | 641.5 | | | | |
| | 3437.2+y | (21$^-$) | | 680.2 | | | | |
| | 3772.9+y | (22$^-$) | | 696.8 | | | | |
| | (4170+y) | (23$^-$) | | (733) | | | | |
| | (4518+y) | (24$^-$) | | (745) | | | | |

$^{170}_{69}Tm_{101}$

| S.No. | E$_{level}$ keV | I$^\pi$ | E$_\gamma$(M1) keV | E$_\gamma$(E2) keV | \|g$_K$-g$_R$\| | B(M1)/B(E2) ($\mu_N$/eb)$^2$ | Keywords | Configuration and Comments |
|---|---|---|---|---|---|---|---|---|
| 1 | 0.0 | 1$^-$ | | | | | 1996HO12 | 1. $\pi1/2[411]\otimes v1/2[521]$ |
| | 38.713(1) | 2$^-$ | 38.714(1) | | | | | 2. Nuclear reaction: |
| | 114.543(1) | 3$^-$ | 75.831(1) | 114.544(1) | | | | $^{169}$Tm(n,γ)$^{170}$Tm, E= 2,24 KeV |
| | 183.189(1) | 4$^-$ | 68.649(0) | 144.480(1) | | | | 2. Half-life of bandhead is 128.6(3) d |
| | 319.326(1) | 5$^-$ | | 204.782(1) | | | | (1968RE02) |
| | 409.4(8) | 6$^-$ | | | | | | 3. Regular band |
| | 616.6(4) | 7$^-$ | | | | | | 4. Signature splitting |
| | | | | | | | | |
| | | | | | | | | |
| | | | | | | | | |
| 2 | 149.718(1) | 0$^-$ | | | | | 1996HO12 | 1. $\pi1/2[411]\otimes v1/2[521]$ |
| | 237.239(1) | 1$^-$ | 87.521(1) | | | | | 2. Irregular band |
| | 219.706(1) | 2$^-$ | | 69.988(1) | | | | 3. Signature splitting |
| | 349.733(1) | 3$^-$ | 130.027(0) | 112.494(2) | | | | 4. E$_\gamma$(M1) = 31.701(10) (1989DU03) |
| | 381.426(1) | 4$^-$ | 31.701(10) | 161.721(1) | | | | |
| | 550.747(2) | 5$^-$ | 169.321(2) | | | | | |
| | | | | | | | | |
| 3 | 189.193(1) | 3$^+$ | | | | | 1996HO12 | 1. $\pi1/2[411]\otimes v7/2[633]$ |
| | 247.148(2) | 4$^+$ | 63.959(4) | | | | | 2. Half-life of bandhead is 4.12(13) μs |
| | 327.1(7) | 5$^+$ | | | | | | (1967AN04) |
| | 426.5(3) | 6$^+$ | | | | | | 3. Regular band |
| | | | | | | | | |
| 4 | 204.448(1) | 2$^-$ | | | | | 1996HO12 | 1. $\pi1/2[411]\otimes v5/2[512]$ |
| | 270.546(1) | 3$^-$ | 66.098(1) | | | | | 2. Regular band |
| | 358.116(1) | 4$^-$ | 87.571(2) | 153.667(1) | | | | |
| | 467.860(1) | 5$^-$ | 109.744(3) | 197.314 | | | | |
| | 598.8(7) | 6$^-$ | | | | | | |

| S.No. | E$_{level}$ keV | I$^\pi$ | E$_\gamma$(M1) keV | E$_\gamma$(E2) keV | \|g$_K$-g$_R$\| | B(M1)/B(E2) ($\mu_N$/eb)$^2$ | Keywords | **Configuration and Comments** |
|---|---|---|---|---|---|---|---|---|
| 5 | 355.048(1) | 4$^+$ | | | | | **1996HO12** | 1. $\pi$1/2[411]$\otimes\nu$7/2[633] |
| | 439.8(4) | 5$^+$ | | | | | | |
| | | | | | | | | |
| 6 | 447.070(1) | 3$^-$ | | | | | **1996HO12** | 1. $\pi$1/2[411]$\otimes\nu$5/2[512] |
| | 539.722(1) | 4$^-$ | 92.654(3) | | | | | 2. Regular band |
| | 655.597(20) | 5$^-$ | | | | | | |
| | 790.1(8) | 6$^-$ | | | | | | |
| | 955.8(13) | 7$^-$ | | | | | | |
| | | | | | | | | |
| 7 | 590.230(2) | 1$^-$ | | | | | **1996HO12** | 1. Competing configurations are |
| | 637.905(2) | 2$^-$ | | | | | | $\pi$7/2[404]$\otimes\nu$5/2[512] |
| | 708.370(4) | 3$^-$ | | | | | | $\pi$3/2[411]$\otimes\nu$1/2[521] |
| | 801(6) | 4$^-$ | | | | | | 2. Regular band |
| | 921(2) | 5$^-$ | | | | | | |
| | 1061(9) | 6$^-$ | | | | | | |
| | 1230(5) | 7$^-$ | | | | | | |
| | | | | | | | | |
| 8 | 603.989(1) | 1$^+$ | | | | | **1996HO12** | 1. $\pi$1/2[541]$\otimes\nu$1/2[521] |
| | 650.372(1) | 2$^+$ | | | | | | 2. Irregular band |
| | 607.886(2) | 3$^+$ | | | | | | |
| | | | | | | | | |
| 9 | 643.8 | 4$^-$ | | | | | **1981DE29** | 1. $\pi$7/2[404]$\otimes\nu$1/2[521] |
| | | | | | | | | 2. Nuclear reaction: |
| | | | | | | | | $^{171}$Yb(t,$\alpha$)$^{170}$Tm, E= 17 MeV |
| | | | | | | | | |
| 10 | 648.746(1) | 1$^-$ | | | | | **1996HO12** | 1. Competing configurations are |
| | 693.286(1) | 2$^-$ | | | | | | $\pi$1/2[411]$\otimes\nu$3/2[521] |
| | 749.849(2) | 3$^-$ | | | | | | $\pi$7/2[404]$\otimes\nu$5/2[512] |
| | 835.4(9) | 4$^-$ | | | | | | $\pi$3/2[411]$\otimes\nu$1/2[521] |
| | | | | | | | | 2. Regular band |
| | | | | | | | | |
| 11 | 661.854(1) | 1$^+$ | | | | | **1996HO12** | 1. $\pi$7/2[523]$\otimes\nu$5/2[512] |
| | 703.629(1) | 2$^+$ | 41.770(2) | | | | | 2. E$_\gamma$(M1) = 41.770(2) (1989DU03) |
| | | | | | | | | |
| 12 | 683.569(3) | 0$^-$ | | | | | **1996HO12** | 1. $\pi$7/2[523]$\otimes\nu$7/2[633] |
| | 782.151(2) | 1$^-$ | 98.583(2) | | | | | 2. Irregular band |
| | 733.812(2) | 2$^-$ | | | | | | |
| | 862.780(2) | 3$^-$ | | | | | | |
| | | | | | | | | |
| 13 | 715.620(1) | 3$^-$ | | | | | **1996HO12** | 1. $\pi$1/2[541]$\otimes\nu$7/2[633] |
| | 756.210(4) | 4$^-$ | | | | | | |
| | | | | | | | | |
| 14 | 715.5 | 2$^-$ | | | | | **1981DE29** | 1. $\pi$5/2[402]$\otimes\nu$1/2[521] |
| | | | | | | | | |
| 15 | 719.263(2) | 1$^+$ | | | | | **1996HO12** | 1. $\pi$1/2[541]$\otimes\nu$1/2[521], K$^\pi$=0$^+$ |
| | 822.395(2) | 2$^+$ | | | | | | 2. Regular band |
| | 891(4) | 3$^+$ | | | | | | |
| | | | | | | | | |
| 16 | 720 | 1$^-$ | | | | | **1981DE29** | 1. $\pi$3/2[411]$\otimes\nu$1/2[521] |
| | 743.6 | 2$^-$ | | | | | | |
| | 800.8 | 3$^-$ | | | | | | |

| S.No. | E_level keV | $I^\pi$ | $E_\gamma$(M1) keV | $E_\gamma$(E2) keV | $|g_K-g_R|$ | B(M1)/B(E2) $(\mu_N/eb)^2$ | Keywords | **Configuration and Comments** |
|---|---|---|---|---|---|---|---|---|
| 17 | 733.8 | $2^-$ | | | | | **1981DE29** | 1. Tentative configuration |
| | 795.8 | $3^-$ | | | | | | $\pi1/2[411]\otimes\nu3/2[521]$ |
| 18 | 758.328(1) | $2^+$ | | | | | **1996HO12** | 1. $\pi1/2[541]\otimes\nu5/2[512]$ |
| 19 | 774.1 | $3^-$ | | | | | **1981DE29** | 1. $\pi7/2[404]\otimes\nu1/2[521]$ |
| | 866 | $(4^-)$ | | | | | | |
| 20 | 775.234(4) | $0^+$ | | | | | **1996HO12** | 1. $\pi7/2[404]\otimes\nu7/2[633]$ |
| | 860.483(2) | $1^+$ | | | | | | 2. Irregular band |
| | 818.508(2) | $2^+$ | | | | | | |
| 21 | 829.4 | $5^+$ | | | | | **1996HO12** | 1. $\pi7/2[523]\otimes\nu1/2[521]$, $K^\pi=4^+$ |
| | 948.5 | $6^+$ | | | | | | |
| 22 | 854.339(3) | $2^-$ | | | | | **1996HO12** | 1. Competing configurations are |
| | 925.273(4) | $3^-$ | | | | | | $\pi1/2[411]\otimes\nu3/2[521]$ |
| | 1014(1) | $4^-$ | | | | | | $\pi3/2[411]\otimes\nu1/2[521]$ |
| | 1131.1(4) | $5^-$ | | | | | | $\pi1/2[411]\otimes\nu1/2[521]-Q_{22}$ |
| | | | | | | | | 2. Regular band |
| 23 | 862.8 | $3^-$ | | | | | **1981DE29** | 1. $\pi5/2[402]\otimes\nu1/2[521]$ |
| | 953.6 | $4^-$ | | | | | | |
| 24 | 863.364(5) | $1^-$ | | | | | **1996HO12** | 1. Competing configurations are |
| | 908.448(3) | $2^-$ | | | | | | $\pi3/2[411]\otimes\nu5/2[512]$ |
| | 979.928(4) | $3^-$ | | | | | | $\pi1/2[411]\otimes\nu5/2[512]-Q_{22}$ |
| | | | | | | | | $\pi1/2[411]\otimes\nu1/2[510]$ |
| | | | | | | | | 2. Regular band |
| 25 | 890.8 | $5^+$ | | | | | **1981DE29** | 1. $\pi7/2[523]\otimes\nu1/2[521]$, $K^\pi=3^+$ |
| | 1011 | $6^+$ | | | | | | |
| 26 | 1213 | $3^-$ | | | | | **1981DE29** | 1. $\pi5/2[413]\otimes\nu1/2[521]$ |
| | 1291 | $4^-$ | | | | | | |
| 27 | 1313 | $1^-$ | | | | | **1981DE29** | 1. Competing configurations are |
| | 1351 | $2^-$ | | | | | | $\pi3/2[411]\otimes\nu1/2[521]+Q_{22}$ |
| | | | | | | | | $\pi1/2[411]\otimes\nu1/2[521]$ (Tentative) |
| 28 | 1382 | $2^-$ | | | | | **1981DE29** | 1. $\pi5/2[413]\otimes\nu1/2[521]$ |
| | 1448 | $3^-$ | | | | | | 2. Regular band |
| | 1539 | $4^-$ | | | | | | |

$^{172}_{69}Tm_{103}$

| S.No. | E$_{level}$ keV | I$^\pi$ | E$_\gamma$(M1) keV | E$_\gamma$(E2) keV | \|g$_K$-g$_R$\| | B(M1)/B(E2) ($\mu_N$/eb)$^2$ | Keywords | Configuration and Comments |
|---|---|---|---|---|---|---|---|---|
| 1 | 0.0 | 2$^-$ | | | | | **2008HU05** | 1. $\pi1/2[411]\otimes\nu5/2[512]$ |
| | 62.6 | 3$^-$ | 62.6 | | | | 1965HA24 | 2. Nuclear reaction: |
| | 145.9 | 4$^-$ | 83.3 | 145.9 | 0.67(4) | | | $^{170}$Er($^7$Li,$\alpha n\gamma$)$^{172}$Tm, E= 30 MeV |
| | 249.8 | 5$^-$ | 103.9 | 187.2 | 0.68(3) | | | 3. Half-life of bandhead is 63.5 h |
| | 374.4 | 6$^-$ | 124.6 | 228.5 | 0.65(2) | | | (1965HA24) |
| | 519.3 | 7$^-$ | (145) | 269.5 | | | | 4. Regular band |
| | (685) | 8$^-$ | (166) | (311) | | | | 5. Predicted g$_K$-g$_R$ for bandhead is -0.42(25) |
| | (871) | 9$^-$ | (186) | (352) | | | | and -0.57(9) by taking g$_\Omega(\pi1/2+[411])$ is |
| | | | | | | | | equal to $-1.75(25)$ and $-1.14(7)$ |
| | | | | | | | | respectively |
| | | | | | | | | 6. Measured g$_K$-g$_R$ for bandhead is 0.67(3) |
| | | | | | | | | 7. Taking Q$_0$ = 7.7eb |
| 2 | 240.0 | 3$^-$ | | | | | **2008HU05** | 1. $\pi1/2[411]\otimes\nu5/2[512]$ |
| | 330.0 | 4$^-$ | 90.0 | | | | 1965HA24 | 2. Regular band |
| | 441.7 | 5$^-$ | 111.7 | 201.7 | 0.75(14) | | | 3. Predicted g$_K$-g$_R$ for bandhead is -0.97(25) |
| | (575) | 6$^-$ | (133) | (245) | | | | and -0.87(9) by taking g$_\Omega(\pi1/2+[411])$ is |
| | (729) | 7$^-$ | (145) | (277) | | | | equal to $-1.75(25)$ and $-1.14(7)$ |
| | | | | | | | | respectively |
| | | | | | | | | 4. Measured g$_K$-g$_R$ for bandhead is 0.75(14) |
| | | | | | | | | 5. Taking Q$_0$ = 7.7eb |
| 3 | 407 | 1$^-$ | | | | | 1965HA24 | 1. $\pi1/2[411]\otimes\nu1/2[521]$ |
| | 446 | 2$^-$ | | | | | | |
| 4 | 475 | 0$^-$ | | | | | 1965HA24 | 1. Competing configurations are |
| | 535 | 1$^-$ | | | | | | $\pi1/2[411]\otimes\nu1/2[521]$ |
| | | | | | | | | $\pi1/2[411]\otimes\nu5/2[512]$-Q$_{22}$ |
| 5 | 610 | 1$^+$ | | | | | 1965HA24 | 1. $\pi7/2[523]\otimes\nu5/2[512]$ |

$^{174}_{69}Tm_{105}$

| S.No. | E$_{level}$ keV | I$^\pi$ | E$_\gamma$(M1) keV | E$_\gamma$(E2) keV | \|g$_K$-g$_R$\| | B(M1)/B(E2) ($\mu_N$/eb)$^2$ | Keywords | Configuration and Comments |
|---|---|---|---|---|---|---|---|---|
| 1 | 252 | 0$^+$ | | | | | **2006CH10** | 1. $\pi7/2[523]\otimes\nu7/2[514]$ |
| | | | | | | | | 2. Nuclear reaction: |
| | | | | | | | | $^{174}$Tm IT decay:2.39 s |
| 2 | 58.5 | (1) | | | | | **1991BE04** | 1. Competing Configurations |
| | | | | | | | | $\pi7/2[523]\otimes\nu5/2[512]$ or |
| | | | | | | | | $\pi7/2[523]\otimes\nu9/2[624]$ |
| | | | | | | | | 2. Nuclear reaction: |
| | | | | | | | | $^{174}$Er $\beta$−decay $^{174}$Tm |
| | | | | | | | | E = 11.6 MeV/nucleon |

| S.No. | E$_{level}$ keV | I$^\pi$ | E$_\gamma$(M1) keV | E$_\gamma$(E2) keV | \|g$_K$-g$_R$\| | B(M1)/B(E2) ($\mu_N$/eb)$^2$ | Keywords | Configuration and Comments |
|---|---|---|---|---|---|---|---|---|
| 3 | (767 or 773) | (1$^-$) | | | | | **1991BE04** | 1. $\pi$1/2[411]$\otimes\nu$1/2[521] |
| 4 | 100.3 | 3$^-$ | | | | | **2013HU08** | 1. $\pi$1/2[411]$\otimes\nu$7/2[514] |
| | | | | | | | | 2. Nuclear reaction: $^{174}$Yb($^{136}$Xe,X$\gamma$)$^{174}$Tm |
| | | | | | | | | 3. Predicted value of \|g$_K$-g$_R$\| for bandhead= +0.21(8) by using (1995ST11,2007BI14) |
| 5 | 0.0 | 4$^-$ | | | 0.32(6) | | **2013HU08** | 1. $\pi$1/2[411]$\otimes\nu$7/2[514] |
| | 106.3 | 5$^-$ | 106.3(2) | | | | 2006CH10 | 2. Half-life of bandhead is 5.4(1) min |
| | 233.6 | 6$^-$ | 127.3(3) | 233.6(3) | 0.37(4) | | 1966WI04 | 3. Predicted value of \|g$_K$-g$_R$\| for bandhead= -0.21(6) by using (1995ST11,2007BI14) |
| | 382 | 7$^-$ | 148.3(3) | 275.7(1) | 0.33(3) | | | |
| | 551.1 | 8$^-$ | 169.2(4) | 317.4(3) | 0.32(5) | | | 4. Regular band |
| | 740.7 | 9$^-$ | 189.4(3) | 358.7(1) | 0.31(3) | | | 5. Energy taken from XUNDL. |
| | 949.5 | 10$^-$ | 209(1) | 398.5(3) | 0.25(11) | | | |
| | 1179.0 | 11$^-$ | 229(1) | 438.3(1) | 0.28(4) | | | |
| | 1426.9 | 12$^-$ | 248(1) | 477.6(3) | 0.48(23) | | | |
| | 1694.3 | 13$^-$ | 267.5(5) | 515.3(3) | 0.44(11) | | | |
| 6 | 191.5 | (4$^+$) | | | 0.56(6) | | **2013HU08** | 1. $\pi$1/2[411]$\otimes\nu$9/2[624] |
| | 288.4 | (5$^+$) | 96.7(2) | | | | | 2. Half-life of bandhead is 15.2(21) ns |
| | 406.2 | (6$^+$) | 117.9(2) | 214.8(3) | 0.50(6) | | | 3. Predicted value of \|g$_K$-g$_R$\| = -0.39(9) by using (1995ST11,2007BI14) |
| | | | | | | | | 4. Regular band |
| 7 | 694.6 | 8$^-$ | | | 0.10(3) | | **2013HU08** | 1. $\pi$7/2[523]$\otimes\nu$9/2[624] |
| | 858.2 | 9$^-$ | 161.6(1) | | | | | 2. Half-life of bandhead is 3.5(14) ns |
| | 1038.5 | 10$^-$ | 180.2(2) | 341.9(2) | 0.07(1) | | | 3. Predicted value of \|g$_K$-g$_R$\| for bandhead= +0.12(9) by using (1995ST11,2007BI14) |
| | 1237.2 | 11$^-$ | 198.7(5) | 378.9(3) | 0.11(3) | | | |
| | 1455.1 | 12$^-$ | 218.0(5) | 416.6(3) | 0.11(3) | | | 4. Regular band |
| | 1690.5 | 13$^-$ | 235.5(4) | 453.3(3) | 0.09(1) | | | |
| | 1947 | 14$^-$ | 256(1) | 492(1) | 0.12(11) | | | |
| 8 | 496.9 | (7$^+$) | | | | | **2013HU08** | 1. $\pi$7/2[523]$\otimes\nu$7/2[514] |
| | 651.5 | (8$^+$) | 154.6(3) | | | | | 2. Half-life of bandhead is 10.4(14) ns |
| | | | | | | | | 3. Predicted value of \|g$_K$-g$_R$\| for bandhead = +0.45(9) by using (1995ST11,2007BI14) |

$^{176}_{69}Tm_{107}$

| S.No. | E$_{level}$ keV | I$^\pi$ | E$_\gamma$(M1) keV | E$_\gamma$(E2) keV | \|g$_K$-g$_R$\| | B(M1)/B(E2) ($\mu_N$/eb)$^2$ | Keywords | Configuration and Comments |
|---|---|---|---|---|---|---|---|---|
| 1 | 0.0 | (4$^+$) | | | | | **1967GU11** | 1. $\pi$1/2[411]$\otimes\nu$9/2[624] |

$^{160}_{71}Lu_{89}$

| S.No. | $E_{level}$ keV | $I^\pi$ | $E_\gamma$(M1) keV | $E_\gamma$(E2) keV | $|g_K-g_R|$ | B(M1)/B(E2) $(\mu_N/eb)^2$ | Keywords | Configuration and Comments |
|---|---|---|---|---|---|---|---|---|
| 1 | (≥0) | (1$^-$) | | | | | **1984AU13** | 1. $\pi 1/2[411]\otimes\nu 3/2[521]$ |
| | | | | | | | | 2. Nuclear reaction: |
| | | | | | | | | $^{144}$Sm ($^{19}$F, 3n)$^{160}$Lu, E= 85 MeV |
| | | | | | | | | 2. Half-life of bandhead is 36.2(3) s. |
| | | | | | | | | |
| 2 | 558.5+x | (9$^-$) | | | | | **2012WA17** | 1. $\pi h_{11/2}\otimes\nu i_{13/2}$ |
| | 734.1+x | (10$^-$) | 175.7 | | | | | 2. Nuclear reaction: |
| | 810.6+x | (11$^-$) | 75.7 | 252.0 | | | | $^{144}$Sm ($^{19}$F, 3n$\gamma$)$^{160}$Lu |
| | 1058.6+x | (12$^-$) | 247.2 | 325.4 | | | | E=90 and 106 MeV |
| | 1217.8+x | (13$^-$) | 159.1 | 407.0 | | | | 3. Regular band |
| | 1522.4+x | (14$^-$) | 304 | 463.9 | | | | 4. Signature splitting |
| | 1754.8+x | (15$^-$) | 232.3 | 537.1 | | | | 5. Level energy is adopted from XUNDL |
| | 2085.4+x | (16$^-$) | 330.1 | 563.1 | | | | |
| | 2371.7+x | (17$^-$) | 286.0 | 617.1 | | | | |
| | 2727.5+x | (18$^-$) | 355.5 | 642.3 | | | | |
| | 3040.8+x | (19$^-$) | 313.1 | 669.2 | | | | |
| | 3426.7+x | (20$^-$) | 385.5 | 699.3 | | | | |
| | 3752.2+x | (21$^-$) | 325.5 | 711.7 | | | | |
| | 4175.5+x | (22$^-$) | 423.1 | 748.4 | | | | |
| | 4520.1+x | (23$^-$) | 344.2 | 768.5 | | | | |
| | 5001.8+x | (24$^-$) | 481.7 | 826.2 | | | | |
| | 5379.8+x | (25$^-$) | | 859.7 | | | | |
| | | | | | | | | |
| 3 | 625.0 | (8$^+$) | | | | | **2012WA17** | 1. $\pi 1/2[411]\otimes\nu 1/2[660]$ |
| | 933.2 | (10$^+$) | | 308.2 | | | | 2. Regular band |
| | 1360.7 | (12$^+$) | | 427.5 | | | | 3. Level energy is adopted from XUNDL |
| | 1854.4 | (14$^+$) | | 493.9 | | | | |
| | 2422.4 | (16$^+$) | | 567.5 | | | | |
| | 3024.2 | (18$^+$) | | 601.8 | | | | |
| | 3657.0 | (20$^+$) | | 632.8 | | | | |
| | 4392.0 | (22$^+$) | | 735.0 | | | | |
| | | | | | | | | |
| 4 | 0.0+x | (6$^+$) | | | | | | |
| | 161.3+x | (7$^+$) | | | | | **2012WA17** | 1. $\pi 5/2[402]\otimes\nu 1/2[660]$ |
| | 323.4+x | (8$^+$) | | 323.4 | | | | 2. Regular band |
| | 463.0+x | (9$^+$) | | 301.7 | | | | 3. Signature Splitting |
| | 705.7+x | (10$^+$) | 243.1 | 382.3 | | | | 4. Level energy is adopted from XUNDL |
| | 857.8+x | (11$^+$) | 152.5 | 394.3 | | | | |
| | 1160.6+x | (12$^+$) | 302.3 | 454.9 | | | | |
| | 1337.8+x | (13$^+$) | 177.5 | 480.3 | | | | |
| | 1661.1+x | (14$^+$) | | 499.8 | | | | |
| | 1898.3+x | (15$^+$) | | 561.0 | | | | |

$^{162}_{71}Lu_{91}$

| S.No. | E$_{level}$ keV | I$^\pi$ | E$_\gamma$(M1) keV | E$_\gamma$(E2) keV | \|g$_K$-g$_R$\| | B(M1)/B(E2) ($\mu_N$/eb)$^2$ | Keywords | Configuration and Comments |
|---|---|---|---|---|---|---|---|---|
| 1 | 0.0 | (1$^-$) or (1$^+$) | | | | | **1980BEYG** **1990SO04** | 1. Competing Configurations: $\pi$1/2[411]⊗$\nu$3/2[521] with (1$^-$) (1980BEYG) $\pi$7/2[523]⊗$\nu$5/2[523] with (1$^+$) (1990SO04) 2. Nuclear reaction: $^{147}$Sm($^{19}$F,4n)$^{162}$Lu 3. Half-life of bandhead is 1.37(2) min |
| 2 | 143.8+x | (10$^-$) | | | | | **1997CA29** | 1. $\pi$h$_{11/2}$⊗$\nu$i$_{13/2}$ |
| | 224.5+x | (11$^-$) | 97.8 | | | | | 2. Nuclear reaction: $^{139}$La($^{28}$Si,5n)$^{162}$Lu, E=160MeV |
| | 420.0+x | (12$^-$) | 195.6 | 293.4 | | | | 3. Regular band |
| | 580.5+x | (13$^-$) | 160.5 | 355.9 | | 2.1(2) | | 4. Signature Splitting |
| | 857.1+x | (14$^-$) | 276.8 | 437.5 | | 1.05(13) | | 5. Signature inversion at I = (21$^-$) |
| | 1091.3+x | (15$^-$) | 234.4 | 511.2 | | 1.50(15) | | 6. Uncertainties in E$_\gamma$ between 0.1 and |
| | 1417.5+x | (16$^-$) | 326.5 | 560.9 | | 1.3(2) | | 0.3 KeV |
| | 1717.2+x | (17$^-$) | 299.9 | 626.3 | | 1.34(11) | | 7. B(M1)/B(E2) ratio calculated in |
| | 2074.0+x | (18$^-$) | 357.2 | 657.1 | | 1.1(2) | | paper by assuming $\delta^2$=0 |
| | 2426.2+x | (19$^-$) | 352.4 | 709.7 | | 1.3(2) | | 8. Level energy is adopted from ENSDF |
| | 2800.3+x | (20$^-$) | 374.0 | 726.2 | | 1.6(3) | | |
| | 3183.1+x | (21$^-$) | 383.0 | 757.0 | | 1.4(2) | | |
| | 3547.4+x | (22$^-$) | 364.4 | 747.0 | | 1.4(4) | | |
| | 3921.9+x | (23$^-$) | 375.3 | 739.2 | | 1.5(4) | | |
| | 4262.2+x | (24$^-$) | 340.4 | 715.5 | | 2.2(4) | | |
| | 4634.5+x | (25$^-$) | 372.1 | 712.6 | | 1.8(5) | | |
| | 4982.2+x | (26$^-$) | 347.6 | 720.2 | | 2.3(5) | | |
| | 5379.8+x | (27$^-$) | 397.6 | 745.6 | | 1.9(7) | | |
| | 5751.2+x | (28$^-$) | 371.6 | 769.2 | | 2.8(9) | | |
| | 6185.4+x | (29$^-$) | 434.7 | 805.9 | | 2.1(7) | | |

$^{164}_{71}Lu_{93}$

| S.No. | E$_{level}$ keV | I$^\pi$ | E$_\gamma$(M1) keV | E$_\gamma$(E2) keV | \|g$_K$-g$_R$\| | B(M1)/B(E2) ($\mu_N$/eb)$^2$ | Keywords | Configuration and Comments |
|---|---|---|---|---|---|---|---|---|
| 1 | 0.0 | 1$^-$ | | | | | **1998GE13** 1977HU02 | 1. $\pi$1/2[411]⊗$\nu$3/2[521] 2. Competing configurations are $\pi$1/2[411]⊗$\nu$3/2[521], I$^\pi$ = (1$^-$) $\pi$7/2[404]⊗$\nu$3/2[521], I$^\pi$ = (2$^-$) (1977HU02) 3. Nuclear reaction: $^{164}$Hf $\epsilon$ decay(111s) 4. Half-life of bandhead is 3.17(3) min (1977HU02) |

| S.No. | $E_{level}$ keV | $I^\pi$ | $E_\gamma$ (M1) keV | $E_\gamma$ (E2) keV | $|g_K-g_R|$ | B(M1)/B(E2) $(\mu_N/eb)^2$ | Keywords | Configuration and Comments |
|---|---|---|---|---|---|---|---|---|
| 2 | 0+x | (10⁻) | | | | | **1996JU01** | 1. π7/2[523]⊗ν5/2[642] |
| | 93.4+x | (11⁻) | 93.4 | | | | | 2. Nuclear Reaction: |
| | 259.0x | (12⁻) | 165.6 | 259.0 | | 1.11(9) | | ¹⁵⁰Sm(¹⁹F,5nγ)¹⁶⁴Lu |
| | 421.3+x | (13⁻) | 162.3 | 327.8 | | 1.30(13) | | 3. Regular band |
| | 658.4+x | (14⁻) | 237.2 | 399.5 | | 1.23(9) | | 4. Signature splitting |
| | 885.0+x | (15⁻) | 226.7 | 463.7 | | 1.16(8) | | 5. Signature inversion occur at $I^\pi$ = (19⁻) |
| | 1170.0+x | (16⁻) | 285.0 | 511.5 | | 1.14(9) | | 6. B(M1)/B(E2) ratio calculated in |
| | 1454.2+x | (17⁻) | 284.2 | 569.2 | | 0.92(12) | | paper by assuming δ²=0 |
| | 1767.2+x | (18⁻) | 313.0 | 597.1 | | 1.04(9) | | 7. Level energy is adopted from XUNDL |
| | 2099.3+x | (19⁻) | 332.1 | 645.2 | | 1.27(12) | | |
| | 2424.3+x | (20⁻) | 325.1 | 657.0 | | 1.55(19) | | |
| | 2788.1+x | (21⁻) | 363.4 | 689.1 | | 1.29(16) | | |
| | 3111.6+x | (22⁻) | 323.5 | 687.3 | | 1.9(4) | | |
| | 3488.9+x | (23⁻) | 377.5 | 700.8 | | 1.0(2) | | |
| | 3809.4+x | (24⁻) | 320.5 | 697.6 | | 2.8(5) | | |
| | 4197.9+x | (25⁻) | 388.5 | 709.1 | | 1.7(4) | | |
| | 4532.4+x | (26⁻) | 334.5 | 723.0 | | 5.0(15) | | |
| | 4946.3+x | (27⁻) | 413.7 | 748.5 | | 2.3(9) | | |
| | 5307.4+x | (28⁻) | 361.0 | 775.0 | | 4.6(18) | | |
| 3 | 0+x | (8⁺) | | | | | **1997CA29** | 1. π$h_{11/2}$⊗ν(3/2[521] or 5/2[523]) |
| | 104.1+x | (9⁺) | 104.1 | | | | 1996JU01 | 2. π7/2[523]⊗ν3/2[521] (1996JU01) |
| | 245.1+x | (10⁺) | 140.9 | 245.1 | | 1.0(2) | | 3. Nuclear reaction: |
| | 413.3+x | (11⁺) | 168.2 | 309.3 | | 0.74(9) | | ¹³⁹La(²⁹Si,4nγ)¹⁶⁴Lu |
| | 623.7+x | (12⁺) | 210.3 | 378.6 | | 0.70(11) | | 4. Regular band |
| | 848.5+x | (13⁺) | 224.7 | 435.2 | | 0.60(7) | | 5. Signature splitting |
| | 1115.4+x | (14⁺) | 267.0 | 491.7 | | 0.55(9) | | 6. B(M1)/B(E2) ratio calculated in |
| | 1382.4+x | (15⁺) | 267.0 | 534.0 | | 0.66(19) | | paper by assuming δ²=0 |
| | 1690.1+x | (16⁺) | 307.8 | 574.6 | | 0.51(14) | | 7. Level energy is adopted from XUNDL |
| | 1988.0+x | (17⁺) | 298.0 | 605.4 | | 0.34(10) | | |
| | 2314.3+x | (18⁺) | 326.3 | 624.3 | | 0.27(6) | | |
| | 2621.4+x | (19⁺) | 307.1 | 633.2 | | 0.79(18) | | |
| | 2919.3+x | (20⁺) | 297.8 | 605.0 | | 2.4(8) | | |
| | 3208.6+x | (21⁺) | 289.2 | 587.2 | | 1.8(5) | | |
| | 3519.2+x | (22⁺) | 310.6 | 600.0 | | 1.2(2) | | |
| | 3836.9+x | (23⁺) | 317.7 | 628.3 | | 0.73(18) | | |
| | 4183.0+x | (24⁺) | 346.1 | 663.8 | | 0.43(11) | | |
| | 4537.5+x | (25⁺) | 354.7 | 700.5 | | 0.60(12) | | |
| | 4920.7+x | (26⁺) | 383.3 | 737.6 | | 0.78(16) | | |
| | 5316.5+x | (27⁺) | | 779.0 | | | | |
| | (5734.7+x) | (28⁺) | | (814) | | | | |
| | (6196.5+x) | (29⁺) | | (880) | | | | |
| 4 | 0+z | (23⁺) | | | | | **1996JU01** | 1. π1/2[541]⊗ν5/2[642] |
| | 344.4+z | (24⁺) | | 344.4 | | | | 2. Regular band |
| | 755.8+z | (25⁺) | | 411.4 | | | | 3. Level energy is adopted from XUNDL |
| | 1227.9+z | (16⁻) | | 472.1 | | | | |
| | 1747.6+z | (18⁻) | | 519.7 | | | | |
| | 2301.5+z | (20⁻) | | 553.9 | | | | |
| | 2885.9+z | (22⁻) | | 584.4 | | | | |
| 5 | 0+z | (9⁺) | | | | | **1996JU01** | 1. π1/2[411]⊗ν5/2[642] |
| | 401.8+z | (11⁺) | | 401.8 | | | | 2. Regular band |
| | 908.3+z | (13⁺) | | 506.5 | | | | 3. Level energy is adopted from XUNDL |
| | 1494.1+z | (15⁺) | | 585.8 | | | | |
| | 2127.5+z | (17⁺) | | 633.4 | | | | |
| | 2777.1+z | | | 649.6 | | | | |

| S.No. | $E_{level}$ keV | $I^\pi$ | $E_\gamma$(M1) keV | $E_\gamma$(E2) keV | $|g_K - g_R|$ | B(M1)/B(E2) $(\mu_N/eb)^2$ | Keywords | Configuration and Comments |
|---|---|---|---|---|---|---|---|---|
| 6 | 9.0+x | 7+ | | | | | 2007BR09 | 1. $\pi g_{7/2} \otimes \nu i_{13/2}$ |
| | 24.6+x | 8+ | | | | | | 2. Nuclear reaction: |
| | 128.7+x | 9+ | 104.0 | 119.0 | | | | $^{121}$Sb($^{48}$Ca,5n$\gamma$)$^{164}$Lu |
| | 269.3+x | 10+ | 140.8 | 244.8 | | | | 3. Regular band |
| | 437.8+x | 11+ | 168.7 | 308.8 | | | | 4. Small signature splitting |
| | 647.9+x | 12+ | 210.1 | 378.7 | | | | 5. Signature inversion occur at $I^\pi = 10^+$ |
| | 873.0+x | 13+ | 225.1 | 435.2 | | | | 6. Level energy is adopted from XUNDL |
| | 1139.8+x | 14+ | 266.8 | 491.9 | | | | |
| | 1407.3+x | 15+ | 267.5 | 534.3 | | | | |
| | 1714.5+x | 16+ | 307.2 | 574.7 | | | | |
| | 2013.1+x | 17+ | 298.6 | 605.8 | | | | |
| | 2339.0+x | 18+ | 325.9 | 624.5 | | | | |
| | 2646.4+x | 19+ | 307.4 | 633.3 | | | | |
| 7 | 0+x | 7− | | | | | 2007BR09 | 1. $\pi h_{11/2} \otimes \nu i_{13/2}$ |
| | 45.0+x | 8− | | | | | | 2. Regular band |
| | 127.0+x | 9− | | | | | | 3. Signature splitting |
| | 199.5+x | 10− | 72.6 | | | | | 4. Signature inversion occur at both |
| | 292.6+x | 11− | 93.1 | 165.5 | | | | $I^\pi = 12^-$ and $19^-$ |
| | 458.0+x | 12− | 165.3 | 258.8 | | | | 5. Level energy is adopted from XUNDL |
| | 620.2+x | 13− | 162.2 | 327.4 | | | | |
| | 857.1+x | 14− | 236.9 | 399.1 | | | | |
| | 1083.5+x | 15− | 226.3 | 463.4 | | | | |
| | 1368.1+x | 16− | 284.6 | 511.1 | | | | |
| | 1652.5+x | 17− | 284.7 | 568.7 | | | | |
| | 1965.3+x | 18− | 312.8 | 597.0 | | | | |
| | 2297.0+x | 19− | 331.7 | 644.7 | | | | |
| | 2622.1+x | 20− | 325.2 | 656.6 | | | | |
| 8 | 2161.7+x | 14− | | | | | 2007BR09 | 1. $\pi i_{13/2} \otimes \nu h_{9/2}$ |
| | 2535.1+x | 16− | | 373.4 | | | | 2. Regular band |
| | 2971.6+x | 18− | | 436.5 | | | | 3. Triaxial Side band |
| | 3468.3+x | 20− | | 496.6 | | | | 4. Level energy is adopted from XUNDL |
| | 4025.9+x | 22− | | 557.5 | | | | |
| | 4642.6+x | 24− | | 616.6 | | | | |
| | 5317.4+x | 26− | | 674.6 | | | | |
| | 6050.2+x | 28− | | 732.7 | | | | |
| | 6840.0+x | 30− | | 789.8 | | | | |
| | 7687.4+x | 32− | | 847.4 | | | | |
| | 8592.8+x | 34− | | 905.4 | | | | |
| | 9556.8+x | 36− | | 964.0 | | | | |
| | 10580.5+x | 38− | | 1023.7 | | | | |
| | 11663.7+x | 40− | | 1083.2 | | | | |
| | 12798.9+x | 42− | | 1135.2 | | | | |
| | 14018.2+x | 44− | | 1219 | | | | |
| | 15274.2+x | 46− | | 1256 | | | | |
| | 16585.2+x | 48− | | 1311 | | | | |
| | (17957.3+x) | 50− | | (1372) | | | | |

| S.No. | E$_{level}$ keV | I$^\pi$ | E$_\gamma$(M1) keV | E$_\gamma$(E2) keV | \|g$_K$-g$_R$\| | B(M1)/B(E2) ($\mu_N$/eb)$^2$ | Keywords | Configuration and Comments |
|---|---|---|---|---|---|---|---|---|
| 9 | 2035.3+x | 13$^+$ | | | | | **2007BR09** | 1. $\pi i_{13/2} \otimes \nu i_{13/2}$ |
| | 2389.1+x | 15$^+$ | | 353.7 | | | | 2. Regular band |
| | 2812.4+x | 17$^+$ | | 423.4 | | | | 3. Triaxial Side band |
| | 3296.4+x | 19$^+$ | | 484.1 | | | | 4. Level energy is adopted from XUNDL |
| | 3842.9+x | 21$^+$ | | 546.6 | | | | |
| | 4449.0+x | 23$^+$ | | 606.2 | | | | |
| | 5113.4+x | 25$^+$ | | 664.5 | | | | |
| | 5834.1+x | 27$^+$ | | 720.9 | | | | |
| | 6610.8+x | 29$^+$ | | 776.8 | | | | |
| | 7442.4+x | 31$^+$ | | 831.6 | | | | |
| | 8328.2+x | 33$^+$ | | 885.8 | | | | |
| | 9268.4+x | 35$^+$ | | 940.2 | | | | |
| | 10263.4+x | 37$^+$ | | 995.0 | | | | |
| | 11313.6+x | 39$^+$ | | 1050.2 | | | | |
| | 12418.0+x | 41$^+$ | | 1104.4 | | | | |
| | 13577.9+x | 43$^+$ | | 1159.9 | | | | |
| | 14788.9+x | 45$^+$ | | 1211.0 | | | | |
| | (16065.9+x) | (47$^+$) | | (1277) | | | | |
| | (17381.9+x) | (49$^+$) | | (1316) | | | | |

$^{166}_{71}Lu_{95}$

| S.No. | E$_{level}$ keV | I$^\pi$ | E$_\gamma$(M1) keV | E$_\gamma$(E2) keV | \|g$_K$-g$_R$\| | B(M1)/B(E2) ($\mu_N$/eb)$^2$ | Keywords | Configuration and Comments |
|---|---|---|---|---|---|---|---|---|
| 1 | 0.0 | (6$^-$) | | | | | **1974DE09** | 1. $\pi 7/2[404] \otimes \nu 5/2[523]$ |
| | | | | | | | | 2. Nuclear Reaction: $^{166}$Lu IT decay (1.41 min) |
| | | | | | | | | 3. Half-life of bandhead is 2.65(10) |
| 2 | 34.37 | (3$^-$) | | | | | **1974DE09** | 1. $\pi 1/2[411] \otimes \nu 5/2[523]$ |
| | | | | | | | | 2. Half-life of bandhead is 1.41(10) |
| 3 | 42.9 | (0$^-$) | | | | | **1974DE09** | 1. $\pi 5/2[402] \otimes \nu 5/2[523]$ |
| | | | | | | | | 2. Half-life of bandhead is 2.12(10) |
| 4 | 57.2 | (1)$^-$ | | | | | **1974DE09** | 1. $\pi 7/2[404] \otimes \nu 5/2[523]$ |
| 5 | 60.5 | (3$^+$) | | | | | **1974DE09** | 1. $\pi 1/2[541] \otimes \nu 5/2[523]$ |
| 6 | 83.5 | (6$^+$) | | | | | **1992HO02** | 1. $\pi 9/2[514] \otimes \nu 3/2[521]$ |
| 7 | 136.0 | 1$^+$ | | | | | **1974DE09** | 1. $\pi 7/2[523] \otimes \nu 5/2[523]$ |
| 8 | 144.8 | (6$^-$) | | | | | **1992HO02** | 1. $\pi 9/2[514] \otimes \nu 3/2[521]$ |

| S.No. | E$_{level}$ keV | I$^\pi$ | E$_\gamma$(M1) keV | E$_\gamma$(E2) keV | \|g$_K$-g$_R$\| | B(M1)/B(E2) ($\mu_N$/eb)$^2$ | Keywords | Configuration and Comments |
|---|---|---|---|---|---|---|---|---|
| 9 | 0.0+x | (7$^+$) | | | | | **2000ZH51** | 1. π7/2[404]⊗ν5/2[642] |
| | 100.18+x | 8$^+$ | 100.1 | | | | 1992HO02 | 2. Nuclear reaction: |
| | 235.12+x | 9$^+$ | 134.8 | 235.1 | | | | $^{152}$Sm($^{19}$F,5nγ)$^{166}$Lu |
| | 400.4+x | 10$^+$ | 165.3 | 300.2 | | | | 3. Regular band |
| | 595.4+x | 11$^+$ | 194.8 | 360.5 | | | | 4. Signature splitting more pronounced at |
| | 813.1+x | 12$^+$ | 217.6 | 412.7 | | | | higher spin |
| | 1058.4+x | 13$^+$ | 245.2 | 463.0 | | | | 5. Signature inversion occur at I$^\pi$ = 19$^+$ |
| | 1319.6+x | 14$^+$ | 261.1 | 506.6 | | | | 6. Level energy is adopted from XUNDL |
| | 1606.4+x | 15$^+$ | 286.7 | 548.1 | | | | |
| | 1902.7+x | 16$^+$ | 296.2 | 583.2 | | | | |
| | 2222.2+x | 17$^+$ | | 615.8 | | | | |
| | 2544.7+x | 18$^+$ | | 642.0 | | | | |
| | 2873.6+x | 19$^+$ | | 651.4 | | | | |
| | 3234.4+x | 20$^+$ | | 689.7 | | | | |
| | 3576.6+x | 21$^+$ | | (703.0) | | | | |
| | 3953.5+x | 22$^+$ | | (719) | | | | |
| | 4323.9+x | 23$^+$ | | (747) | | | | |
| 10 | 0.0+y | (8$^-$) | | | | | **2000ZH51** | 1. π9/2[514]⊗ν5/2[642] |
| | 53.8+y | 9$^-$ | 53.6 | | | | 1992HO02 | 2. Regular band |
| | 138.5+y | 10$^-$ | 84.7 | 138.5 | | | | 3. Signature splitting in ΔE$_\gamma$ vs I |
| | 250.3+y | 11$^-$ | 111.7 | 196.4 | | | | 4. Signature inversion occur at I$^\pi$ = 17$^-$ |
| | 405.3+y | 12$^-$ | 154.7 | 267.0 | | | | 5. Level energy is adopted from XUNDL |
| | 578.2+y | 13$^-$ | 172.8 | 328.0 | | | | |
| | 793.2+y | 14$^-$ | 214.7 | 388.0 | | | | |
| | 1022.1+y | 15$^-$ | 228.8 | 444.0 | | | | |
| | 1283.4+y | 16$^-$ | 261.3 | 490.1 | | | | |
| | 1564.9+y | 17$^-$ | 281.4 | 543.0 | | | | |
| | 1859.5+y | 18$^-$ | 294.5 | 576.0 | | | | |
| | 2189.7+y | 19$^-$ | 330.1 | 625.0 | | | | |
| | 2506.7+y | 20$^-$ | 317.1 | 647.1 | | | | |
| | 2878.0+y | 21$^-$ | 371.4 | 688.3 | | | | |
| | 3204.9+y | 22$^-$ | 327.1 | 698.1 | | | | |
| | 3923.7+y | 24$^-$ | | (718.8) | | | | |
| 11 | 0.0+u | (8$^-$) | | | | | **2000ZH51** | 1. π1/2[541]⊗ν5/2[642] |
| | 147.6+u | 9$^-$ | (148.0) | | | | 1992HO02 | 2. Regular band |
| | 229.7+u | 10$^-$ | | 229.3 | | | | 3. Signature splitting |
| | 428.3+u | 11$^-$ | 198.2 | 281.0 | | | | 4. Signature inversion occur at I$^\pi$ = 18$^-$ |
| | 548.8+u | 12$^-$ | | 319.2 | | | | 5. Level energy is adopted from XUNDL |
| | 787.5+u | 13$^-$ | 238.6 | 359.1 | | | | |
| | 964.7+u | 14$^-$ | | 416.1 | | | | |
| | 1230.0+u | 15$^-$ | 265.4 | 442.4 | | | | |
| | 1474.9+u | 16$^-$ | | 510.2 | | | | |
| | 1752.4+u | 17$^-$ | 277.5 | 522.4 | | | | |
| | 2070.1+u | 18$^-$ | | 595.2 | | | | |
| | 2350.2+u | 19$^-$ | 280.1 | 597.7 | | | | |
| | 2737.7+u | 20$^-$ | | 667.6 | | | | |
| | 3015.9+u | 21$^-$ | | 665.7 | | | | |
| | 3458.2+u | 22$^-$ | | 720.5 | | | | |
| | 3744.4+u | 23$^-$ | | 728.5 | | | | |
| | 4230.7+u | 24$^-$ | | 772.5 | | | | |

| S.No. | E_level keV | $I^\pi$ | $E_\gamma$(M1) keV | $E_\gamma$(E2) keV | $|g_K-g_R|$ | B(M1)/B(E2) $(\mu_N/eb)^2$ | Keywords | Configuration and Comments |
|---|---|---|---|---|---|---|---|---|
| 12 | 0.0+u | (8+) | | | | | **2000ZH51** | 1. $\pi 5/2[402]\otimes \nu 5/2[642]$ |
| | 130.6+u | 9+ | (130.3) | | | | | 2. Regular band |
| | 287.0+u | 10+ | 156.3 | (287.0) | | | | 3. Signature splitting |
| | 470.4+u | 11+ | 183.2 | 339.8 | | | | 4. Signature inversion occur at $I^\pi = 16^+$ |
| | 690.7+u | 12+ | 220.2 | 403.8 | | | | 5. Level energy is adopted from XUNDL |
| | 929.0+u | 13+ | 238.1 | 458.6 | | | | |
| | 1198.1+u | 14+ | 269.1 | 507.6 | | | | |
| | 1486.5+u | 15+ | 288.0 | 557.4 | | | | |
| | 1786.1+u | 16+ | 299.5 | 588.1 | | | | |
| | 2114.5+u | 17+ | (328.0) | 628.0 | | | | |
| 13 | 0.0+z | (6+) | | | | | **2000ZH51** | 1. $\pi 1/2[411]\otimes \nu 5/2[642]$ |
| | 139.3+z | 8+ | | 139.3 | | | | 2. Regular band |
| | 394.7+z | 10+ | | 255.4 | | | | 3. Level energy is adopted from XUNDL |
| | 764.1+z | 12+ | | 269.4 | | | | |
| | 1235.1+z | 14+ | | 471.0 | | | | |
| | 1792.2+z | 16+ | | 557.1 | | | | |
| | 2415.6+z | 18+ | | 623.4 | | | | |
| | 3085.6+z | 20+ | | 670.2 | | | | |
| | 3774.6+z | 22+ | | (689) | | | | |

$^{168}_{71}Lu_{97}$

| S.No. | E_level keV | $I^\pi$ | $E_\gamma$(M1) keV | $E_\gamma$(E2) keV | $|g_K-g_R|$ | B(M1)/B(E2) $(\mu_N/eb)^2$ | Keywords | Configuration and Comments |
|---|---|---|---|---|---|---|---|---|
| 1 | 0 | 6- | | | | | **2021RO11** | 1. $\pi 7/2[404]\otimes \nu 5/2[523]$ |
| | 146.3 | 7- | 146.3 | | | | 1972CH44 | 2. Nuclear reaction: |
| | 313.0 | 8- | 166.7 | 313.0 | | | | $^{123}$Sb($^{48}$Ca,3n$\gamma$)$^{168}$Lu |
| | 512.3 | 9- | 199.3 | 336.0 | | | | 3. Half-life of bandhead is 5.5(1) min (1972CH44) |
| | | | | | | | | 4. Regular band |
| | | | | | | | | 5. 0.5 KeV uncertainties in $E_\gamma$ |
| 2 | 220(130) | 3+ | | | | | 1972CH44 | 1. $\pi 1/2[541]\otimes \nu 5/2[523]$ |
| | | | | | | | | 2. Half-life of bandhead is 6.7(4) min |

| S.No. | $E_{level}$ keV | $I^\pi$ | $E_\gamma$(M1) keV | $E_\gamma$(E2) keV | $|g_K-g_R|$ | B(M1)/B(E2) $(\mu_N/eb)^2$ | Keywords | Configuration and Comments |
|---|---|---|---|---|---|---|---|---|
| 3 | 27.2 | $6^+$ | | | | | **2021RO11** | 1. $\pi7/2[404]\otimes\nu5/2[642]$ |
| | 91.6 | $7^+$ | 64.4(2) | | | | | 2. Regular band |
| | 170.8 | $8^+$ | 79.3(5) | 143.6(5) | | | | 3. Signature splitting more pronounced |
| | 284.8 | $9^+$ | 113.9(2) | 193.2(2) | | | | at higher spin |
| | 429.0 | $10^+$ | 144.2(2) | 258.2(2) | | | | 4. Band crossing from 2qp to 4qp occur |
| | 599.2 | $11^+$ | 170.2(2) | 314.4(2) | | | | at $\hbar\omega_c$ = 0.32 for $\alpha=\pm1/2$ and 4qp to |
| | 792.3 | $12^+$ | 193.0(2) | 363.2(2) | | | | 6qp occur at $\hbar\omega_c$= 0.54 for $\alpha$=+1/2 and |
| | 1007.4 | $13^+$ | 215.2(2) | 408.2(2) | | | | >0.54 for $\alpha$=-1/2 |
| | 1240.3 | $14^+$ | 232.8(2) | 447.9(2) | | | | |
| | 1493.3 | $15^+$ | 253.0(2) | 486.0(2) | | | | |
| | 1758.2 | $16^+$ | 264.9(2) | 517.8(2) | | | | |
| | 2043.0 | $17^+$ | 284.7(2) | 549.9(2) | | | | |
| | 2332.2 | $18^+$ | 289.2(2) | 573.7(2) | | | | |
| | 2642.9 | $19^+$ | 310.7(2) | 600.0(2) | | | | |
| | 2949.0 | $20^+$ | 306.2(2) | 616.8(2) | | | | |
| | 3281.5 | $21^+$ | 332.4(2) | 638.6(2) | | | | |
| | 3602.2 | $22^+$ | 320.7(2) | 653.2(2) | | | | |
| | 3955.8 | $23^+$ | 353.7(2) | 674.1(2) | | | | |
| | 4295.9 | $24^+$ | 340.0(2) | 694.4(2) | | | | |
| | 4674.4 | $25^+$ | 378.6(2) | 718.7(2) | | | | |
| | 5043.1 | $26^+$ | 368.7(2) | 747.2(2) | | | | |
| | 5449.3 | $27^+$ | 406.1(5) | 775.4(2) | | | | |
| | 5852.7 | $28^+$ | | 809.6(2) | | | | |
| | 6289.5 | $29^+$ | | 840.2(2) | | | | |
| | 6728.8 | $30^+$ | | 876.1(2) | | | | |
| | 7197.2 | $31^+$ | | 907.7(2) | | | | |
| | 7671.3 | $32^+$ | | 942.4(2) | | | | |
| | 8171.2 | $33^+$ | | 974.0(2) | | | | |
| | 8676.6 | $34^+$ | | 1005.3(2) | | | | |
| | 9215.9 | $35^+$ | | 1044.7(2) | | | | |
| | 9736.2 | $36^+$ | | 1059.6(2) | | | | |
| | 10311.2 | $(37^+)$ | | 1095.3(5) | | | | |
| | 10818.2 | $(38^+)$ | | 1082.0(5) | | | | |
| | 11457.8 | $(39^+)$ | | 1146.6(5) | | | | |
| | 11901.2 | $(40^+)$ | | 1083.0(5) | | | | |
| | 12637.9 | $(41^+)$ | | 1180.1(5) | | | | |
| | 13030.2 | $(42^+)$ | | 1129.0(5) | | | | |
| | 14206.2 | $(44^+)$ | | 1176.0(5) | | | | |

| S.No. | E$_{level}$ keV | I$^\pi$ | E$_\gamma$(M1) keV | E$_\gamma$(E2) keV | \|g$_K$-g$_R$\| | B(M1)/B(E2) ($\mu_N$/eb)$^2$ | Keywords | Configuration and Comments |
|---|---|---|---|---|---|---|---|---|
| 4 | 192.1 | 6$^-$ | | | | | **2021RO11** | 1. $\pi$1/2[541]$\otimes\nu$5/2[642] |
| | 273.2 | 7$^-$ | 81.1(5) | | | | | 2. Regular nature upto I$^\pi$ = 26$^-$ |
| | 336.5 | 8$^-$ | 63.1(5) | 144.4(2) | | | | 3. Signature splitting |
| | 467.6 | 9$^-$ | 131.0(2) | 194.4(2) | | | | 4. Signature inversion occur at I$^\pi$ = 22$^-$ |
| | 554.7 | 10$^-$ | 87.1(5) | 218.0(2) | | | | 5. Band crossing from 2qp to 4qp occur |
| | 743.3 | 11$^-$ | 188.6(2) | 276.0(2) | | | | at ℏ$\omega_c$ = 0.34 for $\alpha$=+1/2 and 0.44 for |
| | 858.2 | 12$^-$ | 114.9(2) | 303.2(2) | | | | $\alpha$=-1/2 |
| | 1102.5 | 13$^-$ | 244.3(2) | 358.9(2) | | | | |
| | 1251.2 | 14$^-$ | 1488(2) | 393.1(2) | | | | |
| | 1541.3 | 15$^-$ | 290.0(2) | 438.8(2) | | | | |
| | 1733.0 | 16$^-$ | 191.7(2) | 481.7(2) | | | | |
| | 2054.1 | 17$^-$ | 321.2(2) | 513.1(2) | | | | |
| | 2297.8 | 18$^-$ | | 564.3(2) | | | | |
| | 2640.6 | 19$^-$ | 342.8(2) | 586.0(2) | | | | |
| | 2933.4 | 20$^-$ | | 636.4(2) | | | | |
| | 3282.1 | 21$^-$ | 348.8(2) | 641.6(2) | | | | |
| | 3642.3 | 22$^-$ | | 709.1(2) | | | | |
| | 3958.9 | 23$^-$ | 368.7(2) | 676.8(2) | | | | |
| | 4406.0 | 24$^-$ | | 763.7(2) | | | | |
| | 4666.4 | 25$^-$ | | 707.5(2) | | | | |
| | 5222.0 | 26$^-$ | | 816.0(2) | | | | |
| | 5426.8 | 27$^-$ | | 760.4(2) | | | | |
| | 6085.1 | 28$^-$ | | 863.1(2) | | | | |
| | 6250.3 | 29$^-$ | | 823.5(2) | | | | |
| | 6992.1 | 30$^-$ | | 907.0(2) | | | | |
| | 7146.9 | (31$^-$) | | 896.6(2) | | | | |
| | 7946.4 | 32$^-$ | | 954.3(2) | | | | |
| | 8117.4 | (33$^-$) | | 970.4(5) | | | | |
| | 8953.5 | (34$^-$) | | 1007.1(5) | | | | |
| | 9158.7 | (35$^-$) | | 1041.3(5) | | | | |
| | 10025.9 | (36$^-$) | | 1072.4(5) | | | | |
| | 10262.1 | (37$^-$) | | 1103.4(5) | | | | |

| S.No. | E$_{level}$ keV | I$^\pi$ | E$_\gamma$(M1) keV | E$_\gamma$(E2) keV | \|g$_K$-g$_R$\| | B(M1)/B(E2) ($\mu_N$/eb)$^2$ | Keywords | Configuration and Comments |
|---|---|---|---|---|---|---|---|---|
| 5 | 489.0 | 9$^-$ | | | | | **2021RO11** | 1. $\pi 9/2[514] \otimes \nu 5/2[642]$ |
| | 600.0 | 10$^-$ | 111.0(2) | | | | | 2. Regular band |
| | 730.5 | 11$^-$ | 130.4(2) | 241.4(2) | | | | 3. Signature splitting |
| | 891.5 | 12$^-$ | 161.0(2) | 291.4(2) | | | | 4. Band crossing from 2qp to 4 qp occur |
| | 1075.8 | 13$^-$ | 184.3(2) | 345.0(2) | | | | at $\hbar\omega_c$ = 0.32 for $\alpha=\pm 1/2$ and from 4qp |
| | 1287.0 | 14$^-$ | 211.2(2) | 395.2(2) | | | | to 6qp occur at $\hbar\omega_c \approx 0.63$ |
| | 1517.5 | 15$^-$ | 230.4(2) | 441.8(2) | | | | |
| | 1769.5 | 16$^-$ | 252.0(2) | 482.4(2) | | | | |
| | 2040.2 | 17$^-$ | 270.7(2) | 522.7(2) | | | | |
| | 2324.8 | 18$^-$ | 284.6(2) | 555.4(2) | | | | |
| | 2622.0 | 19$^-$ | 297.2(2) | 581.8(2) | | | | |
| | 2939.4 | 20$^-$ | | 614.6(2) | | | | |
| | 3256.4 | 21$^-$ | 317.0(2) | 634.4(2) | | | | |
| | 3590.2 | 22$^-$ | 333.8(2) | 650.8(2) | | | | |
| | 3941.0 | 23$^-$ | | 684.6(2) | | | | |
| | 4277.9 | 24$^-$ | 336.9(2) | 687.8(2) | | | | |
| | 4674.6 | 25$^-$ | 396.7(5) | 733.6(2) | | | | |
| | 5003.7 | 26$^-$ | | 725.8(2) | | | | |
| | 5451.2 | 27$^-$ | | 776.6(2) | | | | |
| | 5779.9 | 28$^-$ | | 776.2(2) | | | | |
| | 6278.1 | 29$^-$ | | 826.9(2) | | | | |
| | 6613.8 | 30$^-$ | | 833.9(2) | | | | |
| | 7158.6 | 31$^-$ | | 880.4(2) | | | | |
| | 7506.3 | 32$^-$ | | 892.4(2) | | | | |
| | 8095.4 | 33$^-$ | | 936.8(2) | | | | |
| | 8456.5 | 34$^-$ | | 950.2(2) | | | | |
| | 9092.8 | 35$^-$ | | 997.4(2) | | | | |
| | 9463.0 | 36$^-$ | | 1006.4(2) | | | | |
| | 10150.8 | 37$^-$ | | 1058.0(2) | | | | |
| | 10524.2 | 38$^-$ | | 1061.2(2) | | | | |
| | 11267.8 | 39$^-$ | | 1116.8(2) | | | | |
| | 11638.1 | 40$^-$ | | 1113.9(2) | | | | |
| | 12438.8 | 41$^-$ | | 1171.2(5) | | | | |
| | 12799.5 | 42$^-$ | | 1161.4(2) | | | | |
| | 13659.6 | 43$^-$ | | 1220.8(5) | | | | |
| | 14008.5 | 44$^-$ | | 1209.0(5) | | | | |
| | 15257.6 | 46$^-$ | | 1250.3(5) | | | | |
| | 16548.3 | 48$^-$ | | 1287.7(5) | | | | |
| | 17881.8 | 50$^-$ | | 1333.5(5) | | | | |

| S.No. | E$_{level}$ keV | I$^\pi$ | E$_\gamma$(M1) keV | E$_\gamma$(E2) keV | \|g$_K$-g$_R$\| | B(M1)/B(E2) ($\mu_N$/eb)$^2$ | Keywords | Configuration and Comments |
|---|---|---|---|---|---|---|---|---|
| 6 | 590.1 | 8$^-$ | | | | | **2021RO11** | 1. π1/2[541]⊗ν5/2[642] |
| | 886.8 | 10$^-$ | | 296.7(2) | | | | 2. Regular band |
| | 1245.3 | 12$^-$ | | 358.4(2) | | | | 3. Signature splitting |
| | 1669.7 | 14$^-$ | | 424.4(2) | | | | 4. Signature inversion occur at I$^\pi$ = (23$^-$) |
| | 2161.3 | 16$^-$ | | 491.6(2) | | | | 5. Band crossing from 2qp to 4 qp occur |
| | 2314.9 | (17$^-$) | | 581.9(2) | | | | at ℏω$_c$ = 0.35 for α=+1/2 and 0.45 for |
| | 2716.2 | 18$^-$ | | 554.9(2) | | | | α=-1/2 |
| | 2949.4 | (19$^-$) | | 634.4(2) | | | | |
| | 3331.4 | 20$^-$ | | 615.2(2) | | | | |
| | 3643.3 | (21$^-$) | | 693.9(2) | | | | |
| | 3997.4 | 22$^-$ | | 666.0(2) | | | | |
| | 4397.6 | (23$^-$) | | 754.3(2) | | | | |
| | 4708.2 | 24$^-$ | | 710.8(2) | | | | |
| | 5196.9 | (25$^-$) | | 799.3(2) | | | | |
| | 5466.4 | 26$^-$ | | 758.2(2) | | | | |
| | 6034.1 | (27$^-$) | | 837.2(2) | | | | |
| | 6275.1 | 28$^-$ | | 808.7(2) | | | | |
| | 6920.3 | (29$^-$) | | 886.2(5) | | | | |
| | 7137.9 | 30$^-$ | | 862.8(2) | | | | |
| | 7859.1 | (31$^-$) | | 938.8(5) | | | | |
| | 8048.5 | 32$^-$ | | 910.6(5) | | | | |
| | 8853.5 | (33$^-$) | | 994.4(5) | | | | |
| | 9001.5 | (34$^-$) | | 953.0(5) | | | | |
| | | | | | | | | |
| 7 | x | (6$^+$) | | | | | **2021RO11** | 1. π1/2[411]⊗ν5/2[642] |
| | 94.8+x | (7$^+$) | 94.8(2) | | | | | 2. Regular band |
| | 220.8+x | (8$^+$) | 126.0(2) | 220.8(5) | | | | 3. Signature splitting |
| | 374.1+x | (9$^+$) | 153.3(2) | 279.3(2) | | | | 4. Band crossing from 2qp to 4qp occur |
| | 547.3+x | (10$^+$) | 173.2(2) | 326.4(2) | | | | at ℏω$_c$ = 0.30 for α=±1/2 and 4qp to |
| | 751.6+x | (11$^+$) | 204.3(2) | 377.5(2) | | | | 6qp occur at ℏω$_c$= 0.54 for α=+1/2 and |
| | 956.4+x | (12$^+$) | 204.8(2) | 409.7(2) | | | | >0.53 for α=-1/2 |
| | 1212.1+x | (13$^+$) | 255.7(2) | 460.4(2) | | | | |
| | 1433.5+x | (14$^+$) | 221.4(2) | 477.1(2) | | | | |
| | 1739.4+x | (15$^+$) | 305.9(2) | 527.3(2) | | | | |
| | 1968.3+x | (16$^+$) | 228.9(2) | 534.8(2) | | | | |
| | 2318.4+x | (17$^+$) | | 579.0(2) | | | | |
| | 2549.2+x | (18$^+$) | | 580.9(2) | | | | |
| | 2928.6+x | (19$^+$) | | 610.1(2) | | | | |
| | 3169.6+x | (20$^+$) | | 620.4(2) | | | | |
| | 3576.6+x | (21$^+$) | | 648.0(2) | | | | |
| | 3833.0+x | (22$^+$) | | 663.4(2) | | | | |
| | 4270.3+x | (23$^+$) | | 693.8(2) | | | | |
| | 4550.2+x | (24$^+$) | | 717.2(2) | | | | |
| | 5020.9+x | (25$^+$) | | 750.5(2) | | | | |
| | 5330.4+x | (26$^+$) | | 780.2(2) | | | | |
| | 5834.9+x | (27$^+$) | | 814.1(2) | | | | |
| | 6176.8+x | (28$^+$) | | 846.4(2) | | | | |
| | 6715.3+x | (29$^+$) | | 880.3(2) | | | | |
| | 7085.8+x | (30$^+$) | | 909.0(2) | | | | |
| | 7653.8+x | (31$^+$) | | 938.5(5) | | | | |
| | 8078.4+x | (32$^+$) | | 992.6(5) | | | | |
| | 8668.3+x | (33$^+$) | | 1014.5(5) | | | | |
| | 9096.0+x | (34$^+$) | | 1017.6(5) | | | | |
| | 9728.8+x | (35$^+$) | | 1060.4(5) | | | | |
| | 10170.0+x | (36$^+$) | | 1074.0(5) | | | | |
| | 10828.8+x | (37$^+$) | | 1100.0(5) | | | | |

| S.No. | E$_{level}$ keV | I$^\pi$ | E$_\gamma$(M1) keV | E$_\gamma$(E2) keV | \|g$_K$-g$_R$\| | B(M1)/B(E2) ($\mu_N$/eb)$^2$ | Keywords | Configuration and Comments |
|---|---|---|---|---|---|---|---|---|
| 8 | y | (8$^+$) | | | | | 2021RO11 | 1. $\pi$5/2[402]$\otimes\nu$5/2[642] |
| | 156.5+y | (9$^+$) | 156.5(5) | | | | | 2. Regular band |
| | 330.0+y | (10$^+$) | 173.5(5) | 330.4(5) | | | | 3. Small signature splitting in $\Delta$E$_\gamma$ vs I |
| | 534.5+y | (11$^+$) | 204.4(5) | 378.0(5) | | | | 4. Signature inversion occur at I$^\pi$ = (17$^+$) |
| | 750.0+y | (12$^+$) | 215.5(5) | 419.2(5) | | | | 5. Band crossing from 2qp to 4qp occur |
| | 1002.0+y | (13$^+$) | 252.0(5) | 467.5(5) | | | | at $\hbar\omega_c$ = 0.28 for $\alpha$=$\pm$1/2 and 4qp to |
| | 1267.0+y | (14$^+$) | 265.0(2) | 517.2(2) | | | | 6qp occur at $\hbar\omega_c$= 0.53 for $\alpha$=+1/2 and |
| | 1556.0+y | (15$^+$) | 289.0(5) | 553.9(5) | | | | >0.54 for $\alpha$=-1/2 |
| | 1852.0+y | (16$^+$) | 296.0(2) | 585.1(2) | | | | |
| | 2141.0+y | (17$^+$) | 289.0(2) | 585.1(2) | | | | |
| | 2441.8+y | (18$^+$) | | 589.8(2) | | | | |
| | 2741.3+y | (19$^+$) | | 600.3(2) | | | | |
| | 3067.1+y | (20$^+$) | | 625.3(2) | | | | |
| | 3390.4+y | (21$^+$) | | 649.1(2) | | | | |
| | 3747.0+y | (22$^+$) | | 679.9(2) | | | | |
| | 4095.2+y | (23$^+$) | | 704.8(2) | | | | |
| | 4481.9+y | (24$^+$) | | 734.9(2) | | | | |
| | 4855.2+y | (25$^+$) | | 760.0(2) | | | | |
| | 5270.5+y | (26$^+$) | | 789.7(2) | | | | |
| | 5671.0+y | (27$^+$) | | 815.8(2) | | | | |
| | 6115.2+y | (28$^+$) | | 844.7(5) | | | | |
| | 6547.7+y | (29$^+$) | | 876.7(2) | | | | |
| | 7021.3+y | (30$^+$) | | 906.1(5) | | | | |
| | 7488.8+y | (31$^+$) | | 941.1(2) | | | | |
| | 7992.5+y | (32$^+$) | | 971.2(5) | | | | |
| | 8495.8+y | (33$^+$) | | 1006.2(2) | | | | |
| | 9022.6+y | (34$^+$) | | 1030.1(5) | | | | |
| | 9541.4+y | (35$^+$) | | 1045.0(5) | | | | |
| | 10094.4+y | (36$^+$) | | 1071.8(5) | | | | |
| | 10586.4+y | (37$^+$) | | 1045.0(5) | | | | |
| | 11198.4+y | (38$^+$) | | 1103.3(5) | | | | |
| | 11659.0+y | (39$^+$) | | 1072.6(5) | | | | |
| | 12339.4+y | (40$^+$) | | 1145.8(5) | | | | |
| | 13523.3+y | (42$^+$) | | 1183.9(5) | | | | |